\documentclass[letterpaper,12pt,]{article}
\usepackage{epsfig}
\usepackage{amssymb}
\usepackage{amsfonts}
\usepackage{amsmath}
\usepackage{color}
\usepackage[stable]{footmisc}
\usepackage{bm}

\usepackage{tikz}
\usepackage{pgf}
\usepackage{xxcolor}

\newskip\humongous \humongous=0pt plus 1000pt minus 1000pt

\newif\ifdtup

\newcommand{\be}{\begin{equation}}
\newcommand{\ee}{\end{equation}}
\newcommand{\beal}{\begin{equation}\begin{aligned}}
\newcommand{\eeal}{\end{aligned}\end{equation}}
\newcommand{\nn}{\nonumber}
\newcommand{\mat}[1]{\begin{pmatrix} #1 \end{pmatrix}}
\newcommand{\parfrac}[2]{\frac{\partial #1}{\partial #2}}

\newcommand{\du}[2]{_{#1}^{\phantom{#1}#2}}
\newcommand{\trans}{\mathsf{T}}
\newcommand{\rep}[1]{\ensuremath{\bm{#1}}}
\newcommand{\anti}[1]{\ensuremath{\overline{\text{#1}}}}
\newcommand{\ie}{{\it i.e.}}
\newcommand{\eg}{{\it e.g.}}
\newcommand{\etc}{{\it etc}}
\newcommand{\cA}{\mathcal{A}}
\newcommand{\cB}{\mathcal{B}}
\newcommand{\cC}{\mathcal{C}}
\newcommand{\cF}{\mathcal{F}}
\newcommand{\cJ}{\mathcal{J}}
\newcommand{\cL}{\mathcal{L}}
\newcommand{\cM}{\mathcal{M}}
\newcommand{\cN}{\mathcal{N}}
\newcommand{\cO}{\mathcal{O}}
\newcommand{\cP}{\mathcal{P}}

\newcommand{\bC}{\mathbb{C}}
\newcommand{\bP}{\mathbb{P}}
\newcommand{\bR}{\mathbb{R}}
\newcommand{\bZ}{\mathbb{Z}}
\newcommand{\so}{\mathfrak{so}}

\newcommand{\Vol}{\mathrm{Vol}}

\DeclareMathOperator{\sign}{sign}
\DeclareMathOperator{\Tr}{Tr}
\DeclareMathOperator{\rank}{rank}

\catcode`\@=11

\@addtoreset{equation}{section}

\def\@normalsize{\@setsize\normalsize{15pt}\xiipt\@xiipt
\abovedisplayskip 14pt plus3pt minus3pt%
\belowdisplayskip \abovedisplayskip
\abovedisplayshortskip \z@ plus3pt%
\belowdisplayshortskip 7pt plus3.5pt minus0pt}

\def\small{\@setsize\small{13.6pt}\xipt\@xipt
\abovedisplayskip 13pt plus3pt minus3pt%
\belowdisplayskip \abovedisplayskip
\abovedisplayshortskip \z@ plus3pt%
\belowdisplayshortskip 7pt plus3.5pt minus0pt
\def\@listi{\parsep 4.5pt plus 2pt minus 1pt
     \itemsep \parsep
     \topsep 9pt plus 3pt minus 3pt}}

\relax

\catcode`@=12

\def\SymBoxes#1#2#3#4{\newdimen\un@t \un@t#3%
\raisebox{#1}{\rule{#2\un@t}{#4}\hskip-#2\un@t% lower horizontal
\@tempdimb\un@t \advance\@tempdimb by-#4\@tempcntb#2\relax%
\@whilenum{\@tempcntb>0}\do{%                         % #2 vertical lines
\rule{#4}{\un@t}\hskip\@tempdimb \advance\@tempcntb by\m@ne}%
\hskip-#2\un@t \rule[\un@t]{#2\un@t}{#4}%
\rule[\un@t]{#4}{#4}\hskip-#4%             % upper horizontal line
\rule{#4}{\un@t}}\hskip-#4}                % rightest vertical line
\renewenvironment{thebibliography}[1]{%
\begin{oldthebibliography}{#1}%
\setlength\itemsep{-.8mm}%
}%
{%
\end{oldthebibliography}%
}

\begin{document}

%%%%%%
\newcommand{\beq}{\begin{equation}}
\newcommand{\eeq}{\end{equation}}
\newcommand{\bea}{\begin{eqnarray}}
\newcommand{\eea}{\end{eqnarray}}
\newcommand{\beas}{\begin{eqnarray*}}
\newcommand{\eeas}{\end{eqnarray*}}
\newcommand{\defi}{\stackrel{\rm def}{=}}
\newcommand{\non}{\nonumber}
\newcommand{\bquo}{\begin{quote}}
\newcommand{\enqu}{\end{quote}}
%%%%%%%%%%%%%%%%
\renewcommand{\(}{\begin{equation}}
\renewcommand{\)}{\end{equation}}
%%%%%%%%%%%%%%%%%%%%%%%%%%%%%%%%%% definitions
\def\IZ{{\mathbb Z}}
\def\IR{{\mathbb R}}
\def\IC{{\mathbb C}}
\def\IQ{{\mathbb Q}}

\def\red{\textcolor{red}}

\def\CM{{\mathcal{M}}}
\def\dCM{{\left \vert\mathcal{M}\right\vert}}

\def \d{\textrm{d}}
\def \p{\partial}

\def \w2{{\omega_2}}

\def \Pf{\rm Pf}

\def \pr{\prime}

\def\half{\frac{1}{2}}

\def \eqn#1#2{\begin{equation}#2\label{#1}\end{equation}}
\def\de{\partial}

\def\H{ \hbox{\rm H}}
\def\HE{ \hbox{$\rm H^{even}$}}
\def\HO{ \hbox{$\rm H^{odd}$}}
\def\K{ \hbox{\rm K}}
\def\Im{ \hbox{\rm Im}}
\def\Ker{ \hbox{\rm Ker}}
\def\const{\hbox {\rm const.}}
\def\o{\over}
\def\im{\hbox{\rm Im}}
\def\re{\hbox{\rm Re}}
\def\bra{\langle}\def\ket{\rangle}
\def\Arg{\hbox {\rm Arg}}
\def\Re{\hbox {\rm Re}}
\def\Im{\hbox {\rm Im}}
\def\exo{\hbox {\rm exp}}
\def\diag{\hbox{\rm diag}}
\def\longvert{{\rule[-2mm]{0.1mm}{7mm}}\,}
\def\a{{\textsl a}}
\def\tq{{\widetilde q}}
\def\p{{}^{\prime}}
\def\W{W}
\def\N{{\cal N}}
\def\hsp{,\hspace{.7cm}}
\newcommand{\C}{\ensuremath{\mathbb C}}
\newcommand{\Z}{\ensuremath{\mathbb Z}}
\newcommand{\R}{\ensuremath{\mathbb R}}
\newcommand{\rp}{\ensuremath{\mathbb {RP}}}
\newcommand{\cp}{\ensuremath{\mathbb {CP}}}
\newcommand{\vac}{\ensuremath{|0\rangle}}
\newcommand{\vact}{\ensuremath{|00\rangle}}
\newcommand{\oc}{\ensuremath{\overline{c}}}

\def\S{\ensuremath{\mathbb S}}
\def\Dt{\rm D3}
\def\Df{\rm D5}
\def\Ds{\rm D7}

\def\M{\mathcal{M}}
\def\F{\mathcal{F}}
\def\d{\textrm{d}}

\def\A{\mathcal{A}}
\def\AI{\mathcal{A}_{\rm I}}
\def\AII{\mathcal{A}_{\rm II}}
\def\FI{F_{\rm I}}
\def\FII{F_{\rm II}}

\def\eps{\epsilon}

\def\smn{$SU(M+N)\times SU(N)\ $}

\def\CM{{\cal M}}
\def\CN{{\cal N}}
\def\CB{{\cal B}}
\def\CC{{\cal C}}
\def\IL{\relax{\rm I\kern-.18em L}}
\def\IH{\relax{\rm I\kern-.18em H}}
\def\IP{\relax{\rm I\kern-.18em P}}
\def\IR{\relax{\rm I\kern-.18em R}}
\def\IS{\relax{\rm \kern-.18em S}}
\def\IC{\relax{\rm I\kern-.56em C}}

\def\mathbbR{{\IR}}
\def\mathbbS{{\IS}}

\def\unit{\relax{\rm 1\kern-.26em I}}

\begin{titlepage}

\begin{flushright}
PU-2389
\end{flushright}

\vspace{18pt}
\begin{center}
{\Large \bf
Comments on the $\bm{\cN=1}$ $\bm{SU(M+p) \times SU(p)}$ \\
quiver gauge theory with flavor}
\end{center}

\vspace{1cm}
\begin{center}
{\large {\bf Francesco Benini}$^a$ and {\bf Anatoly Dymarsky}$^b$}

\vspace{25pt}
\textit{$^a$ Department of Physics, Princeton University,\\ Princeton, NJ 08544, USA}\\
\vspace{6pt}
\textit{$^b$ School of Natural Sciences, Institute for Advanced Study,\\ Princeton, NJ, 08540}\\ \vspace{6pt}
\end{center}

\vspace{1cm}

\begin{center}
\textbf{Abstract}
\end{center}

\noindent
We study supersymmetric vacua of the ${\mathcal N}=1$ cascading ${SU(M+p) \times SU(p)}$ gauge theory  with flavor -- the theory on $p$ D3-branes and $M$ wrapped D5-branes at the tip of the conifold,
and $N_f$ flavor D7-branes wrapping a holomorphic four-cycle inside the conifold. The Coulomb branch of the moduli space is inherited from the pure gauge theory without flavor and was thoroughly studied in the past. Besides, there is a Higgs branch where some D3 and/or D5-branes dissolve in the D7-branes forming the worldvolume gauge instantons.
We study the Higgs branch both from the field theory and the bulk point of view. On the classical level the moduli space is closely related to the one of the $\N=2$ $\bC^2/\Z_2$ orbifold theory, in particular certain vacua of the $\N=1$ theory are related to noncommutative instantons on the resolved $\bC^2/\Z_2$.
On the quantum level the Higgs branch acquires corrections due to renormalization of the K\"ahler potential and non-perturbative effects in field theory. In the bulk this is encoded
in the classical D7-brane geometry. We compute the VEVs of the protected operators and the field theory RG flow and find an agreement with the parallel computations in the bulk.

\end{titlepage}

\renewcommand{\thefootnote}{\arabic{footnote}}

{\small \setlength\parskip{-.5mm} \tableofcontents}

\section{Introduction}

The low energy theory on D3-branes at a conifold singularity, studied by Klebanov and Witten (KW) in \cite{KW}, has attracted significant attention during the last decade. This is the $\cN=1$ $SU(N) \times SU(N)$ gauge theory with bifundamental fields and a superpotential. Although the theory is strongly coupled, it has a simple gravity dual in the sense of the AdS/CFT correspondence \cite{Maldacena, GKPol, Witten98}. Thanks to the simplicity of the original setup and a multitude of possible variations, the KW theory has become a standard arena to study different theoretical and phenomenological phenomena. Let us mention here some of the main features that can be easily engineered within the conifold. The basic KW theory \cite{KW} is conformal. After the gauge group is modified to $SU(N+M) \times SU(N)$, the resulting Klebanov-Strassler (KS) theory exhibits a rich dynamics \cite{KG, KN, KT, KS}. There is a logarithmic RG flow, which is UV complete without a UV fixed point, and the theory has a chiral anomaly \cite{Klebanov:2002gr}. The flow takes place through a ``cascade'' \cite{KS} of Seiberg dualities \cite{Seiberg:1994pq} with many effective descriptions at different scales.
At low energy there is spontaneous chiral symmetry breaking and confinement. The theory asymptotes to pure $SU(M)$ Super-Yang-Mills (SYM) in the IR \cite{KS}. The theory is conjectured to have a meta-stable vacuum that dynamically breaks supersymmetry at an exponentially low scale \cite{Kachru:2002gs}, which might be relevant for phenomenological models \cite{Benini:2009ff}. Moreover this theory is a natural setup for models of cosmological inflation \cite{Baumann:2007np,Baumann:2007ah}.
Adding ``flavors'', \ie{}  fields in the (anti)fundamental representation, modifies the theory such that it asymptotes in the IR to Super-QCD (SQCD) with quartic superpotential, and the moduli space develops a Higgs branch. One can study the Veneziano limit of this theory \cite{Benini:2006hh, Benini:2007gx} which exhibits confinement and screening of charges \cite{Bigazzi:2008zt} in the IR, and a ``duality wall'' in the UV \cite{Benini:2007gx}. The gravity dual setup admits modes with localized wave-functions \cite{Acharya:2006mx} which might be relevant for the Randall-Sundrum scenario \cite{Randall:1999ee, Gherghetta:2000qt}.
This list can go on and on.

In this paper we focus on the flavored theory. In particular we study supersymmetric vacua of the cascading ${SU(N+M) \times SU(N)}$ theory with flavor -- the theory on $N$ D3-branes, $M$ fractional D3-branes (wrapped D5s) and $N_f$ flavor D7-branes inside the conifold, building on a similar analysis of the unflavored case \cite{DKS}. To make use of the known holographic dual to the pure gauge theory,  we keep the number of flavors $N_f$ much smaller (though possibly large) than the number of colors $N+M$ throughout the paper. We focus on the vacua that are directly related to the presence of flavor fields -- the Higgs branch of the moduli space -- which we study using conventional field theory tools as well as the dual holographic description. The $\N=1$ supersymmetry is not enough to protect the Higgs branch from quantum corrections. Although the general structure of the classical moduli space stays intact, particular properties of vacua, such as VEVs of various observables, get quantum corrections. The main goal of this paper is to perform a thorough analysis of Higgs vacua including quantum effects on both sides of the duality and demonstrate how non-perturbative effects in field theory are manifest through the classical geometry in the bulk.

The general structure of the moduli space is clear from the bulk point of view. If the background has mobile D3-branes, there is a Coulomb branch associated with their motion on the conifold. When some D3-branes reach  the D7s, they can dissolve  into worldvolume non-Abelian gauge ``instantons'' \cite{DouglasBWB} with moduli that correspond to the Higgs branch. Besides there could be a (pseudo)-K\"ahler deformation of the conifold metric dual to the baryonic branch of the moduli space. There are also disconnected branches. For instance we can create extra D5 and D3-charge by putting worldvolume flux at the tip of the D7s (to preserve the total charge one would need to adjust the flux at the conifold tip). Also, the mobile D3-branes can turn into Ramond-Ramond (RR) 5-form flux at the price of ``shortening the throat''. We will find that all such configurations have counterpart vacua in field theory.

Our main interest is in the vacua associated with flavors, \ie{} the D7-branes in the bulk.
The D7-branes we consider wrap a holomorphic four-cycle $\Sigma$ which has the same topology and complex structure as the Eguchi-Hanson space, albeit with a non-conventional non-Ricci-flat metric. As outlined above, the Higgs branch(es) are dual to non-trivial supersymmetric worldvolume gauge configurations on the D7s. In many cases we get conventional instantons, \ie{} anti-self-dual gauge field configurations. These instantons on $\Sigma$ bear close resemblance with the conventional instantons on  $\bC^2/\bZ_2$ as the two spaces coincide as complex manifolds, and so should coincide the Higgs branches of the two theories in most cases.
The relation can be seen in field theory: any Higgs branch solution to the $\cN=2$ F- and D-term equations is also a solution to the $\cN=1$ classical vacuum equations. However in certain cases the $\N=1$ supersymmetric gauge configurations satisfy non-linear equations \cite{D7bb}. Such ``instantons'' are not anti-self-dual and a priori we can not say much about their moduli space. Using the dual field theory we show that these ``instantons'' are related to the noncommutative instantons on $\bC^2/\bZ_2$.

Although to explicitly find non-Abelian instantons in the $\cN=1$ case is a difficult task, we find the explicit solutions for the Abelian $U(1)$ instantons together with the corresponding classical and quantum vacua in the dual field theory -- in particular we solve the ADHM equations of the $\N=2$ $\bC^2/\bZ_2$ orbifold theory. Thus we completely cover the case of the field theories with $N_f=1$. This allows us to compute the quantum corrections in field theory and compare the results with gravity. For generic $N_f$ the non-Abelian instantons will emerge but we do not expect this to introduce any qualitatively new feature.

The RG flow of the theory, except in very special situations, is controlled by a cascade of Seiberg dualities \cite{Ouyang:2003df, Benini:2007gx}, in a similar but more articulated way than in the unflavored KS case \cite{KS, Strassler:2005qs}.%
\footnote{Also the $\cN=2$ $\bC^2/\bZ_2$ orbifold theory admits cascading RG flows \cite{Polchinski, Aharony, Argurio:2008mt, Benini:2008ir, Simic:2010ra} although the physics is different than in the $\cN=1$ case.}
A new feature is that, as we change the effective description at each step of the cascade, we also get a non-trivial map between the various Higgs branches of the moduli space. The Seiberg duality on gravity side is manifest through the large gauge transformation of the B-field which nicely reproduces the map between the vacua.

Finally we consider the fully backreacted supergravity solutions for smeared (possibly massive and with worldvolume flux) D7-branes on the conifold \cite{Benini:2006hh, Benini:2007gx, Bigazzi:2008zt, Bigazzi:2008qq}. We exploit such solutions to study the RG flow and show that gravity exactly reproduces the field theory NSVZ $\beta$-functions \cite{Novikov:1983uc} in all vacua.

The paper is organized as follows. In section \ref{sec: review} we review the conformal KW and the cascading KS theories without flavor, gaining enough familiarity to be ready to add probe flavor D7-branes in section \ref{sec: adding d7}. First we digress to consider the $\cN=2$ $\bC^2/\bZ_2$ orbifold theory, which gives us basic intuition about the moduli space of instantons. Then we move to study the D7-branes inside the conifold, and find the solutions to the linearized perturbations of the worldvolume gauge fields thus building the AdS/CFT dictionary in the sense of \cite{GKPol, Witten98}. Finally we explicitly construct the $U(1)$ instantons for D7-branes in all SUSY vacua of the KW/KS theories and calculate the VEVs of the protected operators from the flavor sector. In section \ref{sec: backreaction d7} we go beyond the probe approximation and compute the backreaction of the flavor branes on the geometry in the Veneziano large $N$ limit, with $N_f/N$ small but fixed. To solve the equations we place the D7s in a way that preserves the isometries of the conifold. We read off the RG flow, corrected by the flavors, to compare with the dual gauge theory later in section \ref{sec: comparison}. This ends the gravity analysis. In section \ref{sec: field theory analysis} we study the moduli space using $\N=1$ field theory techniques. We first consider the action of Seiberg duality, then we perform a classical analysis of the moduli space and eventually we include quantum effects. Finally section \ref{sec: comparison} is devoted to the comparison between gravity and field theory results. We draw our conclusions in section \ref{sec: conclusions}. Various computations are exiled to appendices.

\section{Review: pure \smn theory }
\label{sec: review}

This section is a review of the unflavored conifold theories and might be skipped by a reader familiar with  the subject. A thorough discussion of these theories can be found in \cite{Remarks,LH,DKS}.

\subsection{Review of the KW theory}
\label{reviewkw}

Following \cite{KW} we start by placing a stack of $N$ \Dt-branes at the tip of the conical singularity
\be
\label{conifold}
\sum\nolimits_{i=1}^4 z_i^2=0 \;.
\ee
The resulting field theory on the \Dt-branes is an ${\mathcal N}=1$ superconformal quiver gauge theory with gauge group $SU(N)\times SU(N)$ and global symmetry $SU(2)_A\times SU(2)_B\times U(1)_R\times U(1)_\text{baryon}$.
Besides the vector multiplets there are bifundamental fields $A_\alpha,B_{\dot\alpha}$ in the $(\rep{N}, \rep{\overline{N}})$ and $(\rep{\overline{N}},\rep{N})$ representations with charges $(\rep{2},\rep{1}, \frac12 ,1)$ and $(\rep{1},\rep{2}, \frac12,-1)$ under the global symmetry group, and a superpotential
\be
W_{KW} = \frac12 \, h \, \epsilon^{\alpha\beta}\epsilon^{\dot\alpha\dot\beta} \Tr A_\alpha B_{\dot\alpha} A_\beta B_{\dot\beta} \;.
\ee
At the conformal point the theory is always strongly coupled, and the conformal manifold is described by $h(g_1,g_2)$ \cite{Strassler:2005qs}.

The moduli space can be found from the F- and D-flatness conditions.
The former implies the matrix equation
\be
\label{Fflatness}
\epsilon^{\alpha\beta} \epsilon^{\dot\alpha\dot\beta} A_\alpha B_{\dot\alpha}A_\beta B_{\dot\beta} = 0 \;.
\ee
It is convenient to introduce the new variables
\be
\label{definition W}
w_{\dot\alpha\alpha} = \mat{ w_1 & w_3 \\ w_4 & w_2 } = \sqrt h\, B_{\dot\alpha} A_\alpha \;,
\ee
where the prefactor has been introduced for convenience, and rewrite the F-flatness condition in the form $\det w_{\dot\alpha\alpha} =0$. This coincides with the conifold equation (\ref{conifold}). Assuming the matrices $A,B$ are diagonal, the F-flatness condition describes the motion of $N$ \Dt-branes on the singular conifold. The D-flatness condition is
\beal
\label{Dflatness}
A_1 A_1^\dag + A_2 A_2^\dag - B_1^\dag B_1 -B_2^\dag B_2 &= U \, \unit \\
A_1^\dag A_1 + A_2^\dag A_2 - B_1 B_1^\dag - B_2 B_2^\dag &= U \, \unit \;,
\eeal
where both identity matrices are $N \times N$ and $U$ is a constant. For $U=0$ a generic solution -- up to gauge equivalencies -- describes $N$ points on the singular conifold; for $U\neq 0$ the solution describes $N$ points on the resolved conifold. The resolved conifold is the singular conifold with $\S^2$ blown up at the tip. Instead of $\det w_{\dot\alpha\alpha} =0$ the space is described by the equation
\be
(w_{\dot\alpha\alpha}) \mat{ \nu_1 \\  \nu_2} = 0
\ee
with $(\nu_1,\nu_2) \in \C\bP^1$. For $w_i \neq 0$, the space is bi-holomorphic to (\ref{conifold}), while at $w_i=0$ we have a non-trivial $\C\bP^1$.
The resolved conifold has the same complex structure as the singular conifold but different metric \cite{CO}.

The dual geometry in ten dimensions is a warped product of Minkowski space and the Ricci-flat conifold
\be
\label{10d}
ds_{10}^2 = h^{-1/2}dx^2+h^{1/2}d\tilde{s}_6^2 \;.
\ee
In fact there is a one-parameter family of Ricci-flat metrics on (\ref{conifold}). The simplest one is the cone over the $T^{1,1}$ (a homogeneous Sasaki-Einstein space)
\be
\label{sc}
d\tilde{s}^2_6=dr^2 + r^2 \, ds_{T^{1,1}}^2 \;.
\ee
$T^{1,1}$ can be defined as a quotient $\frac{SU(2)\times SU(2)}{U(1)}$ which makes the global $SU(2)_A \times SU(2)_B$ symmetry manifest (it is also invariant under  $U(1)_R$). The remaining global $U(1)_\text{baryon}$ of the field theory is not geometrical. Topologically $T^{1,1} \cong \S^2 \times \S^3$, and one can define the generators of $H^2(T^{1,1},\bZ)$ and $H^3(T^{1,1},\bZ)$
\be
\label{cohomT11}
\int_{\S^2} \omega_2 = 4\pi \;,\qquad\qquad  \int_{\S^3} \omega_3= 8\pi^2 \;.
\ee
Metrically $T^{1,1}$ can be represented as a $U(1)$ fibration over $\S^2 \times \S^2$.

The geometry (\ref{10d})-(\ref{sc}) is the singular conifold. It is invariant under the $\IZ_2$ symmetry that flips the sign of $z_4$.
This symmetry exchanges the two $\S^2$ in the base of $T^{1,1}$. On the field theory side this symmetry exchanges $A_\alpha \leftrightarrow B_{\dot\alpha}^\dag$ accompanied by a charge conjugation. This symmetry flips the sign of $U$ and is spontaneously broken if $U\neq 0$. Hence the singular conifold corresponds to the vacuum with $U=0$. The vacua with $U \neq 0$ correspond to the resolved conifold geometry \cite{KWII}.

The supergravity background is of the GKP type \cite{GKP} and the warp factor depends only on the location of the \Dt-branes on the conifold:
\bea
\label{wf}
- \tilde\nabla^2 h = (4\pi^2\alpha')^2  \sum_{i \,\in\, \text{D3-branes}}^N \delta^{(6)}(x-x_i) \;,
\eea
where tilde corresponds to the unwarped metric $d\tilde s_6^2$.
The $AdS_5 \times T^{1,1}$ solution corresponds to $h = \frac{L^4}{r^4}$, \ie{} all \Dt-branes located at the singularity $r=0$. As
evident from the field theory the \Dt-branes can move anywhere on the conifold. The corresponding background is given by (\ref{10d})-(\ref{wf}) and the RR form $C_4 = h^{-1} dx^0\wedge \dots \wedge dx^3$.

In the case with $U\neq0$ in (\ref{Dflatness}) the D3-branes can still freely move, and the warp factor is determined by (\ref{wf}). If all D3-branes are smeared on the $\S^2$ at the tip, $SU(2)_A \times SU(2)_B$ is preserved but the solution is singular \cite{PZT}; if the D3-branes are localized, the solution is regular but the global symmetries are broken \cite{KM}.

In the dual geometry the gauge couplings are controlled by the value of the dilaton $e^\phi$ and the flux of the $B$-field through $\S^2$
\be
\label{couplingsB}
\frac1{g_1^2} + \frac1{g_2^2} = \frac1{4\pi e^\phi} \;,\qquad\qquad
\frac1{g_1^2} - \frac1{g_2^2} = \frac1{2\pi e^\phi} \Big[ b - \frac12 \ ({\rm mod}\ 1) \Big] \;,
\ee
where we defined $b = \frac1{4\pi^2\alpha'} \int_{\S_2} B_2$. The background with vanishing $B$-field corresponds to $g_1=\infty$ and $b=1/2$ corresponds to $g_1=g_2$.

\subsection{Review of the KS theory}
\label{reviewks}

The conformal $SU(N)\times SU(N)$ KW theory can be generalized to $SU(N+M)\times SU(N)$ gauge group. The theory is no longer conformal but instead experiences a cascade of Seiberg dualities, each decreasing the rank of the gauge groups by $M$. Each description gives rise to a branch of perturbative vacua given by the deformed conifold equation
\be
\label{deformedconifold}
\sum_{i=1}^4 z_i^2 = \det w_{\dot\alpha\alpha} = \epsilon \;,
\ee
where $w_{\dot\alpha\alpha}$ is defined as in (\ref{definition W}) and the constant $\epsilon$ is related to the scales $\Lambda_{1,2}$ of the gauge sector. The eigenvalues of $w_{\dot\alpha\alpha}$ parametrize the locations of \Dt-branes on the  deformed conifold. The chiral $U(1)_R$ symmetry is broken to $\bZ_{2M}$ by the anomaly, and further spontaneously broken to $\bZ_2$ by a gaugino condensate that gives rise to $M$ vacua. The remaining $\IZ_2$ stays unbroken. The whole moduli space is the collection of the mesonic branches \cite{DKS}
\be
\label{modulispace}
\text{Moduli space} = \oplus_{l=0}^k \oplus_{r=1}^M Sym_{N-l M}({\mathcal C}_{r,l}) \;,
\ee
where $k=[N/M]_-$ is the number of steps in the cascade,%
\footnote{We define $[x]_-$ as the largest integer less than or equal to $x$.}
$r$ labels the values of the gaugino condensate and
${\mathcal C}_{r,l}$ is the deformed conifold with the deformation parameter $\epsilon_{r,l} = \epsilon_0 \, e^{2\pi i \frac rM} I^\frac lM$ \cite{DKS,Dymarsky:2011pm}. The RG-invariant parameter $I$ of the field theory is dual to the string coupling constant $I = e^{2\pi i \tau}$. In the regime $g_s M \gg 1$ when supergravity is valid $I^{\frac lM}=1$ at the leading order in $\frac1{g_sM}$. Since all branches with different $r$ are equivalent, in what follows we drop the index of the deformation parameter $r,l$ and
assume real $\epsilon$.

In the special case $N = kM$, the IR gauge group reduces to $SU(2M)\times SU(M)$ and this requires a special treatment. The strongly coupled $SU(2M)$ group has as many colors as flavors, and its moduli space is described by mesons $M_{\dot\alpha\alpha} = B_{\dot\alpha} A_\alpha$ and baryons
\beal
\label{baryons}
\cA &= \frac1{(M!)^2} \, \epsilon_{i_1\cdots i_{2M}}\epsilon^{j_1\cdots j_{M}}\epsilon^{k_1\cdots k_{M}} (A_1)^{i_1}_{j_1} \dots (A_1)^{i_M}_{j_M} (A_2)^{i_{M+1}}_{k_1} \dots (A_2)^{i_{2M}}_{k_M} \\
\cB &= \frac1{(M!)^2} \, \epsilon^{i_1\cdots i_{2M}}\epsilon_{j_1\cdots j_{M}}\epsilon_{k_1\cdots k_{M}} (B_1)_{i_1}^{j_1} \dots (B_1)_{i_M}^{j_M} (B_2)_{i_{M+1}}^{k_1} \dots (B_2)_{i_{2M}}^{k_M} \;,
\eeal
which are singlets of $SU(M)\times SU(2)_A \times SU(2)_B$, subject to the quantum constraint $\det M_{\dot\alpha\alpha} - \cA\cB = \Lambda_1^{4M}$. The constraint can be enforced by a Lagrange multiplier $X$ and the superpotential
\be
\label{npsup}
W_\text{eff} = W_{KW} + X( \det M_{\dot\alpha\alpha} - \cA \cB - \Lambda_1^{4M}) \;.
\ee
There are two distinct branches resulting from (\ref{npsup}).
If $X\neq 0$, F-flatness requires $\cA=\cB=0$ and $w_{\dot\alpha\alpha}$ must satisfy
$\det w_{\dot\alpha\alpha} = \epsilon$. This is one of the mesonic branches discussed before.
If $X=0$, the F-flatness condition requires $M_{\dot\alpha\alpha} = 0$ and hence $\cA \cB = -\Lambda^{4M}_1$. This is the baryonic branch. It has one complex dimension and can be parametrized by the VEV of the baryons.
Therefore in (\ref{modulispace}) the possible factor $Sym_0( \cC_{r,k})$ is assumed to be the baryonic branch $\bC$.

The gravity dual of the \smn theory is the Klebanov-Strassler solution \cite{KS}, possibly generalized by extra mobile D3-branes. It is of the GKP type with metric (\ref{10d}) and RR five-form, where $d\tilde s^2_6$ is the Ricci-flat metric on the deformed conifold. Besides, the solution also has RR and NSNS three-forms.
The solution is engineered by placing $M$ fractional D3-branes  and $N$ regular D3-branes at the conifold singularity and is characterized by
\be
\frac1{4\pi^2 \alpha'} \int_{\S^3} F_3 = M
\ee
while $F_5$ is running. The pure KS solution has no mobile \Dt-branes and it is invariant under the $\IZ_2$ symmetry. Hence it corresponds to the point $\cA = \cB$ of the baryonic branch \cite{Aharony}. The rest of the baryonic branch is given by the BGMPZ solutions \cite{Butti}. They have metric
$ ds^2_{10} = e^{2A} dx^2 + ds^2_6$, where $e^{-2A} ds^2_6$ is some {\it pseudo}-K\"ahler metric on the deformed conifold, running dilaton and the three-form flux is not imaginary-self-dual.
The VEV of the baryons $\cA, \cB$ is related to the D-term parameter $U$. Below the scale of baryon VEV the gauge symmetry is broken to $SU(M)$.
That is why for large $U$ the geometry near the tip
approaches the MN solution \cite{MN,CHV} dual to the $SU(M)$ SYM \cite{Butti,DKS,MM}.

To describe the solutions dual to the mesonic branch we need to introduce mobile \Dt-branes on the conifold. As in the KW case, the extra $p$ \Dt s only affect warp factor and 5-form flux, through the same equation (\ref{wf}) where now $h_\text{tot} = h_{KS} + h$.%
\footnote{Such background can be solved explicitly \cite{Krishnan:2008gx,Pufu:2010ie}.}
While the original solution is dual to $SU\big( (k+1)M \big) \times SU(kM)$, the new one is dual to $SU\big( (k+1)M + p \big) \times SU(kM + p)$. Unless $p = 0 \pmod{M}$, the two theories are different. The new theory
does not have a baryonic branch. If we put D3-branes on the BGMPZ solution SUSY is broken and the baryonic branch is lifted by a potential that returns the system to the vacuum described by the KS solution with mobile \Dt-branes \cite{DKS}. If $p = 0 \pmod{M}$, the new solution describes one of the mesonic branches of the original $SU\big( (k+1)M \big) \times SU(kM)$ theory.

In conclusion let us mention here that besides the regular Klebanov-Strassler gravity background discussed above there is an ``approximate'' version of this background found by Klebanov and Tseytlin (KT) \cite{KT}. This background approaches KS in the UV but is singular in the IR. Although it does not correctly describe physics at low energies it is simpler and gives a good approximation when the scale of energies is much larger than the internal scale $\Lambda$ of the field theory. We will make use of this background in section \ref{sec: backreaction d7} where we discuss the backreaction of the \Ds-branes on the geometry.

\section{D7-branes in probe approximation}
\label{sec: adding d7}

In this section we add probe D7-branes to the conifold backgrounds.
In particular we explicitly construct the Abelian $U(1)$ instantons which are dual to the Higgs vacua in the field theory with $N_f=1$.

\subsection[Warm up: $\Z_2$ orbifold of $\N=4$ SYM with hypermultiplets]%
{Warm up: $\Z_2$ orbifold of $\N=4$ SYM with hypers}
\label{sec: N=2 case}

Before adding D7-branes to the conifold theory, let us consider a simpler but closely related example of the $\N=2$ $\Z_2$ orbifold of $\N=4$ SYM with flavors.
We start with the $\N=4$ $SU(N)$ SYM theory which lives on $N$ D3-branes.
Then we add a small number $N_f \ll N$ of D7-branes \cite{KK} that span the Minkowski space $\bR^{3,1}$ and wrap the holomorphic non-compact cycle $\Sigma = \C^2 \subset \C^3$: they add $N_f$ hypermultiplets in the fundamental representation, and break SUSY to $\cN=2$.
Choosing coordinates $z_1,z_2,z_3$ on $\C^3$, the embedding $\Sigma = \{z_3=m\}$
introduces hypermultiplets of mass $m$. The \Dt-branes are free to move on $\C^3$, and their positions parametrize the Coulomb branch. When $k$ \Dt-branes reach the \Ds s, they can dissolve into them if $N_f >1$ turning to $k$ non-Abelian $U(N_f)$ instantons. This corresponds to Higgsing $SU(N)\rightarrow SU(N-k)$. The field theory analysis of the moduli space relies on the F- and D-term equations
\be
[\Phi_1,\Phi_2] + Q\tilde Q = 0 \;,\qquad\qquad
[\Phi_1,\Phi_1^\dag] + [\Phi_2,\Phi_2^\dag] + QQ^\dag - \tilde Q^\dag \tilde Q = 0
\ee
together with $\Phi_3=m$ for the $k\times k$ block of the $N\times N$ matrices $\Phi_i$. These equations exactly coincide with the ADHM description of the moduli space of $k$ $U(N_f)$ instantons -- the worldvolume gauge instantons on the D7s \cite{WittenADHM,DouglasBWB}. The equivalence between field vacuum equations and the ADHM construction (see \cite{Tong} for a pedagogical review) is at the core of the holographic description of the Higgs branch. Besides the moduli space itself, it can be extended to various observables in field theory: the moduli space metric, chiral operators, \etc \cite{GuralnikH1,GuralnikH2,GuralnikH3,GuralnikH4}.

Then we take a $\Z_2$ orbifold of $\Sigma=\C^2$. The resulting geometry is singular, but it can be smoothened out. One can parametrize $\C^2/\Z_2$ by two complex variables $(w_1,w_2)$ subject to
identification $(w_1,w_2) \sim (-w_1,-w_2)$. Alternatively one can introduce invariant coordinates $z_{1,2} = (w_1^2 \pm w_2^2)/2$, $z_3 = iw_1w_2$, subject to the constraint $\sum_{i=1}^3 z_i^2=0$. The singular orbifold admits a simultaneous deformation of the complex structure
\be
\sum_{i=1}^3 z_i^2 = \epsilon
\ee
and a resolution: both replace the singularity by a finite size $\S^2$. This is the smooth Eguchi-Hanson space. Deformation and resolution are measured by the self-dual forms $\omega^{(2,0)}$ and $J^{(1,1)}$:
\be
\label{resdef}
\int_{\S^2} \omega^{(2,0)} \equiv \xi_{\C} = \epsilon \;,\qquad\qquad \int_{\S^2} J^{(1,1)} = \xi_{\R} \;.
\ee
The resolution and deformation parameters $\xi_{\R}$, $\xi_{\C}$ transform as a triplet under $SU(2)_R$ that rotates the complex structures on the hyper-K\"ahler Eguchi-Hanson space.

Since the deformed/resolved orbifold has an exceptional 2-cycle $\S^2$, it admits $U(1)$ instantons. Hence the orbifold theory has a Higgs branch even for $N_f=1$. In general the $U(N_f)$ instantons on $\C^2/\Z_2$ are characterized by the first and second Chern classes
\be
\label{ch1,2}
{\rm ch}_1 = \frac1{2\pi} \int_{\S^2} \Tr F \;,\qquad\qquad {\rm ch}_2 = \frac1{8\pi^2} \int_{\bC^2/\bZ_2} \Tr F\wedge F
\ee
and the conjugacy class of the monodromy matrix $\hat\rho:\Z_2\rightarrow U(N_f)$. The latter is defined as follows. One considers a radial section $\S^3/\bZ_2$ of the orbifold at infinity, where $F=0$, and computes the holonomy $\hat\rho = \text{Pexp } i \oint_{\partial\Gamma} A \in U(N_f)$ along the generator $\partial\Gamma$ of $\pi_1(\S^3/\bZ_2) = \bZ_2$. Such a matrix must satisfy $\hat\rho^2 = \unit$, and its conjugacy class is a gauge-invariant observable.

The gauge instantons on the D7s' worldvolume are D3-branes on their Higgs branch, dissolved in the D7s. The corresponding moduli space was analyzed from the D-brane point of view in \cite{DnM}, showing that it agrees with the ADHM construction put forward by Kronheimer and Nakajima \cite{KnN} (see also \cite{Bianchi1995,Bianchi1996} for a review). As in the case of pure $SU(N)$ SYM with flavor, we can reproduce the ADHM quiver and equations by analyzing the vacuum equations of the field theory. The $\Z_2$ orbifold gives an $SU(N)\times SU(N)$ quiver theory with $N_{fL}$ left and $N_{fR}$ right flavors, $N_f = N_{fL} + N_{fR}$. Invariance under the $\bZ_2$ orbifold action dictates that only the non-diagonal $N\times N$ blocks of $\Phi_{1,2}$ are non-vanishing, while $\Phi_3$ is block diagonal:
\be
\Phi_\alpha = \mat{0 & A_\alpha \\ \epsilon_{\alpha\beta}B^\beta & 0} \;,\qquad\qquad \Phi_3 = \mat{\phi_3 & 0 \\ 0 & -\tilde \phi_3} \;.
\ee
The fields with index $\alpha=1,2$ are doublets of a flavor $SU(2)$ symmetry, while $SU(2)_R$ acts on $(A, B^\dag)$ as a doublet.
The resulting superpotential is
\be
W = \phi_3 (A_\alpha B^\alpha-Q_L\tilde Q_L) + \tilde \phi_3(B^\alpha A_\alpha + Q_R\tilde Q_R) \;,
\ee
where sum over $\alpha$ is implicit.

\begin{figure}
\centering
\begin{tikzpicture}[node distance=0.5cm, auto]

\node[rectangle,draw=black,thick,right=-2.9cm](flavorL) {$N_{fL}$};
\node[rectangle,draw=black,thick,right=4.5cm](flavorR) {$N_{fR}$};
\node[circle,draw=black, thick](leftgroup) {$k_1$};
\node[circle,draw=black, thick,right=2cm](rightgroup) {$k_2$};

\draw[->>,>=latex, shorten >=2pt, shorten <=2pt, bend left=45, thick]
    (leftgroup.north east) to node[auto, swap] {$A_\alpha$}(rightgroup.north west);
\draw[->>, >=latex, shorten >=2pt, shorten <=2pt, bend left=45, thick]
    (rightgroup.south west) to node[auto, swap] {$B_\alpha$}(leftgroup.south east);

\draw[->, >=latex, shorten >=2pt, shorten <=2pt, bend right=45, thick]
    (flavorR.north west) to node[auto] {$\tilde Q_R$}(rightgroup.north east);
\draw[->, >=latex, shorten >=2pt, shorten <=2pt, bend right=45, thick]
    (rightgroup.south east) to node[] {$Q_R$}(flavorR.south west);

\draw[->, >=latex, shorten >=2pt, shorten <=2pt, bend left=45, thick]
    (flavorL.north east) to node[auto, swap] {$\tilde Q_L$}(leftgroup.north west);
\draw[->, >=latex, shorten >=2pt, shorten <=2pt, bend left=45, thick]
    (leftgroup.south west) to node[auto, swap] {$Q_L$}(flavorL.south east);

\end{tikzpicture}
\caption{ADHM quiver describing instantons on the Eguchi-Hanson space $\C^2/\Z_2$. We will concisely refer to this quiver as $N_{fL} \times k_1 \times k_2 \times N_{fR}$.}
\label{qv2}
\end{figure}
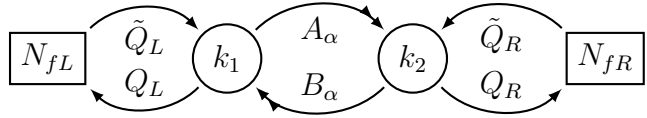

The $N$ \Dt-branes can freely move on $\bC^3/\bZ_2 \times \bC$, realizing the Coulomb branch; when the \Dt s reach the \Ds s they can dissolve turning into instantons, and Higgsing part of the gauge symmetry. Let us denote with $k_{1,2}$ the ranks of the broken symmetry. They might be different, corresponding to the presence of D5-branes dissolved in the D7s and wrapping the 2-cycle of $\bC^2/\bZ_2$. The Higgsed directions of $\Phi_3$, are the $k_1\times k_1$ and $k_2\times k_2$ blocks where $\phi_3,\tilde \phi_3$ are equal to $m$ multiplied by the $k_1\times k_1$ and $k_2\times k_2$ identity matrices to allow non-trivial values of flavor fields. Eliminating $\Phi_3$, the F- and D-term equations describing the Higgs branch effectively represent the quiver in figure \ref{qv2}, which we will concisely denote as $N_{fL} \times k_1 \times k_2 \times N_{fR}$.

To present the F- and D-term equations in a concise form we define the combinations
\be
C^\alpha = \mat{B^\alpha \\ A^{\alpha\dag}} \;,\qquad\qquad P_L = \mat{ -Q_L \\ \tilde Q_L^\dag}, \qquad\qquad P_R = \mat{ Q_R \\ \tilde Q_R^\dag}\ ,
\ee
and use the Pauli matrices $\Gamma_\mu = \big( \begin{smallmatrix} 0 & 1 \\ 1 & 0 \end{smallmatrix}\big)$, $\big( \begin{smallmatrix} 0 & -i \\ i & 0 \end{smallmatrix}\big)$, $\big( \begin{smallmatrix} 1 & 0 \\ 0 & -1 \end{smallmatrix}\big)$ to represent the F-term and D-term equations in a $SU(2)_R$ covariant form
\be
\label{adhm}
C_\alpha^\dag \Gamma_\mu C^\alpha+ P_L \Gamma_\mu P_L^\dag = - \xi^L_\mu \;,\qquad\qquad C^\alpha \Gamma_\mu^* C_\alpha^\dag + P_R \Gamma_\mu^* P_R^\dag = - \xi^R_\mu \;.
\ee
Here $\dag$ acts on gauge indices, while transposition of $SU(2)_R$ indices is implicit.
In general $\xi^L,\xi^R$ should be understood as some parameters of the solution. In the case of the $U(N)\times U(N)$ orbifold theory these are the Fayet-Iliopoulos (FI) terms.
If $k_1=k_2=N$ \ie{} there is no remaining unbroken gauge group $\xi^L$ and $\xi^R$ can be turned on independently. Otherwise $\xi^L=\xi^R=\xi$ as follows from the components of (\ref{adhm}) with trivial $Q$. In the bulk the triplet $\xi_\mu$ controls the resolution/deformation of $\bC^2/\bZ_2$ as seen from  (\ref{resdef}).

In components the equations (\ref{adhm}) are
\beal
\label{adhmcomp}
A_\alpha B^\alpha - Q_L\tilde Q_L&= \xi_\bC\ , \qquad\qquad &
A_\alpha A^{\alpha\dag} - B^\dag_{\alpha} B^{\alpha} + Q_L Q_L^\dag - \tilde Q_L^\dag \tilde Q_L&= \xi_\bR\ , \\
B^\alpha A_\alpha + Q_R\tilde Q_R &= \xi_\bC\ , \qquad\qquad &
B^\alpha B_\alpha^\dag - A^{\alpha\dag} A_\alpha + Q_R Q_R^\dag - \tilde Q_R^\dag \tilde Q_R &= - \xi_\bR\ ,
\eeal
where we defined $\xi_\bR \equiv \xi_3$, $\xi_\bC = - (\xi_1 - i\xi_2)/2$.

The relation between the ranks $N_{fL},N_{fR},k_1,k_2$ in field theory and the properties of the instanton in the bulk is as follows.
The conjugacy class of the monodromy matrix $\hat\rho$ defines splitting of $N_f$ into $N_{fL,R}$. Since $\hat\rho^2 = \unit$, its eigenvalues are $\pm1$ and $N_{fL,R} = \Tr ( \unit \mp \hat\rho)/2$. The ranks $k_1,k_2$ are related to the Chern classes (\ref{ch1,2}) as follows \cite{DnM}:%
\footnote{Our description differs from the one in \cite{DnM} by the sign in the definition of ${\rm ch}_1$ and a $\Z_2$ flip of the quiver.}
\be
{\rm ch}_1 = 2(k_1-k_2) - N_{fL} \;,\qquad\qquad {\rm ch}_2 = k_2 + \frac{N_{fL}}4 \;.
\ee
Finally, the dimension of the moduli space of instantons with given ${\rm ch}_{1,2}$ and $\hat\rho$, \ie{} $k_1,k_2, N_{fL,R}$, is equal to
\be
\dim \M = 4\big( N_{fL} k_1 + N_{fR} k_2 - (k_1-k_2)^2 \big) \;.
\ee

The quiver in figure \ref{qv2} as well as its space of vacuums are invariant under a $\Z_2$ flip that exchanges the left and right groups: $k_1\leftrightarrow k_2$ and $N_{fL} \leftrightarrow N_{fR}$. Although such symmetry is trivial in the field theory, it acts non-trivially on the space of instantons. It multiplies $\hat\rho$ by $-1$ and transforms the Chern classes as
\be
{\rm ch}_1 \;\rightarrow\; -N_f - {\rm ch}_1 \;,\qquad\qquad {\rm ch}_2 \;\rightarrow\; \frac{N_f}4 + (k_1 + k_2) - {\rm ch}_2 \;.
\ee

The relation between geometric properties of instantons and ranks $k_1,k_2,N_{fL,R}$ of the quiver
provides a simple holographic picture.
The splitting of flavors into left and right is determined by the worldvolume gauge field on the \Ds s and the corresponding monodromy matrix $\hat\rho$.
Above the scale $m$ the field theory has gauge group $SU(N)\times SU(N)$ with $N_f$ hypermultiplets while below $m$ the gauge theory is pure $SU(N-k_1)\times SU(N-k_2)$. The resulting low-energy gauge theory is described holographically by
$min(N-k_1,N-k_2)$ D3-branes and $|k_1-k_2|$ D5-branes wrapping homologically non-trivial $\S^2$ of $\C^3/\Z_2$.

To get some intuition about how $k_1,k_2,N_{fL,R}$ are related to the instanton charges, let us consider a simple Abelian instanton of charge $n$, \ie{} a $U(1)$ gauge field with ${\rm ch}_1=n$. The charge $n$ is integer while one finds ${\rm ch}_2= \frac{n^2}4$ and $\hat\rho = (-1)^n$ \cite{DnM}. We have two distinctive cases. When $n = 2r-1$ is odd the quiver is
\be
\label{n2left}
1 \times r^2 \times r(r-1) \times 0 \;,
\ee
while when $n=2r$ is even the quiver is
\be
\label{n2right}
0 \times r(r+1) \times r^2 \times 1 \;.
\ee
The explicit matrices that solve the quiver equations in these cases can be found in section \ref{sec: classical moduli space}.

After we developed some intuition in the $\cN=2$ case we return our attention to the conifold geometry in the next section.

\subsection{Geometry of the D7-brane embedding}
\label{sec: geometry of embedd}

Throughout this paper we consider D7-branes along the so-called Kuperstein embedding \cite{Kuperstein} -- a holomorphic non-compact 4-cycle $\Sigma$ defined by
\be
\label{emb}
z_4 = \frac\mu{\sqrt2} = \text{const} \;.
\ee
If we start with the $\bC^2/\bZ_2 \times \bC$ $\cN=2$ case and the usual \Ds-brane discussed in the previous subsection and introduce the massive deformation of the orbifold theory that leads via the RG flow to the conifold theory the original \Ds-brane result in a \Ds-brane embedded along (\ref{emb}) \cite{Ouyang:2003df}.
This provides us with the field content and superpotential of the flavor sector. A stack of $k$  D7-branes introduces $k$ flavors of hypermutiplets $\tilde Q,Q$ in the fundamental of one of the gauge groups, with superpotential of the form
\be
\label{rough superpotential}
W_\text{flavor} \sim \tilde Q (A_1B_1 + A_2B_2 - \mu )Q + \tilde Q Q \tilde Q Q \;.
\ee
We will be more precise in sections \ref{sec: field theory analysis} and \ref{sec: comparison}, clarifying also under which gauge group the quarks are charged.

The embedding (\ref{emb}) preserves the anti-diagonal $SU(2)_{AB}$ of the global $SU(2)_A \times SU(2)_B$ symmetry. In what follows we assume that the deformation parameter $\epsilon$ and mass $\mu$ are non-zero. The corresponding limits of singular conifold $\epsilon=0$ or zero mass embedding $z_4=0$ are straightforward.

The 4-cycle $\Sigma$ has the complex structure of $\bC^2/\bZ_2$ with deformation parameter $\epsilon - z_4^2$
\be
\label{Sigmagem}
\sum_{i=1}^3 z_i^2 = \epsilon - z_4^2 \;.
\ee
As we discussed in section \ref{sec: N=2 case}, at infinity this space approaches a cone over $\S^3/\bZ_2$, and given a flat bundle on it one can construct a monodromy matrix $\hat\rho = \text{Pexp } i \oint_{\partial \Gamma A}$ whose conjugacy class is a gauge-invariant.
We can also think of $\S^3/\bZ_2$ as a Hopf $S^1$ fibration over $\S^2$, with $\S^1$ shrinking at the tip of $\Sigma$ while $\S^2$ staying finite. Hence $\Sigma$ can support Abelian flux.

We can parametrize $\Sigma$ by the radial coordinate of the conifold
\be
\label{rc}
\sum_{i=1}^4 |z_i|^2 \equiv r^3 \equiv \epsilon \cosh t \;,
\ee
which takes value in the range $r^3 \geq |z_4|^2 + |\epsilon - z_4^2|$, together with some angular coordinates on $\S^3/\bZ_2$. In practice it is convenient to use the one-forms $g_5, dz_i$ of the full conifold geometry pulled-back on $\Sigma$. In terms of the usual conifold coordinates, let us define
\be
g_5 = d\psi - \sum_{i=1,2} \cos\theta_i\, d\varphi_i \;,\qquad\qquad \Vol_i = \sin\theta_i\, d\theta_i \wedge d\varphi_i \;.
\ee
Expressing $z_3$ through $z_1,z_2$ one finds the following useful relations \cite{D7bb}
\be
\label{relation1}
- \frac12 dg_5 \wedge dg_5 \Big|_\Sigma = f(t) dt \wedge g_5 \wedge dg_5 \Big|_\Sigma = \frac{4 |z_4\cosh t-\bar z_4|^2}{\epsilon^2 \sinh^4 t |z_3|^2}dz_1 \wedge d\bar z_1\wedge dz_2\wedge d\bar z_2 \Big|_\Sigma
\ee
where the function
\be
\label{fexpr}
f(t) = - \frac{|z_4\cosh t -\bar z_4|^2}{\sinh t (\epsilon \sinh^2 t-2|z_4^2|\cosh t +z_4^2+\bar z_4^2)}
\ee
is defined through ${\a'/ \a}=-2f$, and $\a(t)$ is a ``volume'' of $\Sigma$ at the given radius $t$
\be
\label{a}
\int_{\S^3/\bZ_2 \text{ at }t}
g_5 \wedge dg_5 = 32\pi^2 \a(t) = 32\pi^2 \frac{\epsilon \sinh^2 t-2|z_4^2|\cosh t +z_4^2+\bar z_4^2}{\epsilon \sinh^2 t} \;.
\ee

Let us remark that on the deformed conifold the 2-form $dg_5$ is singular at the tip, as one can check by computing the norm $|dg_5|^2$ using the inverse metric: the magnitude diverges as $1/t^2$ (whilst $g_5$ is regular). In the case of the massless embedding $z_4 = 0$, the pull-back of $dg_5$ is likewise singular at the tip. Therefore when expressing a gauge field on $\Sigma$, we should be careful to ensure that the coefficient in front of $dg_5$ vanishes. Let us stress that the geometry of $\Sigma$ on the deformed conifold is regular when $z_4 = 0$, and all physical quantities should be continuous in this limit.

A similar subtlety arises in the resolved conifold case. Since at the tip one of the 2-spheres in the base $\S^2 \times \S^2$ of $T^{1,1}$ (as $U(1)$ fibration) vanishes, the corresponding volume form -- say $\Vol_2$ -- diverges. This again can be checked by computing $|\Vol_2|^2$. As a result, both 2-forms
\be
\label{w2}
dg_5 = \Vol_1 + \Vol_2 \;,\qquad\qquad \omega_2 = \frac12 ( \Vol_1 - \Vol_2 )
\ee
are divergent at the tip of the resolved conifold. An easy way to avoid the difficulty is to combine $dg_5$ and $\omega_2$ at the tip into the volume form $\Vol_1 = \frac12 dg_5 + \omega_2$ which is well-defined. We also remark that the limit $z_4 \to 0$ in the resolved conifold case is smooth. The complex equation (\ref{Sigmagem}) of $\Sigma$ has the $\bC^2/\bZ_2$ singularity, however the resolution of the conifold induces a resolution of $\Sigma$, and in fact the blown-up 2-sphere of the conifold coincides with the blown-up 2-sphere of $\Sigma$.

\subsection{Gauge field on the D7-brane and AdS/CFT dictionary}
\label{sec: AdSCFT}

In this section we solve the linearized equations for $SO(3)$ invariant fluctuations of the worldvolume fields on $\Sigma$ and identify them with field theory operators according to the AdS/CFT correspondence. We will be mainly concerned with the UV (large $r$) behavior of the bulk fields, therefore we will work in the singular conifold limit $\epsilon=0$.
We can introduce a set of real coordinates $r,X_I,Y_I$ (with $I=1,\cdots,4$) on the conifold
\be
\label{XY}
z_I = r^{3/2}(X_I+iY_I)
\ee
where $X_I,Y_I$ are subject to the constraints
\be
\label{constr1}
X^2 = Y^2 = \frac12 \;,\qquad\qquad X \cdot Y = 0 \;.
\ee
The base $T^{1,1}$ of the singular conifold is represented as the product of two 3-spheres with an orthogonality condition and metric
\be
\label{metricT11}
ds_{T^{1,1}}^2 = \frac23 (dX^2+dY^2) - \frac29 (XdY-YdX)^2 \;.
\ee

In order to introduce local coordinates on $\Sigma$ and calculate the induced metric we represent the conifold as a foliation of the Kuperstein embeddings parametrized by $X_4,Y_4$, the radial coordinate $r$ and three angular coordinates $t_i$. First we fix
\be
z_4 = \frac{\mu(r)}{\sqrt 2} = r^{3/2}(X_4 + iY_4) \;.
\ee
We can think of this equation with $X_4,Y_4(r)$ as a parametrization of a generic $SO(3)$-invariant embedding $\Sigma$.
Then we arbitrarily choose
\footnote{We chose the parametrization $X_1^{(0)}=\sqrt{1/2-X_4^2},\ X_2^{(0)}=0,\ X_3^{(0)}=0$ and $Y_1^{(0)}=-X_4 Y_4/\sqrt{1/2-X_4^2},\ Y_2^{(0)}=\sqrt{1/2-X_4^2-Y_4^2}/ \sqrt{2(1/2-X_4^2)},\ Y_3^{(0)}=0$.}
$X_I^{(0)},Y_I^{(0)}$, for $I=1,2,3$   such that the constraints (\ref{constr1}) are satisfied. Then we introduce the angular coordinates $t_i$ as the ``Euler angles'' of the $SO(3)$ rotation which transforms the point $X^{(0)} + iY^{(0)}$ into some other point on $\S^3/\bZ_2$. We can use the conventional $3\times 3$ generators $T^i$ of $\so(3)$ embedded into the upper left corner of the $4\times 4$ matrix acting on $z_I$ as follows
\be
X+iY = e^{t_i T^i} (X^{(0)}+iY^{(0)}) \;.
\ee
Clearly this transformation leaves $z_4$ invariant.
Then the tangent vector is
\be
\label{dxdy}
d(X+iY) = dt_i T^i (X+iY) + \frac{\partial (X+iY)}{\partial X_4}dX_4 + \frac{\partial (X+iY)}{\partial Y_4}dY_4 \;.
\ee
The one-forms $dt_i$ are the left-invariant one-forms $e_i$ on $\S^3 \cong SU(2)$ calculated at the origin
\be
e_i = dt_i + \epsilon_{ijk} t_j dt_k + \cO(t^2) \;.
\ee
To obtain the expression valid everywhere on $\S^3/\bZ_2$ we can simply substitute $dt_i$ by $e_i$.
Now, if we substitute (\ref{dxdy}) into the conifold metric $ds_6^2 = dr^2 + r^2 ds_{T^{1,1}}^2$ with (\ref{metricT11}), we obtain the metric in terms of $(r,X_4,Y_4,t_i)$. If we instead interpret $X_4,Y_4$ as radial functions defined by the generic $SO(3)$-invariant embedding $\mu(r)$, we obtain the induced metric on $\Sigma$. In the special case $\mu = \text{const}$ the unwarped metric on $\Sigma$ is
\be
\label{metricSigma}
ds^2_\Sigma = \frac{4r^3-|\mu|^2}{4(r^3-|\mu|^2)} dr^2 + \frac{r^3-|\mu|^2}{3r} e_1^2 + \frac{r^2}3 e_2^2 + \frac{4r^3-|\mu|^2}{9r} e_3^2 \;.
\ee

Our next step is the quadratic action for the fluctuation of the worldvolume gauge field given by
$$
\int d^4x  \ dr \Big[ \frac12 \sqrt{g}\, g^{AB}g^{A'B'}F_{AA'}F_{BB'}-h^{-1}{\Pf}  F\Big] \;.
$$
Here the  indexes $A,B$ run through the Minkowski and internal $r,e_i$ directions.
The induced metric $g_{AB}$ is a warped product of the flat Minkowski metric and the metric (\ref{metricSigma}) on $\Sigma$. The Pfaffian ${\Pf} F$ is calculated with the $4\times 4$ matrix $F_{AB}$ with all indexes taken along the internal directions $r,e_i$.

We are focusing on the lowest $SU(2)$-invariant modes in the KK expansion. The corresponding woldvolume gauge field can always be brought to the form
\be
\label{gaugefield}
A = A_\mu(r,x^\mu)\, dx^\mu + A_i(r,x^\mu)\, e_i \;,
\ee
with vanishing component along $dr$. To fix the residual gauge symmetry we require the Minkowski vector $A_\mu$ to be transverse, $\partial^\mu A_\mu=0$, therefore $A$ splits into a transverse space-time vector and three space-time scalars. The effective Lagrangian (to be integrated over space-time and radius from $r^3=|\mu|^2$ to infinity) for $A_\mu$ is
\be
\label{eqVec}
{\mathcal L}_{A_\mu} = {4(r^3-|\mu|^2) } \, |{\partial_r A_\mu }|^2+ h (4r^3-|\mu|^2) \, |\partial_\nu A_\mu|^2 \;.
\ee
The Lagrangian for the scalars $A_i$ is
\beal
\label{eqB}
{\mathcal L}_{A_i} &= \frac{1}{2h} \big( \rho_i A_i + \rho_i^{-1} A_i' \big)^2 + \frac{(4r^3-|\mu|^2)}{2(r^3-|\mu|^2)} \rho_i^{-2} \, |\partial_\mu A_i|^2 \\
\rho_1^2 &= \frac3{2r} \;,\qquad\qquad
\rho_2^2 = \frac{3r^2}{2(r^3-|\mu|^2)} \;,\qquad\qquad
\rho_3^2 = \frac{(4r^3-|\mu|^2)}{2r(r^3-|\mu|^2)} \;.
\eeal
The linear in derivative term $h^{-1} A_i \partial_r A_i$ in ${\mathcal L}_{A_i}$ comes from the CS term in the action. Eventually the Lagrangian for the perturbation $\delta\mu$ of the geometrical profile, \ie{} $\mu= {\rm const} + \delta \mu$, is
\be
\label{eqmu}
{\mathcal L}_{\delta \mu} = \frac{4r^2(r^3-|\mu|^2)}{(4r^3-|\mu|^2)} \, |\partial_r \delta \mu |^2 + h r^2 \, |\partial_\mu \delta \mu|^2 \;.
\ee

We can now analyze the resulting equations for the fluctuations $A_\mu$, $A_i$, $\delta\mu$ and identify the dual field theory operators. As a by-product we will also find the mass spectra for the corresponding mesons in the KW case. Using the explicit form of the KW warp factor $h=\frac{L^4}{r^4}$
we find the asymptotic static (space-time independent) solutions
\beal
\label{asympt}
A_1 &= c_1 r^{-3/2} + c_2r^{-5/2} \;,\qquad\qquad &
A_\mu &= c_1 + c_2 r^{-2} + \cO(r^{-5}) \;, \\
A_2 &= (r^3-|\mu|^2)^{-1/2}(c_1+c_2 r^{-1}) \;,\qquad\qquad &
\delta\mu &= c_1 + c_2 r^{-1} + \cO(r^{-4}) \;, \\
A_3 &= (r^3-|\mu|^2)^{-1/2}r^{-1/2} \Big[ c_1 + c_2 \Big(\log r + \frac{|\mu|^2}{12 r^3} \Big) \Big] \;.\quad
\eeal
The asymptotic behavior in AdS$_{5}$ of a canonically normalized field $\phi(r)$ dual to an operator of dimension $\Delta$ is%
\footnote{There might be logarithms in (\ref{asymptotic expansion}) as in $A_3$ from (\ref{asympt}) if the two series expansions overlap.}
\be
\label{asymptotic expansion}
\phi(r) \sim c_\text{source} \, r^{\Delta - 4 + s} + c_\text{VEV} \, r^{-\Delta + s}
\ee
where $s=0$ for a scalar and $s=1$ for a vector.
This reveals that the vector $A_\mu$ is dual to an operator of dimension $3$ (the conserved current $J_\mu$ of the flavor $U(1)$ symmetry) and the scalar $A_3$ is dual to an operator of dimension $2$ (the bottom component $|Q^2|-|\tilde Q^2|$ of the $U(1)$ current multiplet). These operators, as we show in section \ref{sec: QM}, are manifestly related in the bulk by a SUSY transformation. We used $Q,\tilde Q$ for the bottom component of the corresponding chiral superfields.

The real and imaginary parts of $\delta\mu$ are degenerate since $\mu$ is dual to a complex chiral superfield in field theory.
It follows from (\ref{asympt}) that $\delta\mu$ corresponds to operators of dimension either $3/2$ or $5/2$. To distinguish between the two \cite{KWII} we notice that, because of the superpotential (\ref{rough superpotential}), $\mu$ couples to the operator $\int d^2\theta\, \tilde QQ$ of  dimension $5/2$, and by AdS/CFT this is the operator dual to $\delta \mu$.
The fluctuations $A_1,A_2$ (which after an appropriate change of variables satisfy the same equation) combine into a complex scalar dual to the bottom component $\tilde QQ$ -- an operator of dimension $3/2$.

The leading asymptotic $A_{1,2}\sim r^{-3/2}$ is dual to the VEV of $\tilde QQ$, while the subleading $A_{1,2} \sim r^{-5/2}$ to the source of  $\tilde QQ$ in the Lagrangian. Was one interested in calculating the mass spectrum of the corresponding meson excitations, such boundary conditions would lead to a complication because one would have to ensure that the subleading asymptotic vanishes. It is more convenient to calculate the spectrum of the superpartner $\delta\mu$, since four-dimensional SUSY guarantees the degeneracy of masses within the multiplet. In the bulk this follows from the SUSY Quantum Mechanics transformation that relates the equations for $\delta \mu$ and $A_{1,2}$ and also for $A_\mu$ and $A_3$.

\subsection{SUSY in the bulk and SUSY QM%
\footnote{We thank D.~Melnikov for his input on the following subsection.}}
\label{sec: QM}

To see how the supersymmetric quantum mechanics works, let us consider a family of one-dimensional effective actions of the form

\be
\label{eqF}
S = \int dr \left(F\psi'^2 - H\psi^2-m^2G\psi^2\right) \;,
\ee
where $F, H,G$ are functions of $r$ and $'$ denotes derivative with respect to $r$.
To bring the corresponding EOM to the canonical form we perform the change of variables $\psi= \frac\phi{\sqrt F}$ resulting in
\be
\label{norm}
\phi'' - V\phi = -m^2 \frac GF \phi \;,\qquad\qquad V = \frac{F''}{2F} - \frac{F'^2}{4F^2} - \frac HF \;.
\ee
In fact the potential $V$ can be expressed as
\be
V = W' + W^2 - \frac HF
\ee
with the function $W$ given by
\be
\label{supQM}
W= \frac12 (\log F)'+\left[ F \, \Big( {\rm const} + \int^r F^{-1} \Big) \right]^{-1} \;.
\ee
In all cases below the ``${\rm const}$'' in the formula above will be infinite and $W= \frac12 (\log F)'$.

If $H=0$, then $V$ is entirely captured by the superpotential $W$. In this case the equation (\ref{norm}) can be written in a form that makes the SUSY QM explicit
\be
\label{susyv1}
Q_1Q_2\phi=-m^2 \phi \;,
\ee
with
\be
\label{Q1Q2}
Q_1 = \alpha \, \left( \frac d{dr} + W - (\log \alpha)' \right) \;,\qquad Q_2 = \alpha \, \left( \frac d{dr} - W \right) \;,\qquad  \alpha^2 = \frac FG \;.
\ee
Clearly equation (\ref{susyv1}) has a superpartner which shares the same mass spectrum
(up to a possible zero-mode $m=0$)
\be
\label{partnereq}
Q_2Q_1 \phi = -m^2 \phi \;.
\ee
This equation can be written in the canonical form  (\ref{norm}) using new functions $\tilde F, \tilde H, \tilde G$. In this case $\tilde{F}/\tilde{G}=F/G$ and the new potential is
\be
\label{newpot}
\tilde{V} = \frac{\tilde{F}''}{2\tilde{F}} - \frac{\tilde{F}'^2}{4\tilde{F}^2} - \frac{\tilde{H}}{\tilde{F}} = (-W'+W^2) + \frac{\alpha''-2W\alpha'}\alpha \;.
\ee

Let us apply this to the equations for $A_\mu,A_i,\delta\mu$. We start with the equations for $A_{1,2}$ and cast them in the form (\ref{norm}).
It turns out that for both modes the potential $V$ vanishes and the equations coincide
\be
\label{A12}
\phi'' = - m^2 L^4 \frac{(4r^3-|\mu|^2)}{4(r^3-|\mu|^2)r^4} \phi \;.
\ee
The equation governing the complex scalar $\delta\mu$ brought to the canonical form is:
\be
\label{eqprof}
\phi'' = - \frac{9r(16r^6+r^3|\mu|^2-8|\mu|^4)|\mu|^2}{4(r^3-|\mu|^2)^2(4r^3-|\mu|^2)^2} \phi - m^2 L^4 \frac{(4r^3-|\mu|^2)}{4(r^3-|\mu|^2)r^4} \phi \;.
\ee
This equation is the SUSY QM partner of (\ref{A12}): if we compute the effective potential for the superpartner of (\ref{eqprof}) using (\ref{newpot})
and
\be
F = \frac{4(r^3-|\mu|^2)r^2}{(4r^3-|\mu|^2)} \;,\qquad H=0 \;,\qquad \alpha^2 = \frac FG = \frac{4(r^3-|\mu|^2)r^4}{(4r^3-|\mu|^2)} \;,
\ee
we find that $\tilde{V}$ vanishes and we arrive at (\ref{A12}).

The equation for $A_\mu$ written in a canonical form is
\be
\label{vec}
\phi'' = \frac{3r(r^3-4|\mu^2|)}{4(r^3-|\mu|^2)^2} \phi - m^2 L^4 \frac{(4r^3-|\mu|^2)}{4(r^3-|\mu|^2)r^4} \phi \;.
\ee
Using $H=0$ and $F=4(r^3-|\mu|^2)$ we can calculate the potential for the superpartner equation arriving at
\be
\label{a3}
\phi'' = - \frac{2(2r^6-10r^3|\mu|^2-|\mu|^4)}{r^2(4r^3-|\mu|^2)^2} \phi - m^2 L^4 \frac{(4r^3-|\mu|^2)}{4(r^3-|\mu|^2)r^4} \phi \;,
\ee
which is the equation for $A_3$ written in a canonical form.

Notice that SUSY QM relates the equations for $(A_\mu,A_3)$ and $(A_{1,2},\delta \mu)$ for any warp factor $h$, because supersymmetry is unbroken for any distribution of D3-branes on the conifold.

Now let us briefly address the question of computing the four-dimensional spectrum of $A_{1,2}$. The leading asymptotic behavior of $\phi$ following from (\ref{A12}) is $\phi=c_1 + c_2 r $.
To calculate the spectrum numerically, say by shooting, one needs to impose the exotic boundary condition that $\phi$ does not have a constant part at infinity while may have the linearly divergent term. In practice this is difficult to control. Instead of dealing with (\ref{A12}) one can calculate the spectrum of equation (\ref{eqprof}). It has the same asymptotic behavior but
the conventional boundary condition, \ie{} $\phi$ may go to a constant at infinity but should not diverge. This gives the following spectrum for $m^2$ (in units of $|\mu|^\frac43 L^{-4}$): $3.6$, $19.3$,\dots
A similar but less severe problem arises while dealing with the equation (\ref{a3}) for the $A_3$ fluctuations. The asymptotic behavior of the wave-function is $\phi= r^{1/2}(c_1+c_2\log r)$.
The subleading term is only logarithmically suppressed and to impose the boundary condition of vanishing  $c_2$ in practice may require a very large cutoff. It is better instead to deal with the equation (\ref{vec}) which results in the asymptotic behavior $\phi=c_1r^{-1/2}+c_2 r^{3/2}$. The boundary condition is simply that $\phi$ vanishes at infinity yielding the spectrum of masses $m^2$ (in units of  $|\mu|^\frac43 L^{-4}$): $6.6$, $24.7$,\dots The lightest mode of the vector multiplet happens to be heaver than the one of the scalar multiplet.

\subsection{Asymptotics of the worldvolume gauge field in KS}

Knowing the asymptotic behavior in the KW case is usually good enough to deal with the KS and BGMPZ solutions as well, because these solutions -- up to logarithmic corrections -- approach the KW background at large radius. The corrections are not important when the leading and subleading asymptotics have two different powers of $r$. This is not the case for $A_3$.
Therefore we repeat the analysis of the UV behavior for this mode in the case of the deformed conifold. The leading UV behavior is not sensitive to the value of $\mu$ and therefore we put it to zero, significantly simplifying the calculation. The cycle $\Sigma$ can be parametrized by the radial coordinate $t$ and the angles $\theta_1=\theta_2 \equiv \theta$, $\phi_1=\phi_2 \equiv \phi$, $\psi$. Using the relation between $e_3$ and $g_5$ (for the singular conifold case)
\be
\label{g5e3relation}
g_5 = r^3 \sqrt\frac{r^3-|\mu|^2}{r^3} \, e_3 \;,
\ee
the gauge field ${\mathcal A}=A_3(r,x^\mu) e_3$ can be written as ${\mathcal A}=\xi(t,x^\mu)g_5$.
The Lagrangian for the static (Minkowski-independent) $\xi(t)$ is
\be
\cL = \int_0^{\infty} dt\ \frac1{2 h(t)} \left(\rho_{\xi} \xi + \rho_{\xi}^{-1} \xi' \right)^2 \;,\qquad\qquad \rho_{\xi}(t) = \sqrt\frac23 \, \frac{\sinh t}{\sqrt{\sinh(t)\cosh(t)-t}} \;.
\ee
The resulting equation (with the restored $x^\mu$-dependence) is a superpartner, in the sense of the SUSY QM discussed before, of the equation for the vector mode discussed in \cite{CDK}, and such relation holds for any warp factor $h(t)$.

The EOM for $\xi$ has two solutions. The subleading solution that corresponds to the VEV of the operator $|Q^2|-|\tilde Q^2|$ represents the anti-self-dual flux on the D7 and does not break supersymmetry
\be
\label{kssusy}
\xi(t) = c_1 \, \big( \cosh(t)\sinh(t)-t \big)^{-1/3} \;.
\ee
The general asymptotic behavior  at infinity in the KS case is
\be
\xi(t) = \left(c_1 + c_2\,(4t-1)^2 \right)e^{-2t/3} + \cO(e^{-5t/3}) \;.
\ee
The leading solution has an extra $t \sim \log r$ compared with (\ref{asympt}), as can be understood in the KT limit from (\ref{eqB}) using the warp factor $h\sim \frac{L^4}{r^4}\log r$.

Something interesting occurs when we turn on the baryonic branch parameter $U$. The corresponding background is the BGMPZ solution that approaches the KS solution at infinity, but this does not guarantee that the asymptotic of the fields on the D7 are the same. The BGMPZ solution approaches the KS solution slowly enough to create a non-trivial source at large $r$ for some fluctuations of the worldvolume fields. This happens to $A_3$, and not to $A_{1,2}$. When $U\neq 0$ the $B$-field acquires an extra term
\be
B_\text{BGMPZ}= B_\text{KS} + \chi' dg_5 + \cO(U^2) \;,\qquad \chi' \rightarrow \frac U2 (t-1)e^{-2t/3} \;.
\ee
The new term has exactly the structure to couple to $\xi$ as both fluctuations correspond to the operators of dimension $2$ -- the bottom components of the $U(1)_{\rm baryon}$ and $U(1)_{\rm flavor}$ currents.
Therefore $\chi$ causes a non-homogeneous term in the linearized equation for $\xi$, and the asymptotic behavior takes the form
\be
\label{assymxiBGMPZ}
\xi = \left(c_1+c_2(4t-1)^2+ \frac{3U}{16}(2t-1)\right)e^{-2t/3} + \cO(e^{-5t/3}) \;.
\ee
This is the bulk manifestation of the mixing between $U(1)_{\rm baryon}$ and $U(1)_{\rm flavor}$.

\subsection{$SO(3)$ invariant flux on the D7-brane}

In this section we will find a general expression for the real $SO(3)$-invariant closed $(1,1)$ two-form  $F^{1,1} = dA$ on $\Sigma:\{z_4={\rm const}\}$, which combines with the pull-back of $B$ to form the gauge-invariant flux $\cF = P[B] + 2\pi \alpha' F$ on the D7. Supersymmetry requires $\cF$ to be of $(1,1)$-type, therefore we require $F$ to be $(1,1)$ as well.

There are four $(1,1)$ $SO(3)$-invariant 2-forms on $\Sigma$ that  can be combined with arbitrary $r$-dependent real coefficients $\zeta_1$, $\zeta_2$, $\lambda_1$, $\lambda_2$
\begin{align}
F^{1,1} = \FI + \FII \;,\qquad\qquad
\label{f1}
\FI &= i \big( \zeta_1 \, dz_i\wedge d\bar z_i+\zeta_2 \, \bar z_i dz_i \wedge z_j d\bar z_j \big) \;, \\
\label{f2}
\FII &= i\epsilon_{ijk}(\bar \lambda z_i-\bar z_i \lambda)\, dz_j \wedge d\bar z_k \;,
\end{align}
where we introduced a complex $\lambda=\lambda_1+i\lambda_2$.
The constraint $dF^{1,1} = 0$ boils down to the two independent equations $d\FI = d\FII = 0$.
The first can be rewritten in terms of the 1-forms $dt$, $g_5$ and $dg_5$ using
\be
\bar z_i \, dz_i= \frac{r^3}2 \Big( 3 \frac{dr}r + ig_5 \Big) \;.
\ee
The only possible closed combination is exact
\be
\label{a1}
\FI = d\AI \;,\qquad\qquad \AI = \xi(t) \, g_5 \;.
\ee
The second constraint implies (now $'$ stands for derivative with respect to $r^3$)
\be
2\lambda + \big[ \lambda' \, (r^3-|z_4|^2) + \bar \lambda' \, (z_4^2 - \epsilon) \big] = 0 \;.
\ee
The general solution is
\be
\label{lambda}
\lambda = - \frac n2 \, \frac{\sqrt{z_4^2-\epsilon}}{\big( r^3 - |z_4|^2 + |z_4^2 - \epsilon| \big)^2} + i \frac m2 \, \frac{\sqrt{z_4^2 - \epsilon}}{\big( r^3 - |z_4|^2 - |z_4^2-\epsilon| \big)^2}
\ee
with $n,m$ real coefficients.

Locally we can express $\FII$ as $\FII = d\AII$, in terms of an $SO(3)$-invariant potential $\AII$. The most general ansatz is
\be
\AII = \sigma \, \epsilon_{ijk} z_i \bar z_j \, dz_k + \text{c.c.}
\ee
The constraint that $d\AII$ be of $(1,1)$-type gives the equation for $\sigma$
\be
\big( (r^3-|z_4|^2)^2 - |z_4^2-\epsilon|^2 \big) \, \sigma' +2(r^3-|z_4|^2)\, \sigma = 0\;,
\ee
with solution
\be
\sigma = \frac{C_0}{(r^3-|z_4|^2)^2-|z_4^2-\epsilon|^2} \;.
\ee
To relate the complex constant $C_0$ to $n,m$ we compute $d\AII$ and cast it in the form (\ref{f2}): $i \lambda = \bar \sigma+(r^3-|z_4|^2) \, \bar\sigma' + (z_4^2-\epsilon)\, \sigma'$.
Eventually comparing with (\ref{lambda}) we find $C_0= \frac12 (m-in) \sqrt{\bar z_4^2 - \bar\epsilon}$.

If $m\neq 0$, $\FII$ is singular at the tip of $\Sigma$ and should be discarded.
If $m=0$, $\FII$ is regular but $\AII$ is still singular at the tip because $\FII$
is cohomologically non-trivial on $\S^2$.
We can parametrize the tip as $z_i = i x_i\sqrt{z_4^2 - \epsilon} $ in terms of real coordinates $x_i$ with $\sum_{i=1}^3 x_i^2 = 1$. Then $\FII = \frac n4 \, \epsilon_{ijk} \, x_i \, dx_j \wedge dx_k$ which gives
\be
\label{quant1}
\int_{\S^2} \FII = 2 \pi n \;.
\ee
Quantization requires $n$ to be integer.
Notice that in the resolved conifold case $\FII$ is proportional to the Betti-form $\omega_2$ on $T^{1,1}$ (\ref{w2})
\be
\label{w2remarks}
\omega_2 = -\frac i{r^6} \epsilon_{ijkl}\, z_i \bar z_j \, dz_k \wedge d\bar z_l
\ee
pulled-back on $\Sigma$: $\FII = \frac n2 P[\omega_2]$. This confirms that $\FII$ is cohomologically non-trivial.

We are interested in the Page D3- and D5-charge induced by the worldvolume gauge field on the D7-brane. The D3-charge is given by the integral of the current $J_\text{D3} = F_5 - B \wedge F_3 + \frac12 B \wedge B \wedge F_1$ on $T^{1,1}$, and the contribution from the D7 is given by the difference between the tip of the D7 at $r_\text{min}$ and very large radius: $(4\pi^2\alpha')^2 N_\text{D3} = \int_{r=\infty} J_\text{D3} - \int_{r = r_\text{min}} J_\text{D3} = \int_\cM dJ_\text{D3}$. Using $dJ_\text{D3} = (2\pi\alpha')^2 \frac12 F \wedge F \wedge \delta_2^\text{D7}$ (where $\delta_2^\text{D7}$ is a 2-form delta-function localized on the D7), we get
\be
N_\text{D3} = \frac1{8\pi^2} \int_\Sigma F \wedge F \;.
\ee
The computation is performed in appendix \ref{app: Page charges comp} and the result is
\be
\label{induced D3-charge}
N_\text{D3} = \frac{n^2}4 \;.
\ee

The D5-charge is given by the integral of the current $J_\text{D5} = F_3 - B \wedge F_1$ on $\S^3 \subset T^{1,1}$, and the contribution from the D7 is $(4\pi^2\alpha') N_\text{D5} = \int_{r = \infty} J_\text{D5} - \int_{r = r_\text{min}} J_\text{D5} = \int_{\S^3 \times \bR_+} dJ_\text{D5}$. Using $dJ_\text{D5} = (2\pi\alpha') F \wedge \delta_2^\text{D7}$ we get
\be
N_\text{D5} = \frac1{2\pi} \int_{\Gamma = \Sigma \cap (\S^3 \times \bR_+)} F \;.
\ee
The computation is performed in appendix \ref{app: Page charges comp} and the result is
\be
\label{induced D5-charge}
N_\text{D5} = \frac n2 \;.
\ee

As we saw in section \ref{sec: N=2 case}, another gauge-invariant is the conjugacy class of the Wilson loop $\hat\rho=\operatorname{Pexp} i \oint \A$, computed at large radius over the non-contractible contour $\partial \Gamma = \S^1$ on $\S^3/\bZ_2$. Note that such class is invariant under regular gauge transformations but can change under large gauge transformations of $B$. To compute $\hat\rho$ we integrate the field strength over $\Gamma$, which for a single D7-brane coincides with the calculation of the \Df-charge (see appendix \ref{app: Page charges comp} for details)
\be
\hat\rho = e^{i \oint \A} = e^{2\pi i N_{\Df}}=(-1)^n \;.
\ee

Eventually we interpret $n$ from the field theory point of view. Expanding $\AII$ at infinity and comparing the leading $r^{-3/2}$ asymptotic with (\ref{asympt}) we find that $n$ corresponds to the VEV
\be
\tilde QQ = n \sqrt{z_4^2 - \epsilon} \;.
\ee
The expectation value of $|Q^2|-|\tilde Q^2|$ depends on the $1/r^2$ asymptotic of $\xi$ and varies in different cases.
Let us note here that identifying the asymptotic behavior with the VEVs of the field theory operators as outlined in section \ref{sec: AdSCFT} is too naive because different operators may have the same quantum numbers and mix. Generically this happens when $\hat\rho$ is non-trivial, so that some extra fields are turned on at the boundary and the AdS/CFT dictionary needs to be corrected.
We will return to this problem in section \ref{sec: matching VEVs}.

Let us now consider in more details various setups and find explicitly the corresponding Abelian $U(1)$ instantons.

\subsection{Singular conifold}
\label{subsec: singular conif}

Consider the singular conifold with a D7-brane along $\Sigma: \{ z_4= \mu/ 2 \}$ and an arbitrary distribution of D3-branes. The latter only affect the warp factor which does not alter the supersymmetry condition for the D7-brane flux:
\be
P[J] \wedge \F = 0 \;,\qquad\qquad \F^{2,0} = 0
\ee
with $\cF = P[B] + 2\pi \alpha' F$.
The K\"ahler form on the singular conifold
\be
\label{KSC}
J = d\left(k g_5\right) \;,\qquad\qquad k= \frac{r^2}6 \;,
\ee
is of the form (\ref{f1}) and is orthogonal to the flux of type (\ref{f2}), see (\ref{orthogonality}). The $B$-field of the KW solution has the form $B=\pi \alpha' b \, \w2$,
where $\w2$ is given in (\ref{w2remarks}) and its pull-back is of the form (\ref{f2}), so it is automatically primitive. Therefore $\FII$ is not constrained and the resulting differential equation for $\xi$ can be easily solved
\be
\xi = \frac{\xi_0}{\a \, k} \;,
\ee
with $\xi_0$ a constant. The resulting $\xi$ is singular at $r_{min}$ either because $\a(r_{min})=0$ when $\mu\neq 0$ or $k(0)=0$ in the massless case $\mu=0$. Hence we must set $\xi_0=0$. The only surviving degree of freedom is the integer $n$ that parametrizes the flux $\FII$ (\ref{lambda}).

Empowered by the AdS/CFT dictionary developed in section \ref{sec: AdSCFT}, we derive that the background with $n$ units of D7 worldvolume flux is dual to a vacuum with VEVs
\be
\label{scvevs}
|Q^2| - |\tilde Q^2| = 0 \;,\qquad\qquad  \tilde{Q}Q= n z_4 \;.
\ee
In fact this is correct only when $\hat\rho = 1$ (so that there are no Wilson lines at the boundary), otherwise we should expect corrections to the AdS/CFT dictionary. Such corrections comes from the mixing of the operators above with other operators with the same quantum numbers, for instance $|Q^2| - |\tilde Q^2|$ can mix with $|A^2| - |B^2|$, while $\tilde QQ$ can mix with $\mu \unit$.
This effect equally applies to all other cases considered below. We will return to the matching of VEVs between the two sides of the duality in section \ref{sec: matching VEVs}.

\subsection{Resolved conifold}

Next consider the resolved conifold with a D7-brane along $\Sigma: \{ z_4= \mu/ 2 \}$ and an arbitrary distribution of D3-branes. The K\"ahler form compared with (\ref{KSC}) contains an extra term $a^2 \Vol_1$
\be
J= d \big( k \, g_5 \big) + \frac{a^2}2\, dg_5 + a^2 \omega_2 \;,\qquad\qquad k = \frac{r^2}6 \;.
\ee
The B-field is
\be
B= \pi \alpha' b \, \Big( \omega_2 + \frac12 d(f_g g_5) \Big)
\ee
where $f_g$ is some radial function. The 2-form $\omega_2$ is singular at the tip of the resolved conifold, and the extra piece makes $B$ regular provided that $f_g(0)=1$ (similarly to $\xi(0)=n/4$ in the massless $z_4=0$ case). At infinity $f_g\rightarrow 0$ to match the singular conifold case. The function $f_g$ is a pure gauge degree of freedom and is not fixed by the EOM. We choose it such that a trivial gauge field $F=0$ preserves supersymmetry
\be
f_g = \frac{a^2}{2k(r) +a^2} \;.
\ee
Since $P[\w2]$ can be expressed as $\FII$ with $n=2,m=0$, we can absorb $f_g$ into $\xi$ and impose primitivity of $\cF$
\be
\label{xirc}
\xi + \frac b2 \, f_g = \frac{a^2}4\, \frac{b + \frac n2}{k + a^2/2} + \a^{-1} \frac{\xi_0}{k + a^2/2} \;.
\ee
If $z_4\neq 0$, the second term is divergent at $r_\text{min}$ due to $\a^{-1}$ and we must set $\xi_0=0$. The first term is non-trivial and we derive
\be
\label{rcvevs}
|Q^2| - |\tilde Q^2| = n \frac{a^2}2 \;,\qquad\qquad  \tilde QQ = n z_4 \;.
\ee

If $z_4 = 0$ we cannot use (\ref{xirc}) because it was obtained by simplifying $\a'$ on both sides and $\a'=0$ in this case: we need to do the analysis anew.
First we set $\xi(0) = n/4$ to avoid a singularity of $F^{1,1}$ at the tip, and now $\int_\text{tip} F^{1,1} = 2\pi n$. Since $\FII$ is zero away from the tip, $P[J] \wedge \F =0$ implies
\be
\label{gfrc}
\xi + \frac b2\, f_g = \frac{\xi_0}{k(r) + a^2/2} \;.
\ee
To satisfy $\xi(0)=n/4$ we choose $\xi_0 = \frac{a^2}4 \big( b + \frac n2 \big)$
and find again (\ref{rcvevs}) with $\mu = 0$.
As discussed at the end of section \ref{sec: geometry of embedd}, the $z_4=0$ limit is smooth.
Moreover the solution  (\ref{gfrc}) is such that the coefficient in front of $dr\wedge g_5$ in $\F$ at $r=0$ vanishes: $(\xi+ \frac b2 f_g)'(0)=0$.

\subsection{Deformed conifold}

In the deformed conifold case, on the KS background the K\"ahler from is
\be
J = d \big( k \, dg_5 \big) \;,\qquad\qquad k(t) = \big( \cosh(t) \sinh(t) - t \big)^{1/3}
\ee
which has the form (\ref{f1}). The B-field is
\be
\label{B0}
B_{KS} = h_2(t) \cosh(t) \, \frac{2i \, \epsilon_{ijkl} z_i \bar z_j dz_k \wedge d\bar z_l}{\epsilon^4 \sinh t\cosh t}
\ee
where $h_2(t)$ is a suitable function, and $B$ has the form (\ref{f2}).
Therefore $n$ is not constrained and $\xi(t)$ must satisfy the differential equation giving
\be
\xi(t) = \frac{\xi_0}{\a(t) \, k(t)} \;,
\ee
which coincides with (\ref{kssusy}) when $\mu = 0$.
This  is singular at $t_\text{min}$ and hence $\xi=0$ , leaving only $n$ as free parameter.
Correspondingly we derive
\be
\label{dcvevs}
|Q^2| - |\tilde{Q}^2| = 0 \;,\qquad\qquad \tilde{Q}Q = n\sqrt{z_4^2-\epsilon} \;.
\ee

\subsection{BGMPZ solutions}
\label{bbbackgrounds}

The BGMPZ solutions \cite{Butti}, based on $SU(3)$-structure geometries, have a more complicated $\kappa$-symmetry condition. The computation for the type I flux (\ref{f1}) was carried out in
\cite{D7bb} -- here we add the type II flux (\ref{f2}). The $\kappa$-symmetry condition reads
\be
\label{kappas}
\frac U2 \big( J\wedge J - \F\wedge \F \big) + e^{2A} J\wedge \F \Big|_\Sigma = 0
\ee
where $U$ is the parameter along the baryonic branch and $A$ is the warp factor. The pseudo-K\"ahler form $J$ is the sum of two terms of type I and II:
\be
e^{2A} J = UB - d\big[ (\lambda +U\chi) g_5 \big] \;,\qquad B = B_{KS} + \chi dg_5 \;,\qquad
\lambda = U \, \frac{e^{2\phi} \, a(t\cosh t-\sinh t)}{2 (a\cosh t + 1)}
\ee
and so is $\F$
\be
\F = B_{KS} + \chi dg_5 + d(\xi \, g_5) + \FII(n) \;.
\ee
Here $B_{KS}$ is given by (\ref{B0}) but with some different function $h_2(t)$.
The term $B_{KS}\wedge\FII$ identically vanishes, and we get a differential equation for $\xi$:
\begin{multline}
\label{xiequation}
- \frac1\a \, \frac d{dt} \bigg[ \a \Big((\xi+\chi)^2 + \frac{2\lambda}U (\xi+\chi) + \Big( e^{-2\phi} \, h_2^2\, \sinh^2 t - \frac{\lambda^2}{U^2} (e^{-2\phi}-1) \Big) \Big) + \\
+ \frac{n^2 |z_4^2-\epsilon|}{8(r^3-|z_4|^2 + |z_4^2 - \epsilon|)} \bigg] = 0 \;.
\end{multline}
The equation can be integrated, in terms of a constant ${\rm c}_0$. At infinity $\lambda$ diverges as $-e^{2t/3} + \frac U2 (t-1) + \cO(e^{-2t/3})$, while $h_2 \sinh t$ remains finite, therefore only one root of the quadratic equation is meaningful
\begin{align}
\label{xisolbutti}
\xi+\chi &= \frac{ -\lambda -e^{-\phi} \sqrt{ \lambda^2 - U^2 h_2^2 \sinh^2 t - e^{2\phi} U^2 \a^{-1} \, {\rm c}}}U \;, \\
\label{valuec}
{\rm c}(t) &= {\rm c}_0 + \frac{n^2}8 \, \frac{|z_4^2-\epsilon|}{(\epsilon\cosh t -|z_4|^2 + |z_4^2 - \epsilon|)} \;.
\end{align}
 At the minimal radius $t_\text{min}$ the functions $\lambda$, $h_2 \sinh t$, $\phi$ are regular
but $\a^{-1}$ is singular, hence to avoid singularities we set ${\rm c}_0 = - n^2/16$ and the large $t$ asymptotic is
\be
\xi \,\rightarrow\, U \, \frac{8t^2+20t+35-2n^2}{64} \, e^{-2t/3} + \cO(e^{-5t/3}) \;,
\ee
in agreement with (\ref{assymxiBGMPZ}).
We interpret the asymptotic with $n=0$ as the one corresponding to the vacuum with $\tilde{Q}=Q=0$ effectively absorbing non-trivial $d(\xi g_5)$ into $B$. Then we derive
\be
\label{bbvevs}
|Q^2| - |\tilde{Q}^2| = - \frac{n^2 U |\epsilon|^{4/3}}{2^{2/3} 32} \;,\qquad\qquad \tilde QQ = n \sqrt{z_4^2 - \epsilon} \;.
\ee

In the special case $z_4 = 0$, although $\a\equiv 1$ and (\ref{xisolbutti}) remains finite at the tip for any ${\rm c}$, $dg_5$ is singular at the tip and we must require $\xi+\chi$ to vanish at $t=0$. This fixes ${\rm c}$ as in (\ref{valuec}).

\section{Backreaction of D7-branes}
\label{sec: backreaction d7}

In order to extract full information about the dynamics of the dual field theory, in particular its RG flow, one has to go beyond the probe approximation and construct a fully backreacted gravity solution. To do so for localized D7-branes is a hard problem. One possibility is to consider smeared solutions.%
\footnote{One could consider smearing orientifold planes as well, as in \cite{Acharya:2006ne}.}
In the Veneziano large $N_c$ limit, with $N_f/N_c \ll 1$ but fixed, the number $N_f$ of D7-branes is large and one can distribute them uniformly along the angular directions. This can be done supersymmetrically, and such a configuration has a precise field theory dual (discussed in section \ref{sec: RG flow}).

One can consider both massless and massive embeddings of \Ds-branes into the KW, KT, and KS backgrounds \cite{Benini:2006hh, Benini:2007gx, Benini:2007kg, Bigazzi:2008zt, Bigazzi:2008qq, Gaillard:2010qg, Nunez:2010sf}. We will consider here massless embeddings in KT \cite{Benini:2007gx}, and move the massive embeddings in KT with extra worldvolume flux to appendix \ref{app: backreacted massive}. We consider an $SU(2)\times SU(2)\times U(1)$ invariant ansatz
\beal
\label{backreacted ansatz}
ds^2 &= h^{-\frac12} dx_{3,1}^2 + h^{\frac12} \Big[ e^{2u} \Big( d\rho^2 + \frac19 g_5^2 \Big) + \frac{e^{2g}}6 \sum \big( d\theta_i^2 + \sin^2\theta_i\, d\varphi_i^2 \big) \Big] \\
J &= \frac{e^{2u}}3 d\rho \wedge g_5 + \frac{e^{2g}}6 \sum \sin\theta_i\, d\theta_i \wedge d\varphi_i \\
\Omega &= \frac16 e^{i\psi + u + 2g} \Big( d\rho + \frac i3 g_5 \Big) \wedge \big( d\theta_1 + i \sin\theta_1\, d\varphi_1 \big) \wedge \big( d\theta_2 + i \sin\theta_2\, d\varphi_2 \big) \\
\delta_2^\text{smeared} &= \frac{N_f}{4\pi} dg_5 =\frac{N_f}{4\pi} \sum \sin\theta_i\, d\theta_i \wedge d\varphi_i \;,\\
F_1 &= \frac{ N_f}{4\pi} g_5\ , \qquad\qquad
B = \alpha' \pi b(\rho) \, \omega_2\ , \qquad\qquad H_3 = \alpha' \pi b'(\rho) \, d\rho \wedge \omega_2\ ,
\eeal
where $u,g,b,h$ are functions of $\rho$ to be determined. $\rho$ is a new radial coordinate, which ranges from $-\infty$ in the deep IR to $0$ at the UV Landau pole. Roughly $\rho \sim \log \frac r{r_{\rm L}}$ where $r_{\rm L}$ is the radius associated to the Landau pole scale.  The smeared charge distribution 2-form $\delta_2^\text{smeared}$ is essentially fixed by symmetries, and $F_1$ has been chosen to satisfy
\be
dF_1 =  \delta_2^\text{smeared} \;.
\ee
The ansatz also includes the $SU(3)$-structure of the conifold: the K\"ahler form $J$ and the $(3,0)$-form $\Omega$ which refer to the 6d unwarped metric. A first set of SUSY equations \cite{Benini:2006hh, Benini:2007gx} is%
\footnote{In particular the $SU(3)$-structure satisfies the relations
$$
dJ = 2(g' - e^{2u-2g}) d\rho \wedge J = 0 \;,\qquad\qquad d\Omega = (2g' + u'-3) d\rho \wedge \Omega = - \frac12 d\phi \wedge \Omega \;.
$$}
\be
\phi' = \frac{3N_f}{4\pi} e^\phi\ , \qquad\qquad u' = 3 - 2e^{2u-2g} - \frac{3N_f}{8\pi} e^\phi\ , \qquad\qquad g' = e^{2u-2g}\ ,
\ee
while the solution with the proper boundary conditions is \cite{Benini:2006hh}
\be
\label{dilaton backreacted}
e^\phi = \frac{4\pi}{3N_f} \, \frac1{(-\rho)}\ , \qquad\qquad e^{2u} = -6\rho(1-6\rho)^{-2/3} e^{2\rho}\ , \qquad\qquad e^{2g} = (1-6\rho)^{1/3} e^{2\rho} \;.
\ee
Another SUSY equation is $H_3 = e^\phi *_6 F_3$, from which we get
\be
F_3 = \frac{ N_f \alpha'}4 (-\rho) \, b'\, g_5 \wedge \omega_2 \;.
\ee
Then we have $d\cF = P[H_3]$, with $\cF^{2,0}=0$ and $\cF \wedge P[ J] = 0$.
In the massless case $P[B]=0$ because $P[\omega_2] = 0$, and the only solution is $\cF = 0$.
That is because any normalizable flux on $\Sigma$ must be supported on the 2-cycle at the tip, while in the massless case, and within the KT approximation, such 2-cycle is shrunk to zero size.
Had we considered the KS setup, worldvolume flux on the massless D7s would be possible.%
\footnote{In section \ref{sec: field theory analysis} we discuss the corresponding field theory. For $\mu =0$, classically there are no vacua corresponding to a non-trivial worldvolume flux. Those vacua reappear, though, in the quantum theory which, on the gravity side, corresponds to the KS background.}
Finally the Bianchi identity $dF_3 = H_3 \wedge F_1$ fixes
\be
\label{b(rho) massless}
b(\rho) = \frac{c_1}{(-\rho)} + c_2 \;,
\ee
where $c_1,c_2$ are integration constants.
The self-dual 5-form flux $F_5$ is fixed by the Bianchi identity $dF_5 = H_3 \wedge F_3$ (where we neglected gravitational corrections on the D7s), and $F_5$ in turn fixes the warp factor via $C_4 = h^{-1} dx^0\wedge \dots \wedge dx^3$.

The integration constant $c_2$ is constrained by quantization of the Page D5-charge
\be
\label{value of c2 massless}
Q_\text{D5} = \frac1{4\pi^2\alpha' } \int_{\S^3} \big( F_3 - B \wedge F_1 \big) = - \frac{N_f c_2}2 \;.
\ee
This charge is sourced by D5-branes and the worldvolume flux on D7-branes, and has to be quantized in terms of the minimal charge in the setup. According to table (\ref{table charges}), it must be semi-integer.%
\footnote{Alternatively, one could compute the charge $\frac1{4\pi^2\alpha' } \int_{S^3} \big( F_3 - B \wedge F_1 - 2\pi \alpha' A \wedge \delta^\text{D7}_2 \big)$ which is sourced by D5-branes only, and needs to be integer.}
The integration constant $c_1$ is free and corresponds to changing the gauge couplings. We will use this solution in section \ref{sec: RG flow} to extract the RG flow.

\section{The conifold field theory with flavors}
\label{sec: field theory analysis}

The field theory dual to (fractional) D3-branes on the conifold,
as reviewed in section \ref{sec: review}, is the $\cN=1$ $SU(M+p) \times SU(p)$ quiver gauge theory \cite{KW,KG,KN,KT,KS}. The left node corresponds to wrapped D5-branes, the right node to \anti{D5}s each with $-1$ unit of worldvolume flux.
The addition of a non-compact D7-brane along the embedding $\Sigma:\{z_4 = \mu/2\}$ introduces a pair of quarks $Q, \tilde Q$ (one ``flavor'') of mass $\sqrt h\, \mu$ (the superpotential coupling $h$ appears because of a choice of normalization). The cycle $\Sigma$ contains a topologically non-trivial $\S^2$ and therefore there are two fractional D7-branes of minimal tension, distinguished by a monodromy $\hat\rho$ at infinity and by the flux at the tip. Similarly to the $\N=2$ $\Z_2$ orbifold case discussed in section \ref{sec: N=2 case}, a pure D7 introduces flavors coupled to the right node, while a D7 with $-1$ units of worldvolume flux introduces flavors to the left node \cite{Bertolini:2001qa, Benini:2007gx}. One way to obtain this result -- as well as the superpotential (\ref{full superpotential}) -- is to start from the $\cN=2$ orbifold $\bC \times \bC^2/\bZ_2$ and follow the RG flow discussed in \cite{KW, Ouyang:2003df}. The precise map between the D-brane charges and ranks in field theory is given in section \ref{sec: comparison}.

\begin{figure}
\centering
\begin{tikzpicture}[node distance=0.5cm, auto]

\node[rectangle,draw=black,thick,right=-2.9cm](flavorL) {$N_{fL}$};
\node[rectangle,draw=black,thick,right=4.5cm](flavorR) {$N_{fR}$};
\node[circle,draw=black, thick](leftgroup) {$N_1$};
\node[circle,draw=black, thick,right=2cm](rightgroup) {$N_2$};

\draw[->>,>=latex, shorten >=2pt, shorten <=2pt, bend left=45, thick]
    (leftgroup.north east) to node[auto, swap] {$A_\alpha$}(rightgroup.north west);
\draw[->>, >=latex, shorten >=2pt, shorten <=2pt, bend left=45, thick]
    (rightgroup.south west) to node[auto, swap] {$B_{\dot\alpha}$}(leftgroup.south east);

\draw[->, >=latex, shorten >=2pt, shorten <=2pt, bend right=45, thick]
    (flavorR.north west) to node[auto] {$\tilde Q_R$}(rightgroup.north east);
\draw[->, >=latex, shorten >=2pt, shorten <=2pt, bend right=45, thick]
    (rightgroup.south east) to node[] {$Q_R$}(flavorR.south west);

\draw[->, >=latex, shorten >=2pt, shorten <=2pt, bend left=45, thick]
    (flavorL.north east) to node[auto, swap] {$\tilde Q_L$}(leftgroup.north west);
\draw[->, >=latex, shorten >=2pt, shorten <=2pt, bend left=45, thick]
    (leftgroup.south west) to node[auto, swap] {$Q_L$}(flavorL.south east);

\end{tikzpicture}
\caption{Quiver of the flavored conifold theory. \label{fig: quiver flavors}}
\end{figure}
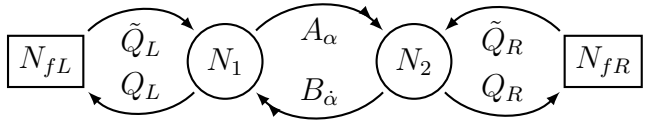

Summarizing, the gauge theory is a quiver with the gauge group $SU(N_1 = M+p) \times SU(N_2 = p)$ (we do not necessarily restrict to $N_1 \geq N_2$) and bifundamental fields $A_\alpha$, $B_{\dot\alpha}$ ($\alpha,\dot\alpha=1,2$) as in the pure conifold theory, with the addition of $N_{fL}$ flavors charged under $SU(N_1)$ and $N_{fR}$ flavors under $SU(N_2)$. We set $N_f = N_{fL} + N_{fR}$. The corresponding quiver in figure \ref{fig: quiver flavors} exactly coincides with the ADHM quiver of the $\N=2$ $\Z_2$ orbifold theory shown in figure \ref{qv2}. To denote this quiver we adapt the same notation as in section \ref{sec: N=2 case}
$$
N_{fL} \times N_1 \times N_2 \times N_{fR} \;.
$$
The full superpotential is (compare with (\ref{rough superpotential}))
\begin{multline}
\label{full superpotential}
\!\!\! W_0 = h (A_1B_1A_2B_2 - A_1B_2A_2B_1) - \sqrt h\, \eta_L \tilde Q_L \Big( A_1B_1 + A_2B_2 - \frac\mu{\sqrt h} \Big) Q_L + \frac{\eta_L^2}2 \tilde Q_L Q_L \tilde Q_L Q_L \\
- \sqrt h\, \eta_R \tilde Q_R \Big( B_1A_1 + B_2A_2 - \frac\mu{\sqrt h} \Big) Q_R - \frac{\eta_R^2}2 \tilde Q_R Q_R \tilde Q_R Q_R
\end{multline}
where trace is implicit.
Various factors of $h$ have been inserted for convenience. The coefficients of the quartic quark terms have specific values, which come from the $\cN=2$ orbifold theory broken to $\cN=1$ \cite{KW, Ouyang:2003df}.%
\footnote{Precisely, the superpotential (\ref{full superpotential}) is obtained from the $\cN=2$ theory with $U(N)$ gauge groups. Starting with $SU(N)$ gauge groups, one obtains other terms with different contraction of flavor indices. The difference is negligible in the large $N$ limit.}
One could consider deforming the theory by the marginal operators $\Tr \tilde Q_L Q_L \tilde Q_L Q_L$ and $\Tr \tilde Q_R Q_R \tilde Q_R Q_R$. These operators contain two traces over color indices and therefore correspond, on the gravity side, to a change of boundary conditions for the modes dual to $\Tr \tilde Q_L Q_L$, $\Tr \tilde Q_R Q_R$ \cite{Witten:2001ua}.
If the superpotential is ignored, the instanton factors related to the 1-loop-exact holomorphic (RG invariant) $\beta$-functions%
\footnote{As in \cite{ArkaniHamed:1997ut} we use holomorphic normalization for the gauge sector, $\frac1{4g^2} F \wedge *F$, and distinguish between holomorphic $\beta$-functions, where chiral matter fields are not renormalized, and physical $\beta$-functions, where chiral matter fields do have anomalous dimensions.} are
\be
\Lambda_1^{3N_1 - 2N_2 - N_{fL}} \equiv \Lambda_1^{b_1} \;,\qquad\qquad \Lambda_2^{3N_2 - 2N_1 - N_{fR}} \equiv \Lambda_2^{b_2} \;.
\ee

\begin{table}[tn]
{\footnotesize
$$
\begin{array}{c|ccccccc|l}
& U(1)_A & U(1)_B & U(1)_{Q_L} & U(1)_{\tilde Q_L} & U(1)_{Q_R} & U(1)_{\tilde Q_R} & U(1)_R & \widetilde{U(1)}_R \\
\hline
A & 1 & 0 & 0 & 0 & 0 & 0 & 1 & 1/2 \\
B & 0 & 1 & 0 & 0 & 0 & 0 & 1 & 1/2 \\
Q_L & 0 & 0 & 1 & 0 & 0 & 0 & 1 & 1/2 \\
\tilde Q_L & 0 & 0 & 0 & 1 & 0 & 0 & 1 & 1/2 \\
Q_R & 0 & 0 & 0 & 0 & 1 & 0 & 1 & 1/2 \\
\tilde Q_R & 0 & 0 & 0 & 0 & 0 & 1 & 1 & 1/2 \\
h & -2 & -2 & 0 & 0 & 0 & 0 & -2 & 0 \\
\eta_L & 0 & 0 & -1 & -1 & 0 & 0 & -1 & 0 \\
\eta_R & 0 & 0 & 0 & 0 & -1 & -1 & -1 & 0 \\
\mu & 0 & 0 & 0 & 0 & 0 & 0 & 1 & 1 \\
\Lambda_1^{b_1} & 2N_2 & 2 N_2 & N_{fL} & N_{fL} & 0 & 0 & 2N_1 & 2N_1 - 2N_2 - N_{fL} \\
\Lambda_2^{b_2} & 2N_1 & 2N_1 & 0 & 0 & N_{fR} & N_{fR} & 2N_2 & - 2N_1 + 2N_2 - N_{fR}
\end{array}
$$
}
\caption{Basis for the global Abelian symmetries. \label{tab: Abelian symm}}
\end{table}

The non-Abelian symmetries (vector-like and non-anomalous) are $SU(N_{fL}) \times SU(N_{fR}) \times SU(2)_{AB}$, where $SU(2)_{AB}$ is the anti-diagonal subgroup of $SU(2)_A \times SU(2)_B$ that preserves $A_\alpha B_{\dot\alpha} \delta^{\alpha\dot\alpha}$. The Abelian symmetries can be analyzed in the basis of table \ref{tab: Abelian symm}, and the exact symmetries are the subgroup under which no coupling or instanton factor is charged. For $\mu = 0$ it is $U(1)_b \times U(1)_{fL} \times U(1)_{fR} \times \Z_{q}$, where $\Z_q \subset \widetilde{U(1)}_R$ and $q = \gcd(2N_1 - 2N_2 - N_{fL}, 2N_1 - 2N_2 + N_{fR})$. $\mu \neq 0$ completely breaks $\widetilde{U(1)}_R$. The generators of the unbroken symmetries are
\beal
U(1)_{fL,fR} &= U(1)_{Q_{L,R}} - U(1)_{\tilde Q_{L,R}} \;,\qquad\qquad\qquad U(1)_b = U(1)_A - U(1)_B \\
\Z_q \subset \widetilde{U(1)}_R &= U(1)_R - \frac12 [ U(1)_A + U(1)_B + U(1)_{Q_L} + U(1)_{\tilde Q_L} + U(1)_{Q_R} + U(1)_{\tilde Q_R} ]
\eeal
where $U(1)_b$ is the usual baryonic symmetry of the conifold. It will also be convenient to define
\be
\label{baryonic 12}
U(1)_1 \equiv U(1)_b + U(1)_{fL} \qquad\qquad U(2)_2 \equiv - U(1)_b + U(1)_{fR}
\ee
which are the ``baryonic symmetries'' of the two $SU$ nodes.
We can form combinations of couplings that are invariant under flavor symmetries. They will correspond to supergravity parameters. First of all we take
\be
L_1 \equiv \Lambda_1^{b_1} h^{N_2} \eta_L^{N_{fL}} \;,\qquad\qquad L_2 \equiv \Lambda_2^{b_2} h^{N_1} \eta_R^{N_{fR}}
\ee
with R-charges $R[L_1] = 2N_1 - 2N_2 - N_{fL}$ and $R[L_2] = -2N_1 + 2N_2 - N_{fR}$. In the massless case we associate the following combinations to supergravity fields:
\be
I \equiv L_1 L_2 \sim e^{2\pi i \tau} \;,\qquad\qquad \frac{L_1}{L_2} \sim \exp \int_{\S^2} \big( B_2 + i C_2 \big)
\ee
as in \cite{DKS}. In the massive case we can construct the dimensionless invariants $\mu^{N_f} L_1 L_2$ and $\mu^{-4N_1 + 4N_2 + N_{fL} - N_{fR}} \frac{L_1}{L_2}$.

By a field redefinition we can take  $\eta_{L,R}=\sqrt h$.  This will lead to a simplified form of the superpotential (\ref{full superpotential}) which we will use in what follows.

\subsection{Seiberg duality, parameters and vacua}
\label{sec: Seiberg duality}

The field theory with superpotential (\ref{full superpotential}) has a remarkable property to be self-similar under Seiberg duality, up to a shift of ranks.
The superpotential is such that left and right quarks become simultaneously massless on the mesonic branch. This must be so, as quarks come from the D3-D7 strings and what distinguishes left quarks from right quarks is the flux on the D7s, not the embedding equation. In fact the coefficients of quartic quark terms are precisely such that the property that left and right flavors become simultaneously massless is invariant under the Seiberg duality. Besides being a map between theories, the duality is also a map between vacua, \eg{} what looks like a simple vacuum in one description may look complicated in another. We analyze here such issues.

Consider the superpotential $W_0$ (\ref{full superpotential}). We perform a Seiberg duality on the right node $SU(N_2)$. The mesons are
\be
M_{\alpha\dot\alpha} = \frac1\Lambda A_\alpha B_{\dot\alpha} \;,\qquad N_\alpha = \frac1\Lambda A_\alpha Q_R \;,\qquad \tilde N_{\dot\alpha} = \frac1\Lambda \tilde Q_R B_{\dot\alpha} \;,\qquad \Phi = \frac1\Lambda \tilde Q_R Q_R
\ee
and the dual quarks are
\be
A_\alpha \to c^\alpha \;,\qquad B_{\dot\alpha} \to d^{\dot\alpha} \;,\qquad \tilde Q_R \to r \;,\qquad Q_R \to \tilde r \;.
\ee
The magnetic gauge group is $SU(2N_1 + N_{fR} - N_2)$. The magnetic superpotential is $W_0$, written in terms of the magnetic variables, plus the extra terms $M_{\alpha\dot\alpha} d^{\dot\alpha} c^\alpha + N_\alpha \tilde r c^\alpha + \tilde N_{\dot\alpha} d^{\dot\alpha} r + \Phi \tilde r r$. On a branch of the moduli space where the fields $M_{\alpha\dot\alpha}$, $N_\alpha$, $\tilde N_{\dot\alpha}$, $\Phi$ are massive they can be integrated out via their F-term equations. We thus obtain the superpotential in the dual magnetic theory
\begin{multline}\nn
W_\text{mag} = \frac1{\Lambda^2 h} (c^1d^1c^2d^2 - c^1d^2c^2d^1) + \frac1\Lambda \tilde Q_L \big( d^1c^1 + d^2c^2 + \Lambda \sqrt h\, \mu \big) Q_L - \frac h2 \tilde Q_L Q_L \tilde Q_L Q_L \\
+ \frac1{\Lambda^2 h} \tilde r \big( c^1d^1 + c^2d^2 + \Lambda \sqrt h\, \mu \big) r + \frac1{2\Lambda^2h} \tilde rr \tilde rr
\end{multline}
where a constant term has been dropped. We can now redraw the quiver, flipping it horizontally and perform a further field redefinition
\beal
c^\alpha &= \epsilon^{\alpha\beta} \sqrt{\Lambda h}\, a_\beta \qquad\qquad &
\tilde r &= \sqrt{\Lambda h}\, \tilde q_L \qquad\qquad &
\tilde Q_L &= \tilde q_R \\
d^{\dot\alpha} &= - \epsilon^{\dot\alpha\dot\beta} \sqrt{\Lambda h}\, b_{\dot\beta} \qquad\qquad &
r &= \sqrt{\Lambda h}\, q_L \qquad\qquad &
Q_L &= q_R \;.
\eeal
The resulting superpotential
\begin{multline}
W_\text{mag} = h (a_1b_1a_2b_2 - a_1b_2a_2b_1) - h\, \tilde q_L \Big( a_1b_1 + a_2b_2 - \frac\mu{\sqrt h} \Big) q_L + \frac h2 \tilde q_L q_L \tilde q_L q_L \\
- h\, \tilde q_R \Big( b_1a_1 + b_2a_2 - \frac\mu{\sqrt h} \Big) q_R - \frac h2 \tilde q_R q_R \tilde q_R q_R
\end{multline}
is manifestly identical to the initial one (\ref{full superpotential}). Notice that the flip exchanges both the gauge ranks and the number of flavors $N_{fR}$ and $N_{fL}$.

To conclude, we can map mesonic gauge-invariant operators with respect to the dualized node from the electric theory to the magnetic one:
\beal
\label{vacua map SD}
b_{\dot\alpha} a_\alpha &= A_\alpha B_{\dot\alpha} - \delta_{\alpha\dot\alpha} Q_L \tilde Q_L \qquad\qquad &
\tilde q_L a_\alpha &= - \epsilon\du{\alpha}{\dot\beta} \tilde Q_R B_{\dot\beta} \qquad\qquad &
q_R &= Q_L \\
\tilde q_L q_L &= \tilde Q_R Q_R - \frac\mu{\sqrt h}\, \unit_{N_{fR}} \qquad\qquad &
b_{\dot\beta} q_L &= \epsilon\du{\dot\beta}{\alpha} Q_\alpha Q_R \qquad\qquad &
\tilde q_R &= \tilde Q_L \;.
\eeal
When a Seiberg duality is performed on the left node, the same formul\ae{} hold by exchanging the electric with the magnetic theory.

Let us now look at the real operators in the bottom component of current supermultiplets. To simplify the discussion, consider SQCD$_{n_c,n_f}$ with quarks $Q,\tilde Q$ and baryons $B = Q^{n_c}, \tilde B = \tilde Q^{n_c}$. The dual description SQCD$_{n_f-n_c,n_f}$ has quarks $q,\tilde q$ and baryons $b = q^{n_f - n_c}, \tilde b = \tilde q^{n_f - n_c}$. The map $b = \Lambda^{n_f - 2n_c} B \cdot \epsilon$, and similarly for tilded quantities, implies the following map for the bottom component of the baryonic current multiplet at weak coupling:
$$
\frac1{n_c}\big( |Q^2| - |\tilde Q^2| \big) = \frac1{n_f - n_c} \big( |q^2| - |\tilde q^2| \big) \;.
$$
Now consider dividing the quarks into two groups: $Q \to (Q_R,P)$ in number $(n_{fR}, n_f - n_{fR})$ (and similarly for tilded quarks). This amounts to considering a subgroup of the global symmetry $SU(n_f) \to SU(n_{fR}) \times SU(n_f - n_{fR}) \times U(1)_\text{aux}$, and defines a splitting of the dual quarks $q \to (q_R,p)$. From $U(1)_\text{baryon}$ and $U(1)_\text{aux}$ we can construct a symmetry $U(1)_{fR}$ that only gives charge $\pm1$ to $Q_R, \tilde Q_R$ respectively. From the charges of quarks and dual quarks we get the map
$$
\frac1{n_{fR}} \big( |Q_R^2| - |\tilde Q_R^2| \big) = - \frac{n_f - n_c - n_{fR}}{n_{fR}(n_f - n_c)} \big( |q_R^2| - |\tilde q_R^2| \big) + \frac1{n_f - n_c} \big( |p^2| - |\tilde p^2| \big) \;.
$$
If we now translate that relation in terms of our quiver, we obtain for the bottom components of the current supermultiplets of $U(1)_{fL,R}$ in the electric description:
\beal
\label{current multiplets map}
|\tilde Q_L^2| - |Q_L^2| &= |\tilde q_R^2| - |q_R^2| \\
(2N_1 - N_2 + N_{fR}) \big( |\tilde Q_R^2| - |Q_R^2| \big) &= (2N_1 - N_2) \big( |\tilde q_L^2| - |q_L^2| \big) + N_{fR} \big( |a^2| - |b^2| \big) \;.
\eeal

\subsection{The classical moduli space}
\label{sec: classical moduli space}

We start our quest of understanding the moduli space with the classical analysis by finding the space of solutions of the F-term and D-term equations, modded out by gauge equivalences. The F-term equations are
\beal
\label{F-term equations}
0 &= B_1A_2B_2 - B_2A_2B_1 - B_1 Q_L \tilde Q_L - Q_R \tilde Q_R B_1 \\
0 &= B_2A_1B_1 - B_1A_1B_2 - B_2 Q_L \tilde Q_L - Q_R \tilde Q_R B_2 \\
0 &= A_2B_2A_1 - A_1B_2A_2 - Q_L \tilde Q_L A_1 - A_1 Q_R \tilde Q_R \\
0 &= A_1B_1A_2 - A_2B_1A_1 - Q_L \tilde Q_L A_2 - A_2 Q_R \tilde Q_R \\
0 &= \Big( A_1B_1 + A_2B_2 - Q_L \tilde Q_L - \frac\mu{\sqrt h} \Big) Q_L = \tilde Q_L \Big( A_1B_1 + A_2B_2 - Q_L \tilde Q_L - \frac\mu{\sqrt h} \Big) \\
0 &= \Big( B_1A_1 + B_2A_2 + Q_R \tilde Q_R - \frac\mu{\sqrt h} \Big) Q_R = \tilde Q_R \Big( B_1A_1 + B_2A_2 + Q_R \tilde Q_R - \frac\mu{\sqrt h} \Big)
\eeal
while the D-term equations following from the canonical K\"ahler potential (at the classical level we disregard corrections to the K\"ahler potential) are
\beal
\label{D-term equations}
\xi_1 \unit_{N_1} &=  A_\alpha A_\alpha^\dag -  B_{\dot\alpha}^\dag B_{\dot\alpha} + Q_L Q_L^\dag - \tilde Q_L^\dag \tilde Q_L \\
\xi_2 \unit_{N_2} &=  B_{\dot\alpha} B_{\dot\alpha}^\dag -  A_\alpha^\dag A_\alpha + Q_R Q_R^\dag - \tilde Q_R^\dag \tilde Q_R \;.
\eeal
Here $\xi_{1,2}$ are free parameters to be determined. If the global symmetries $U(1)_{1,2}$ defined in (\ref{baryonic 12}) are gauged, then $\xi_{1,2}$ become FI terms. In general only one linear combination of $\xi_{1,2}$ can be turned on such that the equations above are satisfied and supersymmetry is preserved, and we will later specify which linear combination depending on the branch of the moduli space.
In the following we use the notations for the (deformed) conifold introduced in section \ref{reviewks}
\be
\cC_\epsilon = \big\{ \det_{\dot\alpha\alpha} w_{\dot\alpha\alpha} = \epsilon \big\} \;.
\ee

The classical moduli space has an intricate structure that we summarize here. First, there are mesonic directions where $A_\alpha$, $B_{\dot\alpha}$ take VEV with $\xi_{1,2} = 0$ and $Q_{L,R} = \tilde Q_{L,R} = 0$. For suitable choices of $N_1$, $N_2$ there can be a baryonic direction where $A_\alpha$, $B_{\dot\alpha}$ take VEV with $\xi_{1,2} \neq 0$ while still $Q_{L,R} = \tilde Q_{L,R} = 0$. These two branches are essentially the same as in the unflavored theory. Second, there are instanton-like directions, when VEVs of $Q_{L,R}$, $\tilde Q_{L,R}$ partially break the gauge group while preserving $N_1 - N_2$. This time $Q_{L,R}$, $\tilde Q_{L,R}$ have moduli and these branches are continuously connected with the mesonic/baryonic directions.
Finally, there are Higgsed mesonic directions (only for $\mu \neq 0$) when $Q_{L,R}$, $\tilde Q_{L,R}$ take VEV and break the gauge group $SU(N_1)\times SU(N_2)$ to two smaller $SU$ factors and changing the difference $N_1 - N_2$. These vacua are disconnected from the previous ones.
For both Higgsed mesonic and instanton-like directions, the low energy theory with the unbroken gauge group usually sits in a mesonic vacuum although in certain cases the parameters $\xi_{1,2}$ can be turned on as well.

\paragraph{Mesonic directions.} Up to gauge transformations, the mesonic vacua are
\beal
\label{meson}
A_\alpha &= \mat{A_\alpha^{(1)} \\ & \ddots \\ & & A_\alpha^{(p)} \\ 0 & \dots & 0 \\ \vdots && \vdots \\ 0 & \dots & 0} \qquad\qquad
B_{\dot\alpha}^\trans &= \mat{B_{\dot\alpha}^{(1)} \\ & \ddots \\ & & B_{\dot\alpha}^{(p)} \\ 0 & \dots & 0 \\ \vdots && \vdots \\ 0 & \dots & 0} \\
\sum_\alpha & |A_\alpha^{(a)}|^2 - \sum_{\dot\alpha} |B_{\dot\alpha}^{(a)}|^2 = 0 \qquad \forall a
\eeal
and $Q_L = \tilde Q_L = Q_R = \tilde Q_R = 0$. Here we assumed $M=N_1-N_2>0$.

At a generic point on the moduli space the gauge group $SU(M+p) \times SU(p)$ is broken to $SU(M) \times U(1)^{p-1} \times \text{Weyl}$ (for $M>0$). The $U(1)$ factors are diagonally embedded, $SU(M) \subset SU(N_1)$ and the Weyl group permutes the $U(1)$s.
The moduli are characterized by the coordinates
\be
\label{complex coordinates}
w_{\dot\alpha\alpha}^a \equiv \sqrt h \, B_{\dot\alpha}^{(a)} A_\alpha^{(a)}
\ee
($\sqrt h$ inserted for convenience) which satisfy $\det_{\dot\alpha\alpha} w_{\dot\alpha\alpha}^a = 0$. This gives a symmetric product (because of the Weyl group) of $p$ copies of the singular conifold
$$
\text{Sym}_p (\cC_0) \;.
$$
At a generic point the low energy spectrum contains the $SU(M)$ gauge multiplet, $3p$ neutral chiral multiplets (parametrizing the moduli) and $p-1$ Abelian vector multiplets. The $U(1)^{p-1}$ groups have $N_f$ flavors, generically of mass $\sqrt h \big( \mu - \Tr_{\dot\alpha\alpha} w_{\dot\alpha\alpha}^a \big)$. $SU(M)$ has $N_{fL}$ flavors of mass $\sqrt h\, \mu$, which become massless for $\mu = 0$, and quartic superpotential. If we start from the origin of the mesonic branch with gauge group $SU(N_1) \times SU(N_2)$ and give large VEV to only one mesonic component, the gauge group is broken to $SU(N_1-1) \times SU(N_2 -1) \times U(1)$ and $U(1)$ gauges the symmetry $U(1)_1 - U(1)_2 = 2U(1)_b + U(1)_{fL} - U(1)_{fR}$ of the low energy theory. At the last step, where we are left with $SU(M)$ and $N_{fL}$ flavors, one linear combination of the $U(1)^{p-1}$ gauges $U(1)_{fL}$.

We have three baryonic symmetries -- $U(1)_b$ and $U(1)_{fL,R}$ -- and we can gauge any linear combination.
For instance if we gauge $U(1)_b$, at low energy we get $p$ $\cN=4$ Abelian vector multiplets (at special points on the mesonic branch there will be massless $\cN=2$ flavors). We can also add a FI term $\xi \equiv \xi_1 = -\xi_2$. If $N_1 = N_2 = N$ we have SUSY vacua describing $p$ symmetrized copies of the resolved conifold
\be
\sum_\alpha |A_\alpha^{(a)}|^2 - \sum_{\dot\alpha} |B_{\dot\alpha}^{(a)}|^2 = \xi \qquad \forall a
\ee
with $Q_{L,R} = \tilde Q_{L,R} = 0$.
To parametrize the tip we need, besides the mesons, the baryons (\ref{baryons}). If $N_1 > N_2$, supersymmetry is broken for $N_f = 0$ but it might be preserved for $N_f>0$ if the ranks $N_1$, $N_2$ allow for a Higgsed mesonic vacuum (discussed below) whose low energy theory is $SU(\tilde N) \times SU(\tilde N)$.

Let us comment here  on the $SU(M)$ non-perturbative dynamics at low energies if $N_1 > N_2$. We will distinguish between the massless and massive cases in what follows.

We start with the massless case $\mu =0$. Since $SU(M) \subset SU(N_1)$, to get the instanton factors by scale matching we give large VEV to $N_2$ components of $A,B$. Each time we turn on one component, the breaking pattern is $SU(N_1) \times SU(N_2) \to SU(N_1-1) \times SU(N_2-1) \times U(1)$. The SQCD$_{n_c, n_f}$ theory goes to SQCD$_{n_c-1, n_f - 2}$ as a result of a VEV $\langle AB \rangle$ and a mass term $h \langle AB \rangle$ from the superpotential. In the final expression the value of VEV cancels out and we are left with instanton factors
$$
\Lambda_{1\,\text{low}}^{3N_1 - 2N_2 - N_{fL} -1} = \Lambda_1^{3N_1 - 2N_2 - N_{fL}} h \;,\qquad\qquad \Lambda_{2\,\text{low}}^{3N_2 - 2N_1 - N_{fR} -1} = \Lambda_2^{3N_2 - 2N_1 - N_{fR}} h \;.
$$
Repeating $N_2$ times, we are left with $SU(M)$ with $N_{fL}$ flavors, instanton factor
\be
\Lambda^{3M - N_{fL}} = \Lambda_1^{3N_1 - 2N_2 - N_{fL}} h^{N_2}
\ee
and a quartic superpotential $W_0 = \frac h2 \tilde Q_L Q_L \tilde Q_L Q_L$. The dynamically generated on-shell superpotential on the mesonic branch%
\footnote{SQCD$_{n_c,n_f}$ with quartic superpotential has an intricate structure \cite{Carlino:2000ff}. For $n_f < n_c$, the number of vacua is $(2n_c - n_f)2^{n_f-1}$, all with the same dynamical scale
$$
W_\text{eff} = \frac{2n_c - n_f}2 \big( \Lambda^{2(3n_c - n_f)} h^{n_f} \big)^{1/(2n_c - n_f)} \;.
$$
}
is
\be
\label{effective superpotential classic}
W_\text{eff}(\text{vacua}) = \frac{2N_1 - 2N_2 - N_{fL}}2 \big( \Lambda_1^{2(3N_1 - 2N_2 - N_{fL})} h^{2N_2 + N_{fL}} \big)^\frac1{2N_1 - 2N_2 - N_{fL}} \;.
\ee

In the massive case $\mu \neq 0$, we can discuss two different scenarios:  large or small mass $\sqrt h\, \mu$. For large mass we integrate out the flavors first and obtain the instanton factors $\Lambda_{1\, \text{low}}^{3N_1 - 2N_2} = (\sqrt h\, \mu)^{N_{fL}} \Lambda_1^{3N_1 - 2N_2 - N_{fL}}$ and $\Lambda_{2\, \text{low}}^{3N_2 - 2N_1} = (\sqrt h\, \mu)^{N_{fR}} \Lambda_2^{3N_2 - 2N_1 - N_{fR}}$. Then we break $SU(N_2)$ by moving on the mesonic branch, while preserving unbroken $SU(M) \subset SU(N_1)$ with instanton factor and on-shell superpotential
\be
\label{vacua large mass no VEV}
\Lambda^{3M} = \Lambda_1^{3N_1 - 2N_2 - N_{fL}} h^{N_2} (\sqrt h\, \mu)^{N_{fL}} \;,\qquad\qquad W_\text{eff} \sim \big( \Lambda^{3M} \big)^\frac1M \;.
\ee
For small mass we break $SU(N_2)$ on the mesonic branch first and obtain a massive quartic SQCD$_{N_1 - N_2, N_{fL}}$ with $\Lambda_\text{low}^{3(N_1 - N_2)-N_{fL}} = \Lambda_1^{3N_1- 2N_2-N_{fL}} h^{N_2}$. For $(\sqrt h\, \mu)^{2n_c - n_f} \ll \Lambda_\text{low}^{3n_c - n_f} h^{n_c}$ the theory is essentially massless and we recover (\ref{effective superpotential classic}). In the opposite limit the theory has vacua where $SU(n_c)$ is broken to $SU(n_c - j)$ and the on-shell superpotential is
\be
\label{low energy other vacua}
W_\text{eff} \sim \big( \Lambda_\text{low}^{3n_c - n_f} h^j (\sqrt h\,\mu)^{n_f - 2j} \big)^{\frac1{n_c - j}} \sim \big( \Lambda_1^{3N_1 - 2N_2 - N_{fL}} h^{N_2 + j} (\sqrt h\, \mu)^{N_{fL} - 2j} \big)^{\frac1{N_1 - N_2 - j}} \;.
\ee
For $j=0$ we recover the vacua in (\ref{vacua large mass no VEV}). For $1 \leq j \leq n_c$ we have Higgsed mesonic vacua, more precisely $j$ blocks with $n=-1$ (discussed below).%

\paragraph{Baryonic direction.}
These vacua are present if $N_1 = (k+1)M$, $N_2 = kM$. Let us first define the \emph{Upper} and \emph{Lower} $(k+1)\times k$ matrices \cite{DKS}
\be
U_k = \mat{\sqrt k & 0 & \dots & 0 & 0 \\ 0 & \sqrt{k-1} & \dots & 0 & 0 \\ \vdots & \vdots & & \vdots & \vdots \\ 0 & 0 & \dots & \sqrt2 & 0 \\ 0 & 0 & \dots & 0 & 1 \\ 0 & 0 & \dots & 0 & 0} \qquad\qquad
L_k = \mat{0 & 0 & \dots & 0 & 0 \\ 1 & 0 & \dots & 0 & 0 \\ 0 & \sqrt2 & \dots & 0 & 0 \\ \vdots & \vdots & & \vdots & \vdots \\ 0 & 0 & \dots & \sqrt{k-1} & 0 \\ 0 & 0 & \dots & 0 & \sqrt k} \;.
\ee
They satisfy the quadratic relations
\be
U_k^\trans U_k + L_k^\trans L_k = (k+1) \unit_k \;,\qquad
U_k U_k^\trans + L_k L_k^\trans = k \unit_{k+1} \;,\qquad
U_{k+1} L_k = L_{k+1} U_k \;.
\ee
Up to a gauge transformation the classical vacua are given by
\be
\label{cbb}
A_1 = C\, U_k \otimes \unit_M \;,\qquad A_2 = C\, L_k \otimes \unit_M \;,\qquad B_1 = B_2 = 0 \;,\qquad Q_{L,R} = \tilde Q_{L,R} = 0 \;.
\ee
There is another set with $A \leftrightarrow B^\trans$. Here $C$ is an arbitrary complex number. The vacua (\ref{cbb}) satisfy the D-term equations with $\xi_1 = k |C|^2$, $\xi_2 = - (k+1)|C|^2$ and  $\xi_1\leftrightarrow -\xi_2$ when $A \leftrightarrow B^\trans$. The branches are parametrized by either the baryon ${\mathcal A}\sim(A_1A_2)^{k(k+1)M/2}$ or the anti-baryon ${\mathcal B}\sim(B_1B_2)^{k(k+1)M/2}$.
The origin of the baryonic branch touches (classically) the origin of the mesonic branch.

For $\tilde p \equiv N_2\  \rm{mod}\ M ( = N_1 - N_2) \neq 0$ there is no baryonic flat direction. One way to see that is to give mesonic VEVs to $\tilde p$ directions. This breaks the gauge group to $SU \big((k+1)M \big) \times SU(kM) \times U(1)^{\tilde p}$. Although this is very close to the theory with the baryonic branch discussed above the low energy bifundamentals are charged under a linear combination of $U(1)^{\tilde p}$. Hence the D-term equations set $C = 0$ and the resulting vacuum belongs to the mesonic direction.

\paragraph{Instanton-like directions.}
This branch is the piece of the Higgs branch continuously connected to the mesonic directions discussed above. The vacua are in one-to-one correspondence with a similar Higgs branch in the $\cN=2$ case, indeed any solution to the $\N=2$ $\bC^2/\Z_2$ ADHM equations (\ref{adhmcomp}) with $\xi_{\C}^L=\xi_{\C}^R={\mu/\sqrt{h}}$ and $\xi^L_{\R}=\xi_1$, $\xi^R_{\R}=\xi_2$ solves the $\N=1$ equations (\ref{F-term equations}), (\ref{D-term equations}). Thus all instantons of the $\N=2$ $\bC^2/\Z_2$ theory are present in the $\N=1$ conifold theory as well, and the two spaces have equal dimension.

The instanton-like vacua have a block diagonal form and can be of the ``Left'' or the ``Right'' type. They are parametrized by two integers $(n_c \geq 1, n_f \geq 2)$ which define the size of the blocks:
\be
\label{d3type}
\text{Left}: n_f \times n_c \times n_c \times 0 \;,\qquad\qquad \text{Right}: 0 \times n_c \times n_c \times n_f \;.
\ee
Moreover solutions for the blocks exist for any choice of $\xi_{1,2}$.
To calculate $\tilde QQ$ one would need to know an explicit solution, but the VEV of $|Q^2|-|\tilde Q^2|$ follows immediately from the equations (\ref{D-term equations}) and the block size:
\be
|Q^2| - |\tilde Q^2| = n_c(\xi_1+\xi_2) \;.
\ee

Depending on the ranks of the unbroken gauge symmetry, $\xi_1$ and $-\xi_2$ can be zero, equal to each other, linearly dependent or arbitrary. The vacua with $\xi_1+\xi_2\neq 0$ correspond to the noncommutative instantons on $\C^2/\Z_2$ in the $\N=2$ case.%
\footnote{Although the field theory of section \ref{sec: N=2 case} admits $\xi_1+\xi_2\neq 0$ only when all gauge symmetry is broken, in the conifold theory more general situations are possible, for instance $\xi_1 + \xi_2 \neq 0$ is found on the baryonic branch.}

The instanton-like vacua describe D3-branes dissolved inside the D7-branes. In general the D3-branes can become point-like instantons and leave the D7s, so these directions touch the mesonic directions (but not the baryonic one).

\paragraph{Higgsed mesonic directions.}
Other disconnected branches of vacua exist in which the two gauge ranks are broken by an unequal amount. Such vacua have a block diagonal form $n_f^L\times n_c^1\times n_c^2\times n_f^R$ with $n_c^1 \neq n_c^2$, and in the classical theory they only exist for $\mu \neq 0$. They are disconnected from the mesonic and baryonic branches discussed before, and for each value of $n_c^1 - n_c^2$ we get a different disconnected branch.

Below we focus on the cases with $n_f^L+n_f^R=1$, which correspond to the Abelian instantons (the more general directions are obtained by ``adding'' non-Abelian instantons).
In this case either $Q_L, \tilde Q_L$ or $Q_R, \tilde Q_R$ acquire VEV and this generically forces  $A_\alpha, B_{\dot\alpha}$ to acquire VEV as well. We parametrize the blocks by an integer $r \in \bZ$, and their dimension is (compare with (\ref{n2left}) and (\ref{n2right}))
\be
\label{block dimensions}
\text{Left: } 1 \times r^2 \times r(r-1) \times 0 \qquad\qquad\qquad \text{Right: } 0 \times r(r+1) \times r^2 \times 1 \;.
\ee
The case $r=0$ coincides with the mesonic flat directions. The quivers (\ref{block dimensions}) can be obtained from the $r=0$ case via a chain of Seiberg dualities discussed in section \ref{sec: comparison}.

Notice that the left gauge rank minus the right gauge rank equals $r$, and there is a symmetry that flips left and right and maps $r \to -r$. We can parametrize both left and right blocks by an integer $n \in \bZ$ defined as
\be
\label{ndef}
n = \left\{ \begin{aligned} &2r-1 \quad &&\text{left} \\ &2r \quad &&\text{right} \end{aligned} \right. \qquad\qquad\Leftrightarrow\qquad\qquad r = \Big[ \frac{n+1}2 \Big]_-
\ee
where $[x]_-$ is the highest integer equal or smaller than $x$.
For $r \neq 0$, $\sign n = \sign r$.
The number $n$ is what appears in the supergravity description.
The blocks contain coefficients $a_1,\dots,a_{K-1}$, and we will define $K$ such that formally $a_K \equiv 0$.
The left blocks for $r \geq 1$ ($1 \times r^2 \times r(r-1) \times 0$) are given by
\beal
\label{blocks Higgsed left}
A_1 &= \beta\alpha \mat{ a_1 U_1^\trans & 0 & \dots \\ a_2 L_2 & a_3 U_3^\trans & \dots \\ 0 & a_4 L_4 & \dots \\ \vdots & \vdots & \ddots} \qquad &
B_1^\trans &= \beta\alpha \mat{ a_1 U_1^\trans & 0 & \dots \\ -a_2 L_2 & a_3 U_3^\trans & \dots \\ 0 & -a_4 L_4 & \dots \\ \vdots & \vdots & \ddots} \\
A_2 &= \beta\alpha \mat{ a_1 L_1^\trans & 0 & \dots \\ -a_2 U_2 & a_3 L_3^\trans & \dots \\ 0 & -a_4 U_4 & \dots \\ \vdots & \vdots & \ddots} &
B_2^\trans &= \beta\alpha \mat{ a_1 L_1^\trans & 0 & \dots \\ a_2 U_2 & a_3 L_3^\trans & \dots \\ 0 & a_4 U_4 & \dots \\ \vdots & \vdots & \ddots} \\
\tilde Q_L &= Q_L^\trans = \alpha \mat{1 & 0 & \dots & 0} \;, & \tilde Q_R &= Q_R = 0 \;.
\eeal
The unknowns%
\footnote{The number of unknowns is really one less, because we could reabsorb $\beta$ into $a_j$. We will fix this redundancy later.}
are $a_1, \dots, a_{2r-2}$, $\alpha$, $\beta$ and $K = 2r-1$.
The right blocks for $r \geq 1$ ($0 \times r(r+1) \times r^2 \times 1$) are given by
\beal
\label{blocks Higgsed right}
A_1 &= \beta\alpha \mat{ a_1 U_1 & a_2 L_2^\trans & 0 & \dots \\ 0 & a_3 U_3 & a_4 L_4^\trans & \dots \\ 0 & 0 & a_5 U_5 & \dots \\ \vdots & \vdots & \vdots & \ddots} \qquad &
B_1^\trans &= \beta\alpha \mat{ -a_1 U_1 & a_2 L_2^\trans & 0 & \dots \\ 0 & -a_3 U_3 & a_4 L_4^\trans & \dots \\ 0 & 0 & a_5 U_5 & \dots \\ \vdots & \vdots & \vdots & \ddots} \\
A_2 &= \beta\alpha \mat{ -a_1 L_1 & a_2 U_2^\trans & 0 & \dots \\ 0 & -a_3 L_3 & a_4 U_4^\trans & \dots \\ 0 & 0 & -a_5 L_5 & \dots \\ \vdots & \vdots & \vdots & \ddots} &
B_2^\trans &= \beta\alpha \mat{ a_1 L_1 & a_2 U_2^\trans & 0 & \dots \\ 0 & a_3 L_3 & a_4 U_4^\trans & \dots \\ 0 & 0 & a_5 L_5 & \dots \\ \vdots & \vdots & \vdots & \ddots} \\
\tilde Q_L &= Q_L = 0 \;, & \tilde Q_R &= Q_R^\trans = \alpha \mat{1 & 0 & \dots & 0} \;.
\eeal
The unknowns are $a_1, \dots, a_{2r-1}$, $\alpha$, $\beta$ and $K = 2r$. The blocks of one kind with $r \leq -1$ are obtained from the blocks of the other kind with $r\geq1$ by taking the transpose of $A_\alpha$, $B_{\dot\alpha}$ and exchanging $Q_L \leftrightarrow Q_R$, $\tilde Q_L \leftrightarrow \tilde Q_R$. The left blocks for $r \leq -1$ ($1 \times |r|^2 \times |r|(|r|+1) \times 0$) have $a_1,\dots,a_{2|r|-1}$ and $K = 2|r|$. The right blocks for $r \leq -1$ ($0 \times |r|(|r|-1) \times |r|^2 \times 1$) have $a_1,\dots,a_{2|r|-2}$ and $K = 2|r|-1$. In all cases $r\neq 0$ the number of $a_j$'s is $\big| n + \frac12 \big| - \frac32$ while $K = \big| n + \frac12 \big| - \frac12$.

The D-term equations are solved by arbitrary $a_j$. From the F-terms we get equations that fix  $a_j$'s through the recursive relation
\be
0 = j\, a_j^2 - a_{j+1}^2 - (j+3) a_{j+2}^2 \qquad \text{for } j = 1,\dots, K-2 \;,\qquad\qquad a_K \equiv 0 \;.
\ee
With some choice of normalization the solution is
\be
\label{solution aj}
a_j^2 = \frac{(2K+1) - (-1)^{j+K}(2j+1)}{j(j+1)}
\ee
and all $a_j$'s are positive. The other unknowns are give by
\be
\label{solution alpha beta}
\beta^2 = \frac{\sign(r)}{a_1^2 + 3a_2^2} = \frac 1{4r}\ , \qquad\qquad \alpha^2 = (-1)^{n+1} \frac\mu{\sqrt h} \, \frac{a_1^2 + 3a_2^2}{a_1^2 - 3a_2^2} = - \frac\mu{\sqrt h} \, r \;.
\ee
From here we can extract the VEV of the quark bilinear
\be
\label{chiral VEVs Higgsed mesonic}
\tilde Q_i Q_i = \alpha^2 = - \frac\mu{\sqrt h} \, r
\ee
where $i=L,R$ depending on the block, while $|Q^2| - |\tilde Q^2|=0$.

Let us note that the explicit solutions above and the ones in appendix \ref{app: Higgsed FI} (discussed below) also solve the $\N=2$ $\bC^2/\Z=2$ ADHM equations (\ref{adhmcomp}).

The Higgs vacua break the theory at scale $(\mu/\sqrt h)^{1/2}$. Each block reduces color and flavor ranks according to its dimension (\ref{block dimensions}). Below the breaking scale the low-energy theory $SU(\tilde N_1) \times SU(\tilde N_2)$ can have mesonic or, if $\tilde N_2 = k(\tilde N_1 - \tilde N_2)$, baryonic directions.
In the massless $\mu = 0$ case all these vacua collapse to the origin of the mesonic directions \ie{} since $\alpha^2\sim \mu$ all fields are zero. We will see that in the quantum theory the vacua described above do not degenerate in the $\mu\rightarrow 0$ limit and survive as independent.

Finally, the Higgsed mesonic vacua correspond to D3 and D5-branes dissolved in the D7s. Because of the D5s, these vacua are not continuously connected with the mesonic/baryonic directions.

\paragraph{Higgsed mesonic directions with resolution.} The Higgsed blocks discussed above can be modified to solve the vacuum equations with generic parameters $\xi_1$ and $\xi_2$. Possible constraints on $\xi_{1,2}$ will come from the remaining components of the D-term equations along the directions with unbroken gauge symmetry.
The explicit solutions generalizing  (\ref{blocks Higgsed left}) and (\ref{blocks Higgsed right}) can be found in appendix \ref{app: Higgsed FI}.
However the VEV of $|Q^2|-|\tilde{Q}^2|$ follows directly from (\ref{D-term equations}) and the size of the blocks:
\beal
\text{L:}\qquad & Q^\dag_L Q_L - \tilde Q_L \tilde Q_L^\dag = r^2 \xi_1 + r(r-1)\xi_2 \;, \\
\text{R:}\qquad & Q^\dag_R Q_R- \tilde Q_R \tilde Q_R^\dag = r(r+1) \xi_1 + r^2\xi_2 \;.
\eeal
Notice that the result is independent of $\mu$, and indeed such vacua remain non-trivial in the $\mu \to 0$ limit.

If the unbroken gauge group is  $SU(N) \times SU(N)$ one can turn on $\xi_1 = - \xi_2$ in the low energy theory  causing the VEV
\be
Q^\dag Q - \tilde Q \tilde Q^\dag = r \, \xi_1 \;.
\ee
If it is $SU((k+1)M) \times SU(kM)$ the low energy theory develops a baryonic branch with $(k+1) \xi_1 = -k \xi_2$ and the VEV
\beal
\label{classical real VEV deformed}
\text{L:}\qquad & Q^\dag_L Q_L - \tilde Q_L \tilde Q_L^\dag = \frac{r^2 - (k+1)r}{k+1} \, \xi_2 \\
\text{R:}\qquad & Q^\dag_R Q_R- \tilde Q_R \tilde Q_R^\dag = \frac{r^2 - kr}{k+1} \, \xi_2 \;.
\eeal

Let us comment on $k$-dependence in (\ref{classical real VEV deformed}). Different $k$ correspond to different steps along the cascading RG flow of the same theory, therefore well-defined physical quantities should not depend on $k$. The reason why (\ref{classical real VEV deformed}) is $k$-dependent is that the definitions of $U(1)_{fL}$ and $U(1)_{fR}$ are not invariant under Seiberg duality -- as we saw in section  \ref{sec: Seiberg duality} -- the precise relation being (\ref{current multiplets map}). It is a simple exercise to show that the VEVs (\ref{classical real VEV deformed}) are a consequence of the map (\ref{current multiplets map}). In short, as we go up in energy and perform a Seiberg duality on the right node, $k^\text{up} = k + 1$ and $n^\text{up} = n + 1$ (exchanging right and left flavors). Moreover $\xi_2^\text{up} = \frac{k+2}{k+1} \xi_2$, as follows analyzing the theory below the Higgsing scale.

\subsection{Quantum moduli space: $2N_2 + N_{fL} < N_1$}
\label{sec: quantum mod 1}

Here we start analyzing how quantum corrections modify the moduli space. In this and the following sections we will consider $M\geq 0$. The case $M<0$ is obtained by flipping the quiver. The quantum moduli space depends on the gauge ranks and number of flavors. We start with $2N_2 + N_{fL} < N_1$, in which case there are no baryonic directions.
The left node goes to strong coupling in the IR while the right node goes to weak coupling. The left node is parametrized by its mesons
\be
\label{mesons def}
\cM = \mat{B_1A_1 & B_1A_2 & B_1Q_L \\ B_2A_1 & B_2A_2 & B_2Q_L \\ \tilde Q_LA_1 & \tilde Q_LA_1 & \tilde Q_LQ_L} = \mat{ M_{11} & M_{12} & N_1 \\ M_{21} & M_{22} & N_2 \\ \tilde N_1 & \tilde N_2 & \Phi} \;.
\ee
First, we study the dynamics of the left node alone as if the right node had zero coupling, and then we gauge the $SU(N_2)$ group and introduce the corresponding D-term equations.

Along the moduli space of the left node there is a dynamically generated Affleck-Dine-Seiberg (ADS) superpotential \cite{Affleck:1983mk}
\be
\label{W_ADS}
W_\text{ADS} = (N_1 - 2N_2 - N_{fL}) \Big( \frac{\Lambda_1^{3N_1 - 2N_2 - N_{fL}}}{\det \cM} \Big)^{\frac1{N_1 - 2N_2 - N_{fL}}} \;.
\ee
The total effective superpotential is a sum of two terms $W_\text{eff} = W_\text{ADS} + W_0$,
\begin{multline}
W_0 =  \Tr \bigg[ h (M_{12}M_{21} - M_{11}M_{22})
- h \Big(\tilde N_1N_1 + \tilde N_2N_2 - \frac\mu{\sqrt h} \Phi \Big) \\
+ \frac h2 \Phi^2 - h \tilde Q_R \Big( M_{11} + M_{22} - \frac\mu{\sqrt h} \Big) Q_R - \frac h2 \tilde Q_RQ_R\tilde Q_RQ_R \bigg] \;.
\end{multline}
It will be convenient to introduce a matrix $\cN$, equal to the variation of the classical superpotential with respect to the mesons
\be
\cN_{ij} \equiv \parfrac{W_0}{\cM_{ji}} = -h \mat{ M_{22} + Q_R \tilde Q_R & - M_{12} & N_1 \\
- M_{21} & M_{11} + Q_R \tilde Q_R & N_2 \\
\tilde N_1 & \tilde N_2 & - \frac\mu{\sqrt h} \unit - \Phi} \;.
\ee
The F-term equations therefore are
\begin{align}
\label{mesonic F-term deformed}
\cN &= \Big( \frac{\Lambda_1^{3N_1 - 2N_2 - N_{fL}}}{\det \cM} \Big)^\frac1{N_1 - 2N_2 - N_{fL}} \cM^{-1} \;, \\
\label{F-terms quark deformed}
0 &= \Big( M_{11} + M_{22} + Q_R \tilde Q_R - \frac\mu{\sqrt h} \Big) Q_R = \tilde Q_R \Big( M_{11} + M_{22} + Q_R \tilde Q_R - \frac\mu{\sqrt h} \Big) \;.
\end{align}
Calling $-\epsilon$ the factor on the right hand side of the first equation and multiplying by $\cM$ on the left and on the right we get $\cM \, \cN = \cN \, \cM = -\epsilon \, \unit_{2N_2 + N_{fL}}$. This is a counterpart of the classical F-term equation with a dynamically generated term. Since the right node is IR free it has a canonical K\"ahler potential and the D-term equation is
\be
\label{D-terms deformed}
[M_{\dot\alpha\alpha},M_{\dot\alpha\alpha}^\dag] + N_{\dot\beta} N_{\dot\beta}^\dag - \tilde N_\beta^\dag \tilde N_\beta + Q_R Q_R^\dag - \tilde Q_R^\dag \tilde Q_R = \xi_2 \unit_{N_2} \;.
\ee

The solutions to these equations form a quantum deformed version of the mesonic and Higgsed mesonic directions of section \ref{sec: classical moduli space}. They have the same block diagonal form, each block describing dissolved \Dt\ and \Df-branes. To illustrate how it works we will find the solutions for the quantum counterparts of the ``Left'' and ``Right'' Abelian Higgs vacua (\ref{block dimensions}) (excluding some special cases, the generic non-Abelian instanton-like directions cannot be presented in a closed form)
\be
\text{Left: } 1 \times r(r-1) \times 0 \;,\qquad\qquad \text{Right: } 0 \times r^2 \times 1 \;.
\ee
Here the ranks refer to $U(N_{fL}) \times SU(N_2) \times U(N_{fR})$ whilst $SU(N_1)$ is confined.
The explicit form of the matrices, for $\xi_2 = 0$, is given in appendix \ref{app: quantum Higgsed}.

The blocks $0 \times 1 \times 0$ that correspond to the mesonic directions along the Coulomb branch (representing mobile D3-branes) only have VEVs of the mesons $M_{\dot\alpha\alpha}$ which satisfy $M_{11}M_{22} - M_{12}M_{21} = \epsilon/h$. In terms of the complex coordinates $w^a_{\dot\alpha\alpha}$ (\ref{complex coordinates}) we have
\be
\det_{\dot\alpha\alpha} w^a_{\dot\alpha\alpha} = \epsilon \;.
\ee
The D3-branes move on a deformed conifold with the deformation parameter $\epsilon$.

We are particularly interested in the quark bilinear. In the ``Left'' and ``Right'' cases it is given by
\be
\label{chiral VEVs deformed Higgsed mesonic}
\Phi = -\sqrt{ \frac{\mu^2/4 - \epsilon}h} \, (2r-1) - \frac\mu{2\sqrt h} \;,\qquad\text{ or }\qquad
\tilde Q_R Q_R = - \sqrt{ \frac{\mu^2/4 -\epsilon}h} \, 2r
\ee
where the branch cut has been chosen to match the $\epsilon\to 0$ limit of section \ref{sec: classical moduli space}.

Finally we determine $\epsilon$. In the massless case $\mu = 0$, all $2N_2 + N_{fL}$ components of $\cM$ are of order $\sqrt{\epsilon/h}$ implying
\be
\label{epsilon massless}
\epsilon \sim \big( \Lambda_1^{2(3N_1 - 2N_2 - N_{fL})} h^{2N_2 + N_{fL}} \big)^{\frac1{2N_1 - 2N_2 - N_{fL}}} \;,
\ee
which agrees with the semiclassical computation (\ref{effective superpotential classic}). This result does not depend on the particular Higgsed vacuum.
For large mass $\mu^2 \gg \epsilon$ and in the trivial vacuum $\cM$ has $2N_2$ components of order $\sqrt{\epsilon/h}$ and $N_{fL}$ of order $\mu/\sqrt h$, so that
\be
\label{epsilon large mu trivial}
\epsilon \sim \big( \Lambda_1^{3N_1 - 2N_2 - N_{fL}} h^{N_2} (\sqrt h\, \mu)^{N_{fL}} \big)^\frac1{N_1 - N_2} \;,
\ee
which agrees with the semiclassical result (\ref{vacua large mass no VEV}).

It is also possible to find the solutions to (\ref{mesonic F-term deformed})-(\ref{D-terms deformed}) for generic values of $\xi_2$, as is done in appendix \ref{app: quantum Higgsed FI}.
The corresponding VEVs in the ``Left'' and ``Right'' cases are
\be
\label{real VEVs resolved deformed}
\sum_{\dot\alpha} |N_{\dot\alpha}^2| - \sum_\alpha |\tilde N_\alpha^2| = r(r-1)\xi_2 \;,\qquad\text{ or }\qquad
|Q_R^2| - |\tilde Q_R^2| = r^2 \xi_2 \;.
\ee
Both agree with the semi-classical computation (\ref{classical real VEV deformed}) for $k=0$ which makes perfect sense as we are considering the IR theory which corresponds to the last step of the cascade. The VEVs of the chiral operators $\tilde Q_R Q_R$ and $\Phi$ are independent of $\xi_2$.

\subsection{Quantum moduli space: $2N_2 + N_{fL} \geq N_1$}
\label{sec: quantum mod 2}

When $2N_2 + N_{fL} \geq N_1$ we have to consider three different cases.

\paragraph{Case $\bm{2N_2 + N_{fL} = N_1}$.}
The left node, which runs to strong coupling in the IR, has baryons and a quantum deformed moduli space, whilst the right node is IR free. We construct the baryons $\cA = A^{2N_2} Q_L^{N_{fL}}$, $\cB = B^{2N_2} \tilde Q_L^{N_{fL}}$ which are singlets of $SU(N_2)$ and $SU(2)_{AB}$. The quantum deformed moduli space is described by the superpotential
\be
W = W_0 + X(\det\cM - \cA \cB - \Lambda_1^{2N_1})
\ee
where $X$ is a Lagrange multiplier. Besides the constraint $\det\cM - \cA \cB = \Lambda_1^{2N_1}$, we also get the F-term equations $0 = \cN + X(\det\cM)\,\cM^{-1}$, $0 = X\cA = X \cB$ together with (\ref{F-terms quark deformed}).

There are two separate branches. The mesonic branch (characterized by $X \neq 0$) where $\cA = \cB = 0$ and therefore $\det\cM = \Lambda_1^{2N_1}$. The solutions along this branch are the same as in the previous section (with the identification $\epsilon = X \det\cM$). The dynamically generated scale $\epsilon$ follows the same formul\ae: (\ref{epsilon massless}) in the massless case, and (\ref{epsilon large mu trivial}) in the case of large mass and trivial vacuum.

The baryonic branch is characterized by $\cA,\cB \neq 0$, while $X=0$ and $\cN = 0$.%
\footnote{The matrix $(\det\cM)\, \cM^{-1}$ is the matrix of cofactors of $\cM$ and therefore is a smooth function of $\cM$.}
In particular $M_{12} = M_{21} = N_{\dot\alpha} = \tilde N_\alpha = 0$, $M_{11} = M_{22} = -Q_R \tilde Q_R$ and $\Phi = - \frac\mu{\sqrt h} \unit$. The D-term equations and (\ref{F-terms quark deformed}) force the eigenvalues of $M_{11}$ to be either 0 or $-\frac\mu{\sqrt h}$, with $\rank M_{11} \leq N_{fR}$. For $N_{fR} < N_2$ -- which will be our focus%
\footnote{For instance, for $N_{fR} = N_2$ there are vacua with $|\det \cM| = (\mu/\sqrt h)^{2N_2 + N_{fL}}$. By suitably tuning $\mu$ one could obtain $\cA \cB = 0$, that is no deformation. Indeed this corresponds in supergravity to a configuration with singular D7 embedding $\mu^2=\epsilon$.}
-- this also implies $\det\cM = 0$ and therefore $\cA \cB = - \Lambda_1^{2N_1}$.
Let us compute the dynamically generated scale in the massless case. Above the scale $\Lambda_1$, the second group $SU(N_2)$ has $2N_1 - N_{fR}$ flavors and the instanton factor $\Lambda_2^{3N_2 - 2N_1 - N_{fR}}= \Lambda_2^{-N_2 - 2N_{fL} - N_{fR}}$. Because of confinement of $SU(N_1)$ below the scale $\Lambda_1$ it has 4 adjoints and $2N_{fL} + N_{fR}$ flavors (plus singlets), and the same instanton factor. All mesons receive mass $h\Lambda_1^2$, therefore the low energy theory is a quartic SQCD$_{N_2, N_{fR}}$ with the instanton factor $\Lambda_1^{8N_2 + 4N_{fL}} \Lambda_2^{-N_2 - 2N_{fL} - N_{fR}} h^{4N_2 + 2N_{fL}}$ and an effective superpotential
\be
W_\text{eff} \sim \big( \Lambda_1^{16N_2 + 8N_{fL}} \Lambda_2^{-2N_2 -4N_{fL} -2N_{fR}} h^{8N_2 + 4N_{fL} + N_{fR}} \big)^\frac1{2N_2 - N_{fR}} \;.
\ee

\paragraph{Case $\bm{2N_2 + N_{fL} = N_1 +1}$.}
The moduli space of the strongly coupled left node is described by mesons and baryons with a superpotential. The baryons $\bar\cA$ and $\bar\cB$ are in the fundamental and anti-fundamental representation of the flavor group $U(2N_2 + N_{fL})$, and we can decompose them as $\bar\cA=(\cA^\alpha,\tilde\cF)$ and $\bar\cB=(\cB^{\dot\alpha},\cF)$ respectively. The superpotential is
\be
W = W_0 - \frac1{\Lambda_1^{3N_1 - 2N_2 - N_{fL}}} \big( \det\cM - \bar\cB \cM \bar\cA \big) \;.
\ee
The F-term equations are $\cN = \Lambda_1^{-(3N_1 - 2N_2 - N_{fL})} (\cM^{-1}\det\cM - \bar\cA \bar\cB)$, $0 = \cM \bar\cB = \bar\cA \cM$ and (\ref{F-terms quark deformed}).

The moduli space has two separate branches. On the mesonic branch $\det\cM \neq 0$, therefore $\bar\cA = \bar\cB = 0$ and $\cN = \frac{\det\cM}{\Lambda_1^{3N_1 - 2N_2 - N_{fL}}} \cM^{-1}$. The solutions have been described in section \ref{sec: quantum mod 1}, and the scale $\epsilon$ is as in (\ref{epsilon massless}) and (\ref{epsilon large mu trivial}).

There is another branch where $\bar\cA, \bar\cB \neq 0$ and $\det\cM = 0$. The F-term equation set
\be
\cN = - \frac1{\Lambda_1^{3N_1 - 2N_2 - N_{fL}}} \, \bar\cA \bar\cB \;.
\ee
In particular, defining $\cA_\alpha = \epsilon_{\alpha\beta} \cA^\beta$, $\cB_{\dot\alpha} = \epsilon_{\dot\alpha\dot\beta} \cB^{\dot\beta}$, we have
\beal
M_{\dot\beta\alpha} + \delta_{\dot\beta\alpha} Q_R \tilde Q_R &= \frac h{\Lambda_1^{3N_1 - 2N_2 - N_{fL}}} \cA_\alpha \cB_{\dot\beta} \;,\qquad &
N_{\dot\alpha} &= \frac h{\Lambda_1^{3N_1 - 2N_2 - N_{fL}}} \delta_{\dot\alpha\alpha} \cA^\alpha \cF \;, \\
- \frac\mu{\sqrt h} \unit - \Phi &= \tilde\cF \cF \;,\qquad &
\tilde N_\alpha &= \frac h{\Lambda_1^{3N_1 - 2N_2 - N_{fL}}} \delta_{\alpha\dot\alpha} \tilde\cF \cB^{\dot\alpha} \;,
\eeal
which formally coincide with (\ref{vacua map SD}). The superpotential has the form
$$
W \sim \frac1{h{\Lambda_1^{3N_1 - 2N_2 - N_{fL}}}} \big[ \det_{\alpha\dot\alpha} \cA_\alpha \cB_{\dot\alpha} + \dots \big]
$$
where the missing terms reproduce $W_0$. We get a $\hat{N_{fL}} \times \hat N_1 \times \hat N_2 \times \hat{N_{fR}}$ theory but with ranks
$$
N_{fR} \times N_2 \times 1 \times N_{fL} \;.
$$
If $2 + N_{fR} < N_2$  we can borrow the results from section \ref{sec: quantum mod 1}. Let us consider the $\mu=0$ case. First we need to match the scales. To that end we canonically normalize the baryons $\hat{\bar \cA} = \bar\cA/\Lambda_1^{N_1 - 1}$, $\hat{\bar \cB} = \bar\cB/\Lambda_1^{N_1 - 1}$ getting the  coefficient in front of the superpotential $\hat h = 1/(h\Lambda_1^2)$. Then we match the scale of $SU(N_2)$. Above $\Lambda_1$ it has the instanton factor $\Lambda_2^{3N_2 - 2N_1 - N_{fR}}$. Below $\Lambda_1$ it has 4 adjoints and $2+2N_{fL} + N_{fR}$ fundamentals, with the instanton factor $\Lambda_\text{low}^{-N_2 -2N_{fL} - N_{fR}} \sim \Lambda_2^{3N_2 - 2N_1 - N_{fR}} \Lambda_1^{-4}$. The 4 adjoints and $2N_{fL}$ fundamentals get mass $h\Lambda_1^2$, so that the scale of the $SU(N_2)$ factor is
$$
\hat\Lambda_1^{3N_2-2-N_{fR}} \sim \Lambda_1^{8N_2 + 4N_{fL} - 4} \Lambda_2^{-N_2 - 2N_{fL} - N_{fR}+2} h^{4N_2 + 2N_{fL}} \;.
$$
Eventually we can plug the hatted quantities in (\ref{epsilon massless}):
\be
\epsilon \sim \big( \Lambda_1^{2(8N_2 + 4N_{fL} - N_{fR} - 6)} \Lambda_2^{2(-N_2 -2N_{fL} - N_{fR} +2)} h^{8N_2 + 4N_{fL} + N_{fR} -2} \big)^\frac1{2N_2 - 2 - N_{fR}} \;.
\ee

\paragraph{Case $\bm{2N_2 + N_{fL} >  N_1 +1}$.}
All the remaining cases are very similar to the previous one. First, the mesonic branch of the $SU(N_1)$ node (which might or might not be strongly coupled) is described by the effective Affleck-Dine-Seiberg superpotential (\ref{W_ADS}). Considering $W_\text{eff} = W_0 + W_\text{ADS}$ plus the D-term equations of $SU(N_2)$ we get the same type of solutions -- (Higgsed) mesonic directions -- as in section \ref{sec: quantum mod 1}. The deformation scale is again given by (\ref{epsilon massless}) or (\ref{epsilon large mu trivial}).

The analysis above however does not exhaust all set of vacua. To find the remaining ones, we dualize the $SU(N_1)$ node to a $SU(2N_2 + N_{fL} - N_1)$ gauge group with mesons $\cM$, dual quarks $\hat A$, $\hat B$ and a superpotential
$$
W = W_0 + \frac1\Lambda \hat B \cM \hat A \;,
$$
where $W_0$ is expressed in terms of $\cM$ and $Q_R, \tilde Q_R$, and the role of the scale $\Lambda$ is explained in \cite{Intriligator:1995au}.
For generic values of $\cM$ the dual quarks are massive, the $SU(2N_2 + N_{fL} - N_1)$ group can be integrated out and we reproduce $W_\text{eff} = W_0 + W_\text{ADS}$ and the mesonic branch above.

On the other hand if we integrate out the massive mesons as in section%
\footnote{In section \ref{sec: Seiberg duality} we dualized the right node, going ``up in energy'', and then flipped the quiver. Proceeding backwards we go ``down in energy'', as here. Besides we used a different normalization of the mesons $\cM$.}
\ref{sec: Seiberg duality}, we reproduce the same theory but with different ranks
\be\nn
N_{fR} \times N_2 \times (2N_2 + N_{fL} - N_1) \times N_{fL} \;.
\ee
This theory has its own (Higgsed) mesonic vacua, plus possibly other vacua obtained by further dualizations. Notice that in the process the quiver is flipped, and mesonic operators are mapped as in (\ref{vacua map SD}). Therefore a vacua labeled by $\hat n$ in the dual theory has a VEV for $\tilde QQ$ corresponding to $n = \hat n + 1$. We will see in section \ref{sec: comparison} what is the supergravity counterpart of this fact.

To get the dynamically generated scale on the mesonic vacua we proceed to match the scales. The left node $SU(N_1)$ is dualized to $SU(\hat N_2)$, with $\hat N_2 = 2N2 + N_{fL} - N_1$. Choosing the normalization scale $\Lambda = \Lambda_1$, we simply have $\hat \Lambda_2 = \Lambda_1$. Integrating out the mesons and rewriting the superpotential in terms of $\hat A$, $\hat B$, we get $\hat h = 1/h\Lambda_1^2$.
The right node $SU(N_2)$ is untouched, so that $\hat N_1 = N_2$. However its dynamical scale gets modified. Above the scale $\Lambda_1$ its instanton factor is $\Lambda_2^{3N_2 - 2N_1 - N_{fR}}$. Below the scale $\Lambda_1$ the gauge group has 4 adjoints and $4N_2 + 4N_{fl} - 2N_1 + N_{fR}$ flavors, with the instanton factor $\Lambda_\text{low}^{2N_1 - 5N_2 -4N_{fL}-N_{fR}} \sim \Lambda_2^{3N_2 - 2N_1 - N_{fR}} \Lambda_1^{4(N_1 - 2N_2 - N_{fL})}$. The adjoints and $2N_{fL}$ fundamentals get mass $h\Lambda_1^2$, so that the low energy $SU(\hat N_1)$ group has an instanton factor $\hat\Lambda_1^{2N_1 - N_2 -2N_{fL} - N_{fR}} \sim h^{4N_2 + 2N_{fL}} \Lambda_1^{4N_1} \Lambda_2^{3N_2 - 2N_1 - N_{fR}}$. Bringing all together we have
\beal
\hat N_{fL} &= N_{fR} \;,\qquad \hat N_1 = N_2 \;,\qquad \hat N_2 = 2N_2 + N_{fL} - N_1 \;,\qquad \hat N_{fR} = N_{fL} \;, \\
\hat h &= \frac1{h\Lambda_1^2} \;,\qquad \hat\Lambda_1^{2N_1 - N_2 -2N_{fL} - N_{fR}} \sim h^{4N_2 + 2N_{fL}} \Lambda_1^{4N_1} \Lambda_2^{3N_2 - 2N_1 - N_{fR}} \;,\qquad \hat \Lambda_2 = \Lambda_1 \;.
\eeal
One could now plug these values in (\ref{epsilon massless}), for instance, to obtain the deformation parameter in the vacua of the theory dualized once. Proceeding in the same way one could obtain $\epsilon$ in all other vacua of the theories dualized multiple times (unfortunately we could not find a closed formula). Notice in particular that at each dualization the parameters of the low energy theory are related to those of the high energy theory by
\be
\hat I = \frac I{(h\Lambda_1)^{N_{fR}}} \;,\qquad\qquad \hat L_1 = \frac{L_1 \, I}{(h\Lambda_1)^{N_{fR}}} \;.
\ee

\section[Comparison: supergravity {\it vs} field theory]{Comparison: supergravity {\it vs}\ \ field theory}
\label{sec: comparison}

The map between the supergravity solutions presented in section \ref{sec: adding d7} and the vacua of the field theories discussed in section \ref{sec: field theory analysis} starts with the UV identification. The parameters that identify the field theory, at some energy scale, are the gauge ranks $N_{1,2}$ and the number of flavors $N_{fL,R}$. In supergravity one can compute the Page charges $Q_\text{D3}$, $Q_\text{D5}$, count the number $N_f(r)$ of D7-branes (we will suppress the dependence on $r$ in the following) and measure the  Wilson line $\hat\rho$ at some cut-off radius representing the UV scale.

The relation between the supergravity charges and the field theory ranks is found with a dictionary. The mutually BPS probe branes on the conifold are: two types of fractional D3-branes (a D5-brane wrapped on the conifold's $\S^2$ and an anti-D5-brane on $\S^2$ with $-1$ units of worldvolume flux $F$) each giving rise to one color (vector multiplet) in the quiver, and two types of fractional D7-branes (both wrapping the $\S^2$, one without and one with $-1$ units of worldvolume flux) each giving rise to one flavor (hypermultiplet). The fact that the D7 without flux gives one flavor coupled to the right node (and vice versa) was first observed in the $\bC^2/\bZ_2$ $\cN=2$ orbifold case in \cite{Bertolini:2001qa} (see our section \ref{sec: N=2 case}). In the conifold case this matches with the expected RG flow (section \ref{sec: RG flow}). The Page charges for different D-branes are
\be
\label{table charges}
\begin{array}{c|cccc}
 & \text{D3}_1 & \text{D3}_2 & \text{D7}_{fL} & \text{D7}_{fR} \\
\hline
Q_\text{D3} & 0 & 1 & \frac14 & 0 \\
Q_\text{D5} & 1 & -1 & -\frac12 & 0 \\
Q_\text{D7} & 0 & 0 & 1 & 1
\end{array}
\ee
The charges of the D7s follow from (\ref{induced D3-charge}) and (\ref{induced D5-charge}) (we neglected gravitational corrections, as we did in section \ref{sec: backreaction d7}). The map is then
\beal
\label{dictionary}
N_{fL} &= \Tr ( \unit - \hat\rho)/2 \qquad & Q_\text{D3} &= N_2 + \frac14 N_{fL} \qquad & N_1 &= Q_\text{D3} + Q_\text{D5} + \frac14 N_{fL} \\
N_{fR} &= \Tr ( \unit + \hat\rho)/2 \qquad & Q_\text{D5} &= N_1 - N_2 - \frac12 N_{fL} \qquad & N_2 &= Q_\text{D3} - \frac14 N_{fL}
\eeal
and we also define $N_f = N_{fL} + N_{fR}$.

This dictionary identifies the field theory description at some energy scale. It is valid only if the NSNS potential $b$ defined after equation (\ref{couplingsB}) is in the range $b\in [0,1]$.%
\footnote{If the B-field is outside the range $[0,1]$, the mutually BPS fractional branes of minimal tension are not the ones used to derive the dictionary, and the dictionary is not correct (\eg{} the holographic formul\ae{} do not give real-valued gauge couplings). One possibility is to construct a corrected dictionary, another is to perform a large gauge transformation of $B$, as proposed in the text.}
If this condition is not met, we can perform a large gauge transformation $b \to b - [b]_-$ which however shifts the Page charges.

Let us compute (see appendix \ref{app: Page charges shift}) how Page charges shift under a large gauge transformation $B \to B + \alpha'\pi \omega_2$, \ie{} $b\rightarrow b+1$
\be
Q_\text{D5}' = Q_\text{D5} - \frac{N_f}2 \;,\qquad\qquad Q_\text{D3}' = Q_\text{D3} - Q_\text{D5} + \frac{N_f}4 \;.
\ee
Since $\cF = P[B] + 2\pi\alpha'F$ is gauge-invariant, the large gauge transformation shifts $F \to F - \frac12 P[\omega_2]$ and it affects the Wilson lines $\hat\rho \to - \hat\rho$. We can then compute the modification of gauge theory ranks associated with such a  shift. Using (\ref{dictionary}) we find that the theory with ranks $(N_{fL}, N_1, N_2, N_{fR})$ is mapped to one with $(N_{fR},\, N_2,\, 2N_2 + N_{fL}- N_1,\, N_{fL})$. It is known \cite{Evslin:2004vs, Benini:2007gx, Benini:2007kg} that a large gauge transformation in the bulk corresponds to a Seiberg duality in field theory. Indeed the shift of ranks agrees with Seiberg duality on the $SU(N_1)$ IR strongly coupled node and a flip of the quiver. Further evidence of our dictionary between the Page charges and the ranks comes from the study of the RG flow in section \ref{sec: RG flow}.

We have identified the field theory (more precisely, the effective description at some energy scale) dual to the gravity background. Now we want to identify the correct vacuum using the following argument. Consider the background with D7-branes, no worldvolume flux and $0<b<1$. Such a configuration corresponds to the field theory with flavors on the right in its trivial vacuum with $\tilde Q_R Q_R = 0$.%
\footnote{This is in agreement with the naive holographic map since there is no worldvolume flux.}
Let us now crank up the value of $b$, that is we change the gauge couplings. When $b=1$ the right node is at infinite coupling and we can move to a dual description by performing Seiberg duality on the right node ($b \to b-1$). We also flip the quiver (so that the largest gauge rank is always on the left) and thus the flavors are now on the left. According to (\ref{vacua map SD}) the theory is in a non-trivial vacuum with $\tilde Q_L Q_L = - \mu/\sqrt h$. This is the Higgsed vacuum labeled by $n=1$ in section \ref{sec: classical moduli space}.

Seiberg duality corresponds to a large gauge transformation in supergravity which includes a shift $F \to F + \frac12 P[\omega_2]$. The new background has one unit of worldvolume flux on each of the D7s while in the dual field theory the flavors are on the left. We learn that the background with $n=1$ units of worldvolume Abelian flux corresponds to a theory in the $n=1$ Higgsed vacuum. Repeating the argument (possibly in the opposite direction as well) we recover the quiver dimensions (\ref{block dimensions}) and the VEVs of $\tilde Q Q$ (\ref{chiral VEVs Higgsed mesonic}),%
\footnote{Notice that to derive the VEVs (\ref{chiral VEVs deformed Higgsed mesonic}) of the quantum theory one would have to include the non-perturbative superpotential $W_\text{ADS}$ in the analysis of section \ref{sec: Seiberg duality}.}
and conclude that a background with $n$ units of Abelian worldvolume flux $F$ corresponds to the Higgsed vacuum labeled by $n$ (\ref{ndef}).

\subsection{Matching of operator VEVs}
\label{sec: matching VEVs}

Let us compare the expectation values of protected operators from the flavor sector computed in field theory and in supergravity. Here we restrict for the moment to the case with $N_f=1$.

We start with the singular conifold, discussed from the gravity point of view in section \ref{subsec: singular conif}. The D7-brane affects the background above $\mu$, while below $\mu$ the background is unperturbed and the low-energy theory is on the mesonic branch. The vacua of the theory above $\mu$ are the classical (Higgsed) mesonic vacua of section \ref{sec: classical moduli space}, with
$$
\tilde Q_iQ_i = - \frac\mu{\sqrt h}\, r \;,\qquad\qquad Q_i^\dag Q_i - \tilde Q_i \tilde Q_i^\dag = 0 \;,
$$
where $r=\left[ \frac{n+1}2 \right]_-$ and $i=R(L)$ for $n$ even(odd). In the case of even $n$ we have an exact matching with the supergravity computation (\ref{scvevs}) up to an overall normalization factor $h^{-1/2}$. This is a universal factor for the operator $\tilde QQ$ in all vacua. Such normalization factors are anyway unavoidable as the kinetic term of $Q$ is not explicitly known in field theory.

In the case of odd $n$ we cannot directly compare with supergravity: the background has a non-trivial worldvolume connection $A$ (\ie{} $\bZ_2$ Wilson line $\hat\rho=-1$) at the boundary, and the AdS/CFT dictionary requires modification. We can however overcome this problem by exploiting symmetries. First we flip the quiver by mapping $N_1\leftrightarrow N_2$, $N_{fL}\leftrightarrow N_{fR}$ and $\tilde QQ \to - \tilde QQ$.%
\footnote{The flip  transformation maps $A \leftrightarrow B$, and invariance of the superpotential \ref{full superpotential} with $\eta_{L,R} = \sqrt h$ requires $h \to -h$, $\tilde QQ \to -\tilde QQ$ and $\sqrt h\, \mu \to - \sqrt h\, \mu$, so that $\mu/\sqrt h$ is invariant.}
On the gravity side this corresponds to changing the sign of $F_3$, $B$ and $F$. Moreover, since $-b$ is outside the range $[0,1]$, we perform a large gauge transformation $-b \rightarrow -b+1$. As a result $n \to n' = -n-1$, which turns odd $n$ into even $n'$ and $r \to r'=-r$. Since for even $n'$ field theory and supergravity match, we have established agreement for odd $n$ as well.

Let us try to understand what exactly happens when $n$ is odd. In this case the gravity computation gives a result which is shifted, compared to the field theory VEV, by an $n$-independent number
\be
\label{corrected AdS/CFT dictionary}
\tilde Q_L Q_L \Big|_\text{field theory} = \tilde Q_L Q_L \Big|_\text{gravity} - \frac\mu{2{\sqrt h}} \;.
\ee
The interpretation is that, for odd $n$, the gravity field $F$ is dual to a mix of the operator $\tilde Q_L Q_L$ with the unity operator multiplied by $\mu/\sqrt h$, which has the same dimension and R-charge.
We saw in section \ref{sec: Seiberg duality} that Seiberg duality mixes the operators: one has to introduce the shift above to make this mixing compatible with the large gauge transformation in the bulk.

In the resolved conifold case the field theory VEVs are
$$
\tilde Q_iQ_i = - \frac\mu{\sqrt h} \, r \;,\qquad\qquad Q_i^\dag Q_i - \tilde Q_i \tilde Q_i^\dag = r \xi_1
$$
whilst the gravity result is $Q^\dag Q - \tilde Q \tilde Q^\dag = a^2n/2$ (\ref{rcvevs}). For even $n$ we have agreement, up to a universal overall coefficient, as the resolution parameter $a^2$ is proportional to $\xi_1$. We have agreement for odd $n$ as well, by flipping the quiver and noticing that it maps $\xi_1 \leftrightarrow \xi_2=-\xi_1$.

As before, for odd $n$ the gravity computation differs from the field theory VEV by an $n$-independent constant shift
\be
\label{rcanswer}
Q^\dag_L Q_L - \tilde Q_L \tilde Q_L^\dag \Big|_\text{field theory} = Q^\dag_L Q_L - \tilde Q_L \tilde Q_L^\dag \Big|_\text{gravity} + \frac{a^2}2 \;.
\ee
Therefore the operators $\tilde Q_L Q_L$ and $|Q_L^2| - |\tilde Q_L^2|$, which in the $\cN=2$ case form an $SU(2)_R$ triplet, mix with the deformation/resolution parameters $\mu$ and $a^2(\xi)$ of $\Sigma = \bC^2/\Z_2$.

In the deformed conifold case, depending on whether or not there are mobile D3-branes, below $\mu$ the low-energy theory is either on the mesonic branch or at the $\bZ_2$-invariant point of the baryonic branch. The quantum analysis of sections \ref{sec: quantum mod 1} and \ref{sec: quantum mod 2} gives us for the theory above scale $\mu$  (\ref{chiral VEVs deformed Higgsed mesonic})
$$
\Phi = -\sqrt{ \frac{\mu^2/4 - \epsilon}h} \, (2r-1) - \frac\mu{2\sqrt h} \;,\qquad\text{ or }\qquad
\tilde Q_R Q_R = - \sqrt{ \frac{\mu^2/4 -\epsilon}h} \, 2r
$$
and $|N_{\dot\alpha}^2| - |\tilde N_\alpha^2|=0$ or $|Q_R^2|-|\tilde Q_R^2| = 0$. For even $n$ we have a perfect agreement with the gravity result (\ref{dcvevs}): remarkably the non-perturbative field theory effects encoded in $\epsilon$ are precisely reproduced by the geometry of $\Sigma$ embedded in the deformed conifold. For odd $n$ we cannot flip the quiver, because the quantum analysis of sections \ref{sec: quantum mod 1}, \ref{sec: quantum mod 2} assumes that the left node goes to strong coupling. On the other hand we can exploit the dictionary (\ref{corrected AdS/CFT dictionary}), derived for the classical vacua, which is an identification in the UV that does not rely on the IR effects. Again we find a perfect agreement.

Finally, the resolved deformed conifold (or BGMPZ background) describes the KS theory on the baryonic branch. From the gravity computation (\ref{bbvevs}) the VEV of $|Q^2|-|\tilde Q^2|$ grows as $n^2$, as opposed to the linear growth (\ref{rcvevs}) in the resolved conifold case. The quadratic in $n$ behavior is indeed what we find in (\ref{real VEVs resolved deformed}) from the quantum field theory analysis (for odd and even $n$ correspondingly)
\bea
\sum_{\dot\alpha} |N_{\dot\alpha}^2| - \sum_\alpha |\tilde N_\alpha^2| = {n^2-1\over 4}\xi_2 \;,\qquad\text{ and }\qquad
|Q_R^2| - |\tilde Q_R^2| = {n^2\over 4} \xi_2 \;,
\eea
where in the left case we again find an $n$-independent shift in the gravity result.
The gravity computation is done in the gauge such that $b=0$ at the tip of the deformed conifold. Hence this calculation refers to the lowest step in the cascade of Seiberg dualities, $k=0$.
Indeed the quantum VEVs above match the semi-classical computation (\ref{classical real VEV deformed}) with $k=0$.

\subsection{Theories with $N_f>1$ and noncommutative instantons}

When $N_f>1$, the moduli space includes the instanton-like directions (section \ref{sec: classical moduli space}), which represent the mobile D3-branes dissolved into the D7s forming the conventional gauge instantons with continuous moduli. On the field theory side this picture is backed by the fact that any solution to the $\N=2$ $\bC^2/\Z_2$ ADHM equations -- \ie{} vacuum equations in field theory -- is also a solution to the classical F- and D-term equations in the $\N=1$ conifold case. On the quantum level this relation holds in the $\epsilon\rightarrow 0$ limit as the solution to the classical equations solves the quantum equations in this case as well (see the comment after (\ref{deformed right ansatz})).
On the gravity side, the cycle $\Sigma$ has the same complex structure as the deformed/resolved $\bC^2/\bZ_2$, hence the moduli spaces of instantons in the two cases share the same complex structure. Therefore the parallel with the $\N=2$ $\bC^2/\Z_2$ theory provides a good qualitative understanding of these vacua.

It is interesting however to consider D7-branes embedded in the BGMPZ background. In this case the SUSY condition for the worldvolume gauge field is not anti-self-duality, but rather a non-linear deformation of it (see section \ref{bbbackgrounds}) so that a parallel to $\cN=2$ instantons is less straightforward. On the other hand the classical field theory analysis of section \ref{sec: classical moduli space} is not sensitive to the low-energy theory and is valid in all cases. Since the baryonic vacua of the low energy theory require $(k+1)\xi_1 + k\xi_2 = 0$, \ie{} $\xi_1 + \xi_2 \neq 0$, the ``non-linear instantons'' in the BGMPZ background are related to the noncommutative instantons on $\C^2/\Z_2$. This relation helps understanding why the non-linear instantons in question cannot shrink to zero size and leave the D7s and become the mobile D3-branes in the bulk. We know that this indeed must be the case, because the mobile D3-branes are not SUSY on the BGMPZ background \cite{DKS} (see sections \ref{reviewks} and \ref{sec: classical moduli space} for a field theory explanation), but this is not apparent from the SUSY condition (\ref{kappas}) itself. The relation to noncommutative instantons partially clarifies this point, as the noncommutative instantons cannot shrink to zero size and leave the larger brane as well \cite{Seiberg:1999vs}. It would be interesting to study the moduli space of the non-linear instantons satisfying (\ref{kappas}) and provide an explicit map, in the spirit of the Seiberg-Witten map \cite{Seiberg:1999vs}, to the noncommutative instantons on $\Sigma$.

Another interesting question is related to the vacua that completely break the whole gauge symmetry of the field theory. These vacua admit generic values of $\xi_1$, $\xi_2$ and the form of the F- and D-term equations (or at least the $\cN=2$ ADHM equations) suggest that we are dealing with the noncommutative instantons. This comment equally applies to the $\N=2$ $\bC_2/\Z_2$ orbifold theory and to the $\N=4$ theory broken to $\N=2$ by flavor. It is tempting to attribute the appearance of the noncommutative instantons to the presence of a self-dual $B$-field in the bulk \cite{Nekrasov:1998ss}, however such $B$-field is not normalizable \cite{DnM} and therefore cannot describe a branch of vacua of the field theory. In fact since the whole gauge symmetry is broken, \ie{} all D3-branes are dissolved in the D7s, the probe approximation for D7s breaks down and we have no control over the geometric description.

\subsection{The RG flow}
\label{sec: RG flow}

Eventually we want to compare the RG flow of gauge couplings in the large $N$ limit, computed in field theory with the NSVZ beta-function formula \cite{Novikov:1983uc}, with the backreacted supergravity solutions of section \ref{sec: backreaction d7}. Those solutions represent an $SU(2) \times SU(2) \times U(1)$ invariant smeared distribution of D7-branes, which describe a precise large $N$ field theory dual \cite{Benini:2006hh}. In the Veneziano limit $N_f/N_c = \text{fixed}$, the number $N_f$ of D7-branes is large. Let us parametrize them with a flavor index $U \in U(2)$
which takes $N_f$ values in $U(2)$. Each D7-brane has a different embedding and correspondingly a different superpotential, \eg{} in the ``Right'' case:
\be
\Sigma_U = \{ U^{\alpha\dot\alpha} w_{\dot\alpha\alpha} = \mu \} \qquad\qquad W_0 \;\supset\; -h \tilde Q_R^U \Big( U^{\alpha\dot\alpha} B_{\dot\alpha} A_\alpha - \frac\mu{\sqrt h} \Big) Q_R^U \;.
\ee
In the $N_f \to \infty$ limit the index $U$ becomes continuous and, if we uniformly distribute the $N_f$ values on $U(2)$, the theory acquires an extra $U(2) \cong SU(2) \times U(1)$ symmetry. Since the running of gauge couplings does not depend on the details of the superpotential, it will be the same as in the original theory in (\ref{full superpotential}).

At large $N_{1,2}$ the field theory is quasi-conformal and the anomalous dimensions (at scales much larger than $\sqrt h \, \mu$) are fixed by the quartic superpotential (\ref{full superpotential}) together with charge conjugation symmetry
\be
\gamma[A] = \gamma[B] = \gamma[Q_{R,L}] = \gamma[\tilde Q_{R,L}] = - \frac12 \;.
\ee
At scale $\Lambda$ where the effective description has ranks $N_{fL} \times N_1 \times N_2 \times N_{fR}$ the NSVZ formula
\be
\frac{\partial}{\partial \log\Lambda} \, \frac{8\pi^2}{g^2} = 3T[G] - \sum_{\text{chiral } i} T[r_i] (1-\gamma_i)
\ee
gives
\be
\label{NSVZ result}
\frac{\partial}{\partial \log\Lambda} \, \frac{8\pi^2}{g_1^2} = 3\Big( N_1 - N_2 - \frac{N_{fL}}2 \Big) \;,\qquad \frac{\partial}{\partial \log\Lambda} \, \frac{8\pi^2}{g_2^2} = 3\Big( N_2 - N_1 - \frac{N_{fR}}2 \Big) \;.
\ee

Let us extract the RG flow from the backreacted supergravity solution. First, consider the massless solution ($\mu=0$) characterized by the dilaton $e^\phi$ (\ref{dilaton backreacted}) and the B-field $b(\rho)$ (\ref{b(rho) massless}). The gauge couplings can be extracted with the holographic formul\ae{} \eqref{couplingsB}
\be
\label{holographic formulae specific}
\frac{8\pi^2}{g_1^2} = \frac{2\pi}{ e^\phi} \, b \;,\qquad\qquad \frac{8\pi^2}{g_2^2} = \frac{2\pi}{ e^\phi} (1-b) \;.
\ee
We obtain
\be
\frac{\partial}{\partial\rho} \, \frac{8\pi^2}{g_1^2} = - 3 \frac{N_fc_2}2 \;,\qquad\qquad \frac{\partial}{\partial\rho} \, \frac{8\pi^2}{g_2^2} = 3 \frac{N_f(c_2 -1)}2 \;.
\ee
The holographic formul\ae{} can be applied in a gauge where $b \in [0,1]$. If this is not the case, $B$ should be shifted by a large gauge transformation to meet the condition, and the Page charges shift accordingly. In such a gauge, from (\ref{value of c2 massless}) and (\ref{dictionary}) we obtain $-N_fc_2/2 = N_1 - N_2 - \frac{N_{fL}}2$ in terms of the effective description. After identifying $\rho = \log\Lambda$, supergravity precisely reproduces the NSVZ beta-function.
We stress that the fully backreacted solution  is necessary to reproduce the exact NSVZ result.

Finally consider the backreacted supergravity solution for massive D7-branes with $n$ units of worldvolume flux (in the KT approximation) detailed in appendix \ref{app: backreacted massive}. Again we need the dilaton (\ref{dilaton backreacted massive}) and the B-field (\ref{B-field backreacted massive}). Below the scale $\mu$ the dilaton is constant, the B-field is logarithmically running and supergravity reproduces the beta-functions $\frac\partial{\partial\rho} \frac{8\pi^2}{g_{1,2}^2} = \pm 3(N_1 - N_2)$. At the scale $\mu$ (that we called $\rho = \rho_0$) the page charges in (\ref{charges backreacted massive}) jump by $\delta Q_\text{D5} = \frac n2$ and $\delta Q_\text{D3} = \frac{n^2}4$, according to the breaking pattern of the Higgsed vacuum. Above the scale $\mu$, we can use the holographic formul\ae{} (\ref{holographic formulae specific}) writing the result first in terms of the Page charges (\ref{charges backreacted massive}) and then in terms of the ranks using the dictionary (\ref{dictionary}), to exactly reproduce the NSVZ result (\ref{NSVZ result}).

\section{Discussion}
\label{sec: conclusions}

In this paper we studied the supersymmetric vacua of the $\N=1$ $SU(M+N)\times SU(N)$ theory with bifundamental and flavor matter.
In the limit $N_f \ll N+M$ we used the dual geometries with probe D7-branes and worldvolume gauge configurations (``instantons'') to describe various Higgs vacua.
In the $\cN=1$ case, as opposed to $\cN=2$, supersymmetry is not powerful enough to prevent quantum corrections to the Higgs branch.
On the gravity side in most cases the quantum corrections
arise from the deformations of the geometry and of the K\"ahler potential in the bulk, affecting the VEVs of the protected operators from the flavor sector. On the field theory side, instead, the quantum corrections arise from the non-perturbative contributions to the superpotential and the change of degrees of freedom: when a gauge group confines, the original microscopic flavor degrees of freedom are not relevant anymore and one has to use the low energy meson variables.

In the $\cN=2$ case there is a direct relation between the bulk description (\ie{} the world-volume instantons) and the field theory description (\ie{} the F- and D-term equations), given by the ADHM construction.
Clearly this relation does not rely on the AdS/CFT correspondence. The opposite is also true: the AdS/CFT duality predicts a one-to-one correspondence between the field theory vacua and the configurations in the bulk, but does not outline in details how to construct the map. The fact that such a direct relation is known in the $\N=2$ case is a nice bonus. It is not immediately clear if such a relation can be found for $\N=1$ theories: although the instantons in the bulk do have some version of the ADHM construction, the corresponding matrix equations are different from the quantum version of the vacuum equations in field theory.

So far we mainly discussed the quantum corrections from the field theory point of view. In fact the $\N=1$ case can be drastically different from the $\N=2$ case in the bulk as well.
When the underlying $\N=1$ background has a complicated structure, the SUSY condition for the wold-volume gauge fields becomes nonlinear \cite{Martucci:2005ht, D7bb}. We saw
that when this happens on the conifold, the resulting nonlinear instantons are related to noncommutative instantons on the same space. It would be very interesting to study the moduli space of these nonlinear instantons systematically and investigate if one can find such configurations with some sort of matrix equations in the spirit of the ADHM construction.

\subsection*{Acknowledgments}

We thank F. Fucito, S. Kachru, I. R. Klebanov, Z. Komargodski,  J. Maldacena, N. Nekrasov, N. Seiberg, Y. Tachikawa, H. Verlinde, and E. Witten for numerous discussions.
F.B. would like to thank the Aspen Center for Physics, the Galileo Galilei Institute for Theoretical Physics in Florence, the High Energy group at the Weizmann Institute and the Simons Center for Geometry and Physics, for hospitality and support during the course of this work. A.D. would like to thank the Aspen Center for Physics, and the High Energy group at the Tel-Aviv University for hospitality.
The work of F.B. was supported in part by the US NSF under Grants No. PHY-0844827 and PHY-0756966.
The research of A.D. was supported by the DOE grant DE-FG02-90ER40542, by Monell Foundation, and in part by the grant RFBR 07-02-00878 and the Grant for Support of Scientific Schools NSh-3035.2008.2.

\appendix

\section{Higgsed vacua}
\label{app: Higgsed}

In this appendix we give the explicit form of various vacua discussed in the main text.

\subsection{Classical Higgsed mesonic directions with resolution}
\label{app: Higgsed FI}

\begin{table}[tn]
\beal\nn
& \begin{aligned}
A_1 &= \beta\alpha \mat{ c_1 a_1 U_1^\trans & 0 & \dots \\ c_2 a_2 L_2 & c_3 a_3 U_3^\trans & \dots \\ 0 & c_4 a_4 L_4 & \dots \\ \vdots & \vdots & \ddots}  &
B_1^\trans &= \beta\alpha \mat{ c_1^{-1} a_1 U_1^\trans & 0 & \dots \\ - c_2^{-1} a_2 L_2 & c_3^{-1} a_3 U_3^\trans & \dots \\ 0 & - c_4^{-1} a_4 L_4 & \dots \\ \vdots & \vdots & \ddots} \\
A_2 &= \beta\alpha \mat{ c_1 a_1 L_1^\trans & 0 & \dots \\ - c_2 a_2 U_2 & c_3 a_3 L_3^\trans & \dots \\ 0 & - c_4 a_4 U_4 & \dots \\ \vdots & \vdots & \ddots} &
B_2^\trans &= \beta\alpha \mat{ c_1^{-1} a_1 L_1^\trans & 0 & \dots \\ c_2^{-1} a_2 U_2 & c_3^{-1} a_3 L_3^\trans & \dots \\ 0 & c_4^{-1} a_4 U_4 & \dots \\ \vdots & \vdots & \ddots}
\end{aligned} \\
& Q_L^\trans = \alpha c_0 \mat{1 & 0 & \dots & 0} \;,\qquad
\tilde Q_L = \alpha c_0^{-1} \mat{1 & 0 & \dots & 0} \;,\qquad
\tilde Q_R = Q_R = 0 \;.
\eeal
\caption{Higgsed mesonic vacua with resolution, left blocks. \label{tab: Higgsed FI left}}
\end{table}

\begin{table}[tn]
{\small
\beal\nn
& \!\!\!\!\!\!\!\! A_1 = \beta\alpha \mat{ c_1a_1 U_1 & c_2a_2 L_2^\trans & 0 & \dots \\ 0 & c_3a_3 U_3 & c_4a_4 L_4^\trans & \dots \\ 0 & 0 & c_5a_5 U_5 & \dots \\ \vdots & \vdots & \vdots & \ddots} \qquad\quad
A_2 = \beta\alpha \mat{ - c_1a_1 L_1 & c_2a_2 U_2^\trans & 0 & \dots \\ 0 & - c_3a_3 L_3 & c_4a_4 U_4^\trans & \dots \\ 0 & 0 & - c_5a_5 L_5 & \dots \\ \vdots & \vdots & \vdots & \ddots} \\
& \!\!\!\!\!\!\!\!  B_1^\trans = \beta\alpha \mat{ - c_1^{-1} a_1 U_1 & c_2^{-1} a_2 L_2^\trans & 0 & \dots \\ 0 & - c_3^{-1} a_3 U_3 & c_4^{-1} a_4 L_4^\trans & \dots \\ 0 & 0 & c_5^{-1} a_5 U_5 & \dots \\ \vdots & \vdots & \vdots & \ddots}
B_2^\trans = \beta\alpha \mat{ c_1^{-1} a_1 L_1 & c_2^{-1} a_2 U_2^\trans & 0 & \dots \\ 0 & c_3^{-1} a_3 L_3 & c_4^{-1} a_4 U_4^\trans & \dots \\ 0 & 0 & c_5^{-1} a_5 L_5 & \dots \\ \vdots & \vdots & \vdots & \ddots} \\
& \tilde Q_L = Q_L = 0 \;,\qquad
Q_R^\trans = \alpha c_0 \mat{1 & 0 & \dots & 0} \;,\qquad
\tilde Q_R = \alpha c_0^{-1} \mat{1 & 0 & \dots & 0} \;.
\eeal
}
\caption{Higgsed mesonic vacua with resolution, right blocks. \label{tab: Higgsed FI right}}
\end{table}

The blocks (\ref{blocks Higgsed left}), (\ref{blocks Higgsed right}) of the classical Higgsed mesonic directions can be generalized to incorporate arbitrary parameters $\xi_1$ and $\xi_2$.
The left blocks for $r \geq 1$, of dimension $1 \times r^2 \times r(r-1) \times 0$, are in table \ref{tab: Higgsed FI left}.
The variables $a_1, \dots, a_{2r-2}$ ($K = 2r-1$), $\alpha$, $\beta$ are the same as in the $\xi_{1,2} = 0$ case (\ref{solution aj}) and (\ref{solution alpha beta}) and the unknowns are  $c_0, \dots, c_{2r-2}$. The right blocks for $r \geq 1$, of dimension $0 \times r(r+1) \times r^2 \times 1$, are in table \ref{tab: Higgsed FI right}.
Again, $a_1, \dots, a_{2r-1}$ ($K= 2r$), $\alpha$, $\beta$ are the same as before and the new unknowns are $c_0, \dots, c_{2r-1}$.
The blocks of one kind with $r\leq -1$ are obtained from the blocks of the other kind with $r\geq 1$ by taking the transpose of $A_\alpha$, $B_{\dot\alpha}$ and by exchanging $Q_L \leftrightarrow Q_R$, $\tilde Q_L \leftrightarrow \tilde Q_R$.

The F-term are solved by (\ref{solution aj}), (\ref{solution alpha beta}) for any choice of $c_j$'s. It is convenient to define the quantities
\be
x_j \equiv |\alpha\beta|^2 a_j^2 (c_j^2 - c_j^{-2}) \;.
\ee
From the D-term equations we get for a left (right) block the recursive equations
\beal
\xi_1 &= j \, x_j + (j+2) x_{j+1} \qquad & &j \text{ even(odd)} \;,\qquad\qquad &
\xi_K &\equiv 0 \;, \\
-\xi_2 &= j \, x_j + (j+2) x_{j+1} \qquad & &j \text{ odd(even)} \;,\qquad\qquad &
\xi_{1(2)} &= |\alpha|^2 (c_0^2 - c_0^{-2}) \pm 2x_1 \;.
\eeal
The solution is
\be
x_j = \frac{(-1)^{j+K}(2K+1) - (2j+1)}{8j(j+1)} (\xi_2 - \xi_1) - (-1)^{j+n} \frac{K(K+1) - j(j+1)}{4j(j+1)} (\xi_1 + \xi_2) \;,
\ee
and the resulting quark bilinears are
\beal
\text{L:}\qquad & Q^\dag_L Q_L - \tilde Q_L \tilde Q_L^\dag = |\alpha|^2(c_0^2 - c_0^{-2}) = \xi_1 - 2x_1 = r^2 \xi_1 + r(r-1)\xi_2 \\
\text{R:}\qquad & Q^\dag_R Q_R- \tilde Q_R \tilde Q_R^\dag = |\alpha|^2(c_0^2 - c_0^{-2}) = \xi_2 + 2x_1 = r(r+1) \xi_1 + r^2\xi_2 \;.
\eeal

Besides solving the vacuum equations in the $\N=1$ case, the matrices above solve the $\N=2$ ADHM equations and describe the noncommutative Abelian instantons on $\C^2/\Z_2$
(also see \cite{Braden:1999zp}).

\subsection{Quantum deformed Higgsed mesonic directions}
\label{app: quantum Higgsed}

\begin{table}[tn]
\beal
\label{deformed right ansatz}
M_{11} &= \beta^2\alpha^2 \mat{ -\eta_1 U_1^\trans U_1 & - \eta_{12} U_1^\trans L_2^\trans & 0 & \dots \\ \eta_{12} L_2 U_1 & \eta_2 L_2 L_2^\trans - \eta_3 U_3^\trans U_3 & - \eta_{34} U_3^\trans L_4^\trans & \dots \\ 0 & \eta_{34} L_4 U_3 & \eta_4 L_4 L_4^\trans - \eta_5 U_5 U_5^\trans & \dots \\ \vdots & \vdots & \vdots & \ddots} \\
M_{12} &= \beta^2\alpha^2 \mat{ \eta_1 U_1^\trans L_1 & - \eta_{12} U_1^\trans U_2^\trans & 0 & \dots \\ -\eta_{12} L_2 L_1 & \eta_2 L_2 U_2^\trans + \eta_3 U_3^\trans L_3 & - \eta_{34} U_3^\trans U_4^\trans & \dots \\ 0 & -\eta_{34} L_4 L_3 & \eta_4 L_4 U_4^\trans + \eta_5 U_5 L_5^\trans & \dots \\ \vdots & \vdots & \vdots & \ddots} \\
M_{21} &= \beta^2\alpha^2 \mat{ \eta_1 L_1^\trans U_1 & \eta_{12} L_1^\trans L_2^\trans & 0 & \dots \\ \eta_{12} U_2 U_1 & \eta_2 U_2 L_2^\trans + \eta_3 L_3^\trans U_3 & \eta_{34} L_3^\trans L_4^\trans & \dots \\ 0 & \eta_{34} U_4 U_3 & \eta_4 U_4 L_4^\trans + \eta_5 L_5 U_5^\trans & \dots \\ \vdots & \vdots & \vdots & \ddots} \\
M_{22} &= \beta^2\alpha^2 \mat{ -\eta_1 L_1^\trans L_1 & \eta_{12} L_1^\trans U_2^\trans & 0 & \dots \\ -\eta_{12} U_2 L_1 & \eta_2 U_2 U_2^\trans - \eta_3 L_3^\trans L_3 & \eta_{34} L_3^\trans U_4^\trans & \dots \\ 0 & -\eta_{34} U_4 L_3 & \eta_4 U_4 U_4^\trans - \eta_5 L_5 L_5^\trans & \dots \\ \vdots & \vdots & \vdots & \ddots} \\
\tilde Q_R &= Q_R^\trans = \alpha \mat{ 1 & 0 & \dots & 0} \qquad\qquad N_i = \tilde N_i = \Phi = 0 \;.
\eeal
\caption{Higgsed mesonic directions with deformation, right blocks. \label{tab: Higgsed deformed right}}
\end{table}

When the left gauge group goes to strong coupling, we describe it using gauge-invariants. For $2N_2 + N_{fL} < N_1$ there are only mesons, defined in (\ref{mesons def}), while for $2N_2 + N_{fL} \geq N_1$ there are also baryons. Mesons are always good coordinates on mesonic branches.
The right blocks, of dimension $0 \times r^2 \times 1$, are in table \ref{tab: Higgsed deformed right}.
Let us take $r>0$, although the same ansatz gives the solution for both $r$ and $-r$.
The unknowns are $\eta_1, \dots, \eta_{2r-1}$, $\eta_{12}, \dots, \eta_{2r-3,2r-2}$, $\beta$, $\alpha$. Setting $\eta_i = a_i^2$ and $\eta_{ij} = a_i a_j$ we simply have $M_{\dot\alpha\alpha} = B_{\dot\alpha} A_\alpha$ and the mesons solve the underformed equations. That would correspond to the classical theory, where mesons are products of elementary fields. In the quantum theory -- as a result of confinement -- the mesons are independent fields.

The D-term equations (\ref{D-terms deformed}) with $\xi_2 = 0$ are identically solved. From the F-term equations we find
\beal
&\begin{aligned}
0 &= j \, \eta_j - \eta_{j+1} - (j+3) \eta_{j+2} \qquad\qquad & j &= 1 , \cdots, 2r - 3 \\
0 &= (\eta_{2k} + \eta_{2k+1})^2 - (2k-1) \eta_{2k-1,2k}^2 + (2k+3) \eta_{2k+1,2k+2}^2 & k &= 1 ,\cdots, r-1 \\
0 &\equiv \eta_{2r-1,2r}
\end{aligned} \\
&\, 0 = 1 - \frac{2\eta_1}{\eta_1 + 3 \eta_2} - \frac{\mu}{\sqrt h \, \alpha^2} \qquad\quad
\beta^2 = \frac1{\eta_1 + 3 \eta_2} \qquad\quad
\epsilon = - 3h\alpha^4 \frac{\eta_1 \eta_2 - \eta_{12}^2}{(\eta_1 + 3\eta_2)^2} \;.
\eeal
The recursive equation for $\eta_j$ is the same as in the classical case, but the last equation for $j=2r-2$ is missing. As a result the boundary condition is different.

In the massless $\mu=0$ case we have to impose $\eta_1 - 3\eta_2 = 0$, and after arbitrarily fixing a multiplicative constant by $\eta_1 + 3\eta_2 \equiv 1$, we get $\eta_j = \frac1{j(j+1)}$ and $\eta_{2k-1,2k}$ as given below with $C_1=1$, $C_2 = 0$. Fixing $\alpha$ in terms of $\epsilon$ we get $\alpha^4 = - \epsilon h^{-1} (2r)^2$.
In the case with generic $\mu$ we proceed as follows: The general solution to the recursive equations is
\be
\eta_j = \frac{C_1 + C_2 \big[ 1 - (-1)^j(2j+1) \big]}{j(j+1)} \qquad\qquad \eta_{2k-1,2k}^2 = C_1^2 \, \frac{r^2 - k^2}{4r^2k^2(4k^2 - 1)} \;.
\ee
Then we determine $\alpha$ and $\epsilon$:
\be
\alpha^2 = - \frac\mu{\sqrt h} \, \frac{\eta_1 + 3\eta_2}{\eta_1 - 3\eta_2} \qquad\qquad \epsilon = -3\mu^2 \, \frac{\eta_1 \eta_2 - \eta_{12}^2}{(\eta_1 - 3\eta_2)^2} \;.
\ee
Notice that only the ratio $C_1/C_2$ affects the solution, while the overall normalization drops out. We should fix $C_1/C_2$ to match $\epsilon$, and then determine the full solution and $\alpha^2$ as a function of $\epsilon$. However one can directly verify that
\be
\alpha^4 = \frac{- \epsilon + \mu^2/4}h \, (2r)^2 = \frac{- \epsilon + \mu^2/4}h \, n^2 \;.
\ee
We take the branch cut in the square root such that
\be
\tilde Q_R Q_R = \alpha^2 = - \sqrt{ \frac{-\epsilon + \mu^2/4}h} \, 2r
\ee
which matches with the $\epsilon \to 0$ limit of section \ref{sec: classical moduli space}. Notice that for each choice of matrices, whose size is fixed by $|r|$, the equations have two solutions corresponding to $r$ and $-r$.

\begin{table}[tn]
\beal
\label{deformed left ansatz}
M_{11} &= \beta^2 \alpha^2 \mat{ \eta_1 U_1 U_1^\trans - \eta_2 L_2^\trans L_2 & - \eta_{23} L_2^\trans U_3^\trans & \dots \\ \eta_{23} U_3 L_2 & \eta_3^2 U_3 U_3^\trans - \eta_4 L_4^\trans L_4 & \dots \\ \vdots & \vdots & \ddots} \\
M_{12} &= \beta^2 \alpha^2 \mat{ \eta_1 U_1 L_1^\trans + \eta_2 L_2^\trans U_2 & - \eta_{23} L_2^\trans L_3^\trans & \dots \\ -\eta_{23} U_3 U_2 & \eta_3^2 U_3 L_3^\trans + \eta_4 L_4^\trans U_4 & \dots \\ \vdots & \vdots & \ddots} \\
M_{21} &= \beta^2 \alpha^2 \mat{ \eta_1 L_1 U_1^\trans + \eta_2 U_2^\trans L_2 & \eta_{23} U_2^\trans U_3^\trans & \dots \\ \eta_{23} L_3 L_2 & \eta_3^2 L_3 U_3^\trans + \eta_4 U_4^\trans L_4 & \dots \\ \vdots & \vdots & \ddots} \\
M_{22} &= \beta^2 \alpha^2 \mat{ \eta_1 L_1 L_1^\trans - \eta_2 U_2^\trans U_2 & \eta_{23} U_2^\trans L_3^\trans & \dots \\ -\eta_{23} L_3 U_2 & \eta_3^2 L_3 L_3^\trans - \eta_4 U_4^\trans U_4 & \dots \\ \vdots & \vdots & \ddots} \\
\tilde N_1 &= N_1^\trans = \beta \alpha^2 \mat{ \zeta U_1^\trans & 0 & \dots} \qquad\qquad
\Phi = \alpha^2 \mat{1} \\
\tilde N_2 &= N_2^\trans = \beta \alpha^2 \mat{ \zeta L_1^\trans & 0 & \dots} \qquad\qquad
\tilde Q_R = Q_R = 0 \;.
\eeal
\caption{Higgsed mesonic directions with deformation, left blocks. \label{tab: Higgsed deformed left}}
\end{table}

The left blocks, of dimension $1 \times r(r-1) \times 0$, are in table \ref{tab: Higgsed deformed left}.
Let us take $r>0$, although the same ansatz gives the solution for both $r$ and $-r+1$.
The unknowns are $\eta_1,\dots,\eta_{2r-2}$, $\eta_{23},\dots,\eta_{2r-4,2r-3}$, $\zeta$, $\beta$, $\alpha$. Setting $\eta_i = a_i^2$, $\eta_{ij} = a_i a_j$, $\zeta = a_1$ we have $M_{\dot\alpha\alpha} = B_{\dot\alpha} A_\alpha$, $\tilde N_\alpha = \tilde Q_L A_\alpha$, $N_{\dot\alpha} = B_{\dot\alpha} Q_L$ and $\Phi = \tilde Q_L Q_L$ as in the undeformed equations.

The D-term equations with $\xi_2=0$ are identically solved. From the F-term equations we find
\beal
&\begin{aligned}
0 &= j \, \eta_j - \eta_{j+1} - (j+3) \eta_{j+2} & j &= 1 , \cdots, 2r-4 \\
0 &= (\eta_{2k+1} + \eta_{2k+2})^2 - 2k \, \eta_{2k,2k+1}^2 + (2k+4)\eta_{2k+2,2k+3}^2 & k &= 1,\cdots, r-2
\end{aligned} \\
&\begin{aligned}
0 &= 1 - \frac{2\eta_1}{\eta_1 + 3\eta_2} + \frac{\mu}{\sqrt h \, \alpha^2} &
\beta^2 &= \frac1{\eta_1 + 3\eta_2} &
& 0 &\equiv \eta_{2r-2,2r-1} \\
\epsilon &= -2h\alpha^4 \frac{\eta_1\eta_2 - \eta_2^2 - 4\eta_{23}^2}{(\eta_1 + 3\eta_2)^2} \qquad &
\zeta^2 &= \frac{(\eta_1 + \eta_2)^2 + 4\eta_{2,3}^2}{\eta_1 + 3\eta_2} \;. \quad
\end{aligned}
\eeal
In the massless $\mu = 0$ case we have to impose $\eta_1 - 3\eta_2 = 0$, and after arbitrarily fixing a multiplicative constant by $\eta_1 + 3\eta_2 \equiv 1$, we get $\eta_j = \frac1{j(j+1)}$ and $\eta_{2k,2k+1}$ as given below with $C_1 = 1$, $C_2 = 0$. Fixing $\alpha$ in terms of $\epsilon$ we get $\alpha^4 = - \epsilon h^{-1} (2r-1)^2$. In the case with generic $\mu$ we first write down the general solution of the recursive equations:
\be
\eta_j = \frac{C_1 + C_2 \big[ 1 - (-1)^j(2j+1) \big]}{j(j+1)} \;, \eta_{2k,2k+1}^2 = (C_1 + 2C_2)^2 \frac{r(r-1) - k(k+1)}{(2r-1)^2(2k+1)^2(k^2+k)}
\ee
Then we determine $\alpha$ and $\epsilon$:
\be
\alpha^2 = \frac\mu{\sqrt h} \, \frac{\eta_1 + 3\eta_2}{\eta_1 - 3\eta_2} \qquad\qquad \epsilon = - 2 \mu^2 \, \frac{\eta_1 \eta_2 - \eta_2^2 - 4\eta_{23}^2}{(\eta_1 - 3\eta_2)^2} \;.
\ee
One can verify the following relation:
\be
\Big(\alpha^2 + \frac\mu{2\sqrt h} \Big)^2 = \frac{-\epsilon + \mu^2/4}h \, (2r-1)^2 \;.
\ee
We take the square root as
\be
\Phi = \alpha^2 = -\sqrt{ \frac{-\epsilon + \mu^2/4}h} \, (2r-1) - \frac\mu{2\sqrt h}
\ee
which matches the $\epsilon \to 0$ limit. Notice that for each choice of matrices, whose size is fixed by $\big| r-\frac12 \big| + \frac12$, the equations have two solutions corresponding to $r$ and $-r+1$.

\subsection{Quantum deformed Higgsed directions with resolution}
\label{app: quantum Higgsed FI}

The blocks of the previous section can be generalized to solve the D-term equation (\ref{D-terms deformed}) with generic $\xi_2$.

The right blocks ($n$ even) are constructed by taking the ansatz (\ref{deformed right ansatz}) and adding new variables $c_{ij}$ in front of $\eta_{ij}$ below the diagonal, $c_{ij}^{-1}$ in front of $\eta_{ij}$ above the diagonal, $c_0$ in front of $Q_R$ and $c_0^{-1}$ in front of $\tilde Q_R$. The new variables cancel out of the F-term equations. Let us define
\be
\tilde x_{2k-1,2k} \equiv |\alpha\beta|^4 \eta_{2k-1,2k}^2 \big( c_{2k-1,2k}^2 - c_{2k-1,2k}^{-2} \big) \;.
\ee
From the D-term equations we obtain the system:
\beal
\xi_2 &= (2k-1)2k \, \tilde x_{2k-1,2k} - (2k+2)(2k+3) \, \tilde x_{2k+1,2k+2} \qquad && k=1,\cdots,r-1 \\
\xi_2 &= |\alpha|^2 (c_0^2 - c_0^{-2}) - 6 \tilde x_{12} \;, && \tilde x_{2r-1,2r} \equiv 0 \;.
\eeal
The solution is
\be
\tilde x_{2k-1,2k} = \xi_2 \, \frac{r^2 - k^2}{2k(4k^2-1)}
\ee
from which we extract $|Q_R|^2 - |\tilde Q_R|^2 = r^2 \xi_2$.

The left blocks ($n$ odd) are constructed by taking the ansatz (\ref{deformed left ansatz}) and adding new variables $c_{ij}$ in front of $\eta_{ij}$ below the diagonal, $c_{ij}^{-1}$ in front of $\eta_{ij}$ above the diagonal, $c_{01}$ in front of $N_i$ and $c_{01}^{-1}$ in front of $\tilde N_i$. Let us define
\be
\tilde x_{01} = |\alpha^2 \beta \zeta|^2 (c_{01}^2 - c_{01}^{-2}) \;,\qquad \tilde x_{2k,2k+1} = |\alpha\beta|^4 \eta_{2k,2k+1}^2 \big( c_{2k,2k+1}^2 - c_{2k,2k+1}^{-2} \big) \;.
\ee
From the D-term equations we get the system:
\beal
\xi_2 &= 2k(2k+1) \, \tilde x_{2k,2k+1} - (2k+3)(2k+4) \, \tilde x_{2k+2,2k+3} \qquad && k=1,\cdots,r-2 \\
\xi_2 &= \tilde x_{01} - 12 \tilde x_{23} \;, && \tilde x_{2r-2,2r-1} \equiv 0 \;.
\eeal
The solution is
\be
\tilde x_{2k,2k+1} = \xi_2 \, \frac{r(r-1) - k(k+1)}{4k(k+1)(2k+1)}
\ee
from which we extract $\sum_{i=1,2} \big( |N_i|^2 - |\tilde N_i|^2 \big) = r(r-1)\xi_2$.

\section{Page charges}
\label{app: Page charges comp}

Here we compute the Page D3- and D5-charges on the D7-brane. At some fixed radius $r$ in the bulk the Page charges are given by (we keep $g_s=1$ everywhere in text)
\beal
Q_\text{D3}(r) &= \frac1{(4\pi^2\alpha')^2} \int_{T^{1,1}\text{ at }r} \!\! F_5 - B \wedge F_3 + \frac12 B \wedge B \wedge F_1 \;, \\
Q_\text{D5}(r) &= \frac1{4\pi^2\alpha' } \int_{\S^3\text{ at }r} \!\! F_3 - B \wedge F_1 \;.
\eeal
It will be useful to call the integrands $J_\text{D3}$ and $J_\text{D5}$ ``Page currents''. Using the Bianchi identities
\beal
dF_1 &=  \delta_2^\text{D7} \\
dF_3 &= 4\pi^2\alpha' \delta_4^\text{D5} + H_3 \wedge F_1 + \cF \wedge \delta_2^\text{D7} \\
dF_5 &= (4\pi^2\alpha')^2 \delta_6^\text{D3} + H_3 \wedge F_3 + \frac12  \cF \wedge \cF \wedge \delta_2^\text{D7} \;,
\eeal
in the absence of non-dissolved  \Dt\ and \Df-branes we find
\be
dJ_\text{D3} = (2\pi\alpha')^2 \, \frac12 F \wedge F \wedge \delta_2^\text{D7} \;,\qquad dJ_\text{D5} = (2\pi \alpha') \, F \wedge \delta_2^\text{D7} \;,\qquad dJ_\text{D7} = \delta_2^\text{D7} \;.
\ee
Here $\delta_2^\text{D7}$ is a delta 2-form localized on (and orthogonal to) the D7s.

The D3-charge is given by the integral of $J_\text{D3}$ on $T^{1,1}$ and using Gauss law it reduces to
$$
N_{\Dt} = \frac1{8\pi^2} \int_\Sigma F \wedge F \;.
$$
An important observation is that for any functions $\xi(t)$, $\lambda(t)$  from (\ref{a1},\ref{f2})
\be
\label{orthogonality}
\FI \wedge \FII \big|_\Sigma = 0
\ee
and  therefore the integral splits into two parts. The first part is a full derivative that can be computed at the boundary:
$\frac1{8\pi^2} \int_\Sigma \FI \wedge \FI = 4\a \xi^2 \big|^{r= \infty}_{r=r_\text{min}}$.
Since $\xi \rightarrow r^{-2}$ for large $r$, the contribution at infinity is zero. If we require regularity of $\xi$ at the tip, the contribution at $r=r_\text{min}$ vanishes as well because $\a(t_\text{min})=0$. The only exception is the case $z_4={\mu/\sqrt{2}}=0$ when $\a \equiv 1$. Then the integral gives $4\xi(0)^2$. In the deformed conifold case $\xi(0)$ must vanish because $g_5$ is not well-defined at the tip; in the resolved conifold case only the combination $\frac12 dg_5 + \omega_2$ is regular at the tip, hence $\xi(0)=\frac n4$. To calculate the second part we notice that
\be
\label{d3}
\frac{\FII\wedge \FII}2 = \frac{n^2}4 \, \frac{|z_4^2-\epsilon|}{(r^3 - |z_4|^2 + |z_4^2-\epsilon|)^2} \, \frac{dz_1 \wedge d\bar z_1 \wedge dz_2 \wedge d\bar z_2}{|z_3|^2} \;,
\ee
and using (\ref{relation1}) and (\ref{a}) and integrating over $t$ we get $\frac1{8\pi^2} \int_\Sigma \FII \wedge \FII = \frac{n^2}4$. The only exception is the resolved conifold case $\epsilon=0$ with $z_4 = 0$. In this case $\FII$ vanishes everywhere except at the tip and, as follows from (\ref{d3}), the second part is zero. We conclude that, in all cases,
\be
N_\text{D3} = \frac{n^2}4 \;.
\ee

The D5-charge is given by the integral of $J_\text{D5}$ on $\S^3 \subset T^{1,1}$ and using Gauss law it reduces to
$$
N_\text{D5} = \frac1{2\pi} \int_\Gamma F \;,\qquad\qquad \Gamma = \Sigma \cap (\S^3 \times \bR_+) \;.
$$
$\Gamma$ is a two-submanifold inside $\Sigma$ whose radial sections are circles $\S^1 = \Sigma \cap \S^3$. Since $F = \FI+\FII = d(\AI+\AII)$, we easily compute $2\pi N_{\Df} = \int_{\partial \Gamma} (\AI+\AII)$.
The boundary $\partial \Gamma$ is the difference between an $\S^1$ at large radius and an $\S^1$ at the tip $r_\text{min}$, where $\S^1$ shrinks into a point. $\int_{\partial \Gamma} \AI$ vanishes at infinity because $\xi$ goes to  zero, and since $\xi$ is regular at $r_\text{min}$, the contribution there vanishes as well (with the exception $\epsilon=z_4=0$). Similarly $\int_{\partial \Gamma} \AII$ does not contribute at infinity, but it does at $r_\text{min}$. Although $\S^1$ shrinks, the potential $\AII$ is singular.
To get the answer we compute the integral $\int_{\S^1}\AII$ at radius $r$ and take $r$ to $r_\text{min}$. To do that we need to define $\S^1$ more explicitly. We use the coordinates (\ref{XY}) but now on the deformed conifold $X^2 = \frac12 (1+\epsilon \, r^{-3})$, $Y^2 = \frac12 (1-\epsilon \, r^{-3})$, $X \cdot Y=0$. We define $\S^3$ at the given radius $r$ as follows: we take a point $(X,Y)$ and consider its orbit under the global symmetry $SU(2)_L$. There are many different $\S^3$ corresponding to different initial points, but since $F^{1,1}$ is closed, $N_{\Df}$ will not depend on the choice of $\S^3$. To understand how $\S^3$ intersects $\Sigma$, let us start with $(X,Y)$ that actually belongs to $\Sigma$ \ie{} $z_4 = \mu/\sqrt2$. There is one particular $U(1)\subset SU(2)_L$ that keeps $z_4$ invariant, and its orbit is the desired $\S^1$ which is  the homologically non-trivial path on $\S^3/\bZ_2$. Such $U(1)$ acts on $z_i$ as a rotation around the constant vector $n_i$, $i=1,2,3$
\be
\label{dzthroughphi}
dz_i = \frac{\epsilon_{ijk} n_j z_k}{|\vec{n}|} d\phi \;,\qquad
n_i = -i (z_i \bar z_4-\bar z_i z_4 +\epsilon_{ijk} z_j \bar z_k) \;,\qquad |\vec{n}|^2 = r^6 - |\epsilon|^2 \;.
\ee
Indeed $dn_i=0$. Using the explicit form of $\sigma$ we get
\be
\label{a2}
\AII = - \frac n2 \, \frac{z_4\sqrt{\bar z_4^2 - \bar \epsilon}}{\sqrt{r^6-|\epsilon|^2}} \, d\phi  + \text{c.c.}
\ee
Hence the integral over $\S^1$ at the minimal radius $r_\text{min}^3 = |z_4|^2+|z_4^2-\epsilon|$ gives $N_\text{D5}^\text{II} = n/2$.
This result is valid unless $z_4=\epsilon=0$ when (\ref{a2}) vanishes.

Now we can return back to the contribution of $\int_{\partial\Gamma(r_\text{min})} \AI$. Using (\ref{dzthroughphi}) we find
\be
g_5 = 2 \frac{[(r^3-|z_4|^2)^2-|\epsilon-z_4^2|^2]}{r^3\sqrt{r^6-|\epsilon|^2}} \, d\phi \;.
\ee
The integral of $\AI=\xi(r)g_5$ over $\S^1$ parametrized by $\phi$ located at the minimal radius $r_\text{min}$ vanishes, unless $z_4=0$ in which case the expression for $\int_{S^1} \AI$ takes the form
$4\pi\xi(r)\sqrt{\frac{r^6-|\epsilon|^2}{r^3}}$ and in the resolved conifold case $\epsilon=0$ we simply get $N^{\rm I}_{\Df}=2\xi(0)$. Taking into account that $\xi(0)=n/4$ and that (\ref{a2})
and hence $N^{\rm II}_{\Df}$ vanish in this case, we get in all cases
\be
N_{\Df}= \frac n2 \;.
\ee

\subsection{Shift of Page charges}
\label{app: Page charges shift}

Let us compute the shift of Page charges under $B \to B + \pi\alpha'\omega_2$ (accompanied by $F \to F - \frac12 P[\omega_2]$ \ie{} $n\rightarrow n-1$). First
\be
Q_\text{D5}(r) - Q_\text{D5}(0) = \frac1{4\pi^2\alpha'} \int_{S^3\times I} dJ_\text{D5} = \frac{N_f}{2\pi} \int_{\S^3\times I} F \wedge \delta_2^\text{D7} = \frac{N_f}{2\pi} \int_{\Sigma \cap (\S^3 \times I)} F
\ee
where $I$ is the interval $[0,r]$ in $r$, and we have included the dependence on the number of branes $N_f$. For $r\to0$ the D7s have no effect, therefore $\delta Q_\text{D5}(0) = 0$. We conclude that
\be
\delta Q_\text{D5}(r) = - \frac{N_f}{4\pi} \int_{\Sigma \cap (\S^3 \times I)} \omega_2 = - \frac{N_f}2
\ee
where in the last equality we exploited the computations of the previous section and took the $r\to\infty$ limit.
Then
\be
Q_\text{D3}(r) - Q_\text{D3}(0) = \frac1{(4\pi^2\alpha')^2} \int_{T^{1,1}\times I} dJ_\text{D3} = \frac{N_f}{8\pi^2} \int_{\Sigma \cap (T^{1,1}\times I)} F \wedge F
\ee
and its variation under a large gauge transformation is
\be
\delta Q_\text{D3}(r) - \delta Q_\text{D3}(0) = \frac{N_f}{8\pi^2} \int_{\Sigma \cap (T^{1,1} \times I)} \Big( - \omega_2 \wedge F + \frac14 \omega_2 \wedge \omega_2 \Big) \;.
\ee
However this time the variation at $r=0$ does not vanish. Using the fact that for every closed $g_3$ form, $\int_{T^{1,1}} \omega_2 \wedge g_3 = 4\pi \int_{\S^3} g_3$, we get
\be
\delta Q_\text{D3}(0) = - \frac1{16\pi^3\alpha'} \int_{T^{1,1}} \omega_2 \wedge F_3 = - \frac1{4\pi^2\alpha'} \int_{\S^3} F_3 = - Q_\text{D5}(0) \;.
\ee
Finally we use that for every closed $g_2$ form with compact support on $\Sigma$, $\int_{\Sigma \cap (T^{1,1} \times I)} \omega_2 \wedge g_2 = 4\pi \int_{\Sigma \cap (\S^3 \times I)} g_2$. Therefore
\be
\delta Q_\text{D3}(r) = - Q_\text{D5}(r) + \frac{N_f}{8\pi^2} \int_{\Sigma \cap (T^{1,1} \times I)} \frac14 \omega_2 \wedge \omega_2 = - Q_\text{D5}(r) + \frac{N_f}4 \;.
\ee
Again, in the last equality we took the $r\to\infty$ limit.

\section{Backreacted solution with massive flavors and worldvolume flux}
\label{app: backreacted massive}

We can generalize the solutions of section \ref{sec: backreaction d7} to the case of a massive embedding $\mu\neq 0$, possibly with worldvolume flux $F$ (the solution without flux has been found in \cite{Bigazzi:2008qq}). The $SU(2)\times SU(2)\times U(1)$ invariant ansatz is the same as in (\ref{backreacted ansatz}), but the number of flavors $N_f$ is substituted by a radial function $N_f(\rho)$
\beal
\label{massive backreacted ansatz}
ds^2 &= h^{-\frac12} dx_{3,1}^2 + h^{\frac12} \Big[ e^{2u} \Big( d\rho^2 + \frac19 g_5^2 \Big) + \frac{e^{2g}}6 \sum \big( d\theta_i^2 + \sin^2\theta_i\, d\varphi_i^2 \big) \Big] \\
F_1 &= \frac{ N_f(\rho)}{4\pi} \, g_5 \;,\qquad
B_2 = \alpha' \pi b(\rho) \, \omega_2 \;,\qquad
H_3 = \alpha' \pi b'(\rho) \, d\rho \wedge \omega_2 \;.
\eeal
The unwarped metric is K\"ahler and hence a SUSY embedding must be holomorphic. To construct holomorphic coordinates on the backreacted background (\ref{massive backreacted ansatz}) we proceed as follows. From the K\"ahler form $J$ and the metric in (\ref{backreacted ansatz}) we construct the complex structure and the holomorphic projector
\be
J = \frac12 J_{ab} \, dx^a \wedge dx^b \;,\qquad g = g_{ab} \, dx^a \otimes dx^b \;,\qquad \cJ = J g^{-1} \;,\qquad \cP = \frac{\cJ + i \unit}{2i} \;.
\ee
One can check that given an expression for holomorphic coordinates $z_j(r,\psi,\theta_i, \varphi_i)$ on the usual singular conifold, the substitution $r \to e^\rho$ provides holomorphic coordinates on the backreacted background that satisfy $\cP\, dz_i = dz_i$ and $\cP\, d\bar z_i = 0$.

The embeddings we consider are $z_4 = \mu/2$ and the ones obtained by the action of $SU(2)\times SU(2) \times U(1)$. Let us compute the smeared charge distribution. The symmetries dictate the form of $F_1$ and therefore
\be
\delta^\text{smeared}_2 = {dF_1} = \frac{N_f'(\rho)}{4\pi}\, d\rho \wedge g_5 + \frac{N_f(\rho)}{4\pi} \sum \sin\theta_i\, d\theta_i \wedge d\varphi_i \;.
\ee
To determine the function $N_f(\rho)$, we consider a single localized embedding in the ensamble, \eg{} $z_4 = \mu/2$, and integrate an invariant 4-form, \eg{} $\omega_2 \wedge \omega_2$, on it up to radius $\rho$. We get
\be
\int_\text{D7}^\rho \omega_2 \wedge \omega_2 =
8\pi^2 \big( 1 - 2|\mu|^2 e^{-3\rho} \big) = \int_{\frac13 \log 2|\mu|^2}^\rho 48\pi^2 |\mu|^2 e^{-3\rho} d\rho \;.
\ee
On the other hand, integrating the same 4-form with the charge distribution $\delta_2^\text{smeared}$ we get
\be
\int^\rho \omega_2 \wedge \omega_2 \wedge \delta_2^\text{smeared} = \int^\rho 8\pi^2 N_f'(\rho) d\rho \;.
\ee
Comparing and solving the differential equation (and multiplying by the number $\bar N_f$ of D7-branes) we get
\be
N_f(\rho) = \bar N_f \big( 1 - 2|\mu|^2 e^{-3\rho} \big) \equiv \bar N_f \big( 1- e^{-3(\rho - \rho_0)} \big) \;.
\ee
We defined $\rho_0 = \frac13 \log 2|\mu|^2$, which is the tip of  D7-branes in the  coordinate $\rho$. 

The SUSY equations are the same as before. For dilaton and metric we find
\be
\phi' = \frac{3 N_f(\rho)}{4\pi} e^\phi \; ,\qquad\qquad u'= 3 - 2e^{2u-2g} - \frac{3 N_f(\rho)}{8\pi} e^\phi \; ,\qquad\qquad g' = e^{2u-2g} \;.
\ee
The solution for the dilaton with the boundary condition $\phi(\rho \to 0^-) = +\infty$ is
\be
\label{dilaton backreacted massive}
e^\phi = \frac{4\pi}{ f(\rho)} \;,
\ee
we we introduced the function
\be
f(\rho) = \left\{ \begin{aligned} & \bar N_f \big[ -3\rho + e^{3\rho_0} - e^{-3(\rho - \rho_0)} \big] \qquad && \text{for } \rho_0 \leq \rho < 0 \\ & \bar N_f \big[ -3\rho_0 - 1 + e^{3\rho_0} \big]  = \text{const} \equiv f_0 \qquad && \text{for } \rho \leq \rho_0 \end{aligned} \right.
\ee
Notice that $f(\rho) \geq 0$ for $\rho \leq 0$, and $f(0) = 0$. Moreover $f'(\rho) = -3 N_f(\rho)$, so that $f'(\rho \leq \rho_0) = 0$ and $f(\rho)$ is continuous with its first derivative, while its second derivative jumps. For $e^{-(\rho-\rho_0)} \ll 1$ and $e^{3\rho_0} \ll 1$ we get $f(\rho) \simeq - 3 \bar N_f \rho$.

Also $u$ and $g$ can be analytically solved
\beal
e^{2u} &= \left\{ \begin{aligned}
& c \frac{-6\rho + 2e^{3\rho_0} - 2 e^{-3(\rho-\rho_0)}}{\big[ 1-6\rho+2 e^{3\rho_0} - 4 e^{-3(\rho - \rho_0)} + e^{-6(\rho-\rho_0)} \big]^{2/3}} \, e^{2\rho} \quad\; &\text{for } & \rho_0 \leq \rho < 0 \\
& e^{2\rho} & \text{for } & \rho \leq \rho_0
\end{aligned} \right. \\
e^{2g} &= \left\{ \begin{aligned}
& c \big[ 1-6\rho+2 e^{3\rho_0} - 4 e^{-3(\rho - \rho_0)} + e^{-6(\rho-\rho_0)} \big]^{1/3} e^{2\rho} \qquad &\text{for } & \rho_0 \leq \rho < 0 \\
& e^{2\rho} & \text{for } & \rho \leq \rho_0
\end{aligned} \right.
\eeal
even though we will not need them. We imposed $e^{2u} = e^{2g}$ at $\rho = \rho_0$, whilst there is still one multiplicative integration constant $c$ which should be fixed by continuity. One can check that both functions are positive for $\rho_0 \leq \rho <0$. For $\rho \leq \rho_0$, $u=g=\rho$.

From the SUSY equation $H_3 = e^\phi *_6 F_3$ we get
\be
F_3 = \frac{\alpha'}{12} f(\rho) b'(\rho) \, g_5 \wedge \omega_2 \;.
\ee
Then the Bianchi identity $dF_3 = H_3 \wedge F_1 +  \cF \wedge \delta_2^\text{smeared}$, taking into account that $F = \frac n2 P[\omega_2]$, gives
\be
\frac13 \, (f\, b')' = N_f b' + N_f'(b+n)
\ee
for $\rho_0 \leq \rho < 0$, where $n$ is the number of flux units, and $(f\, b')' = 0$ for $\rho < \rho_0$. The equation can be solved on both sides of $\rho_0$ giving
(here $\Theta$ is the Heaviside step function $\Theta(x)=0$ for $x<0$ and $\Theta(x)=1$ for $x>1$)
\be
b^\pm = c_1^\pm \frac{1}{f(\rho)} + c_2^\pm \frac{\rho}{f(\rho)} - n \Theta(\pm1) \;.
\ee
If we impose continuity of $b$ and $b'$, we get $c_1^+ = c_1^- + n\, f_0$ and $c_2^+ = c_2^-$.
Now we put everything together
\be
\label{B-field backreacted massive}
b(\rho) = \frac{(c_1 + c_2 \rho) f_0 - \big( f(\rho) - f_0 \big)n}{f(\rho)}
\ee

For $\rho \leq \rho_0$, $b(\rho) = c_1 + c_2\rho$ which coincide with the B-field of the KT solution \cite{KT}. Here $c_2 = 6Q_\text{D5}^\text{low}/f_0$ is related to the integer number $Q_\text{D5}^\text{low}$ of fractional D3-branes at the tip, while $c_1$ is a free parameter related to the difference of gauge couplings (which imposes a constraint on the 5-form flux by integrality of the Page $Q_\text{D3}$). For $\rho_0 \leq \rho$ we can compute the Page charges
\be
\label{charges backreacted massive}
Q_\text{D5} = \frac{c_2f_0}6 + \frac n2\, N_f(\rho) \;,\qquad\qquad Q_\text{D3} = Q_\text{D3}(\rho = \rho_0) + \frac{n^2}4 \, N_f(\rho)
\ee
where partial integration and the SUSY equations have been used.

\bibliographystyle{JHEP}
\bibliography{paper}

\end{document}